\newcommand{\beq}{\begin{equation}}
\newcommand{\eeq}{\end{equation}}
\newcommand{\bea}{\begin{eqnarray}}
\newcommand{\eea}{\end{eqnarray}}

\newcommand{\ra}{\rightarrow}

\newcommand{\gsim}{\lower.7ex\hbox{$
\;\stackrel{\textstyle>}{\sim}\;$}}
\newcommand{\lsim}{\lower.7ex\hbox{$
\;\stackrel{\textstyle<}{\sim}\;$}}

\newcommand{\GeV}{\,\mbox{GeV}}
\newcommand{\MeV}{\,\mbox{MeV}}
\newcommand{\matel}[3]{\langle #1|#2|#3\rangle}

\def\op{{\bf P}}
\def\oc{{\bf C}}
\def\ot{{\bf T}}
\def\cp{{\bf CP}}
\def\cpt{{\bf CPT}}

\documentstyle[12pt,epsfig,cite]{article}
\begin{document}

\thispagestyle{empty}
\vspace*{-22mm}

\begin{flushright}
UND-HEP-05-BIG\hspace*{.08em}05\\
LPT-ORSAY 06-06\\
LAL 06-04\\
hep-ph/0601167\\

\end{flushright}
\vspace*{1.3mm}

\begin{center}
{\LARGE{\bf
Flavour Dynamics \& \cp~Violation in the \vspace*{2mm}\\
SM$^*$: A Tale in Five Parts of Great \vspace*{2mm}\\ 
Successes, 
Little Understanding -- \vspace*{2mm}\\ and Promise for the Future!
}}
\vspace*{19mm}

{\Large{\bf I.I.~Bigi}} \\
\vspace{7mm}

{\sl Department of Physics, University of Notre Dame du Lac}
\vspace*{-.8mm}\\
{\sl Notre Dame, IN 46556, USA}\\
{\sl email: ibigi@nd.edu}

\vspace*{10mm}

{\bf Abstract}\vspace*{-1.5mm}\\
\end{center}

\noindent
Our knowledge of flavour dynamics has undergone a `quantum jump' since just before the 
turn of the millenium: direct \cp~violation has been firmly {\em established} in $K_L \to \pi \pi$ 
decays in 1999; the first \cp~asymmetry outside $K_L$ decays has been discovered in 2001 in 
$B_d \to \psi K_S$, followed by $B_d \to \pi^+\pi^-$ and $B \to K^{\pm}\pi^{\mp}$, 
the latter establishing direct \cp~violation also in the beauty sector. Furthermore CKM dynamics 
allows a description of \cp~insensitive and sensitive $B$, $K$ and $D$ transitions that is 
impressively consistent also on the quantitative level. Theories of flavour dynamics that could 
serve as {\em alternatives} to CKM have been ruled out. Yet these novel successes of the Standard Model (SM) do not invalidate any of the theoretical arguments for the incompleteness of the SM. In 
addition we have also more direct evidence for New Physics, namely neutrino oscillations, the 
observed baryon number of the Universe, dark matter and dark energy. While the New Physics anticipated at the TeV scale is not likely to shed any light on the SM's mysteries of flavour, 
detailed and comprehensive studies of heavy flavour transitions will be essential in diagnosing salient 
features of that New Physics. Strategic principles for such studies will be outlined.

\setcounter{page}{0}
\vfill

\tableofcontents

In my lecture series I will sketch the past evolution of central concepts of the Standard 
Model (SM), which are of particular importance for its flavour dynamics. The reason is not primarily of 
a historical nature. I hope these sketches will illuminate the main message I want to convey, namely 
that we find ourselves in the midst of a great intellectual adventure: even with the recent novel successes of the SM the case for New Physics at the TeV and at higher scales  
is as strong as ever; yet at the same time we cannot count on such New Physics having a 
{\em massive} impact on $B$ decays. Furthermore while there is a crowd favourite for the TeV scale 
New Physics, namely some implementation of Supersymmetry (SUSY) -- an expectation I 
happen to share -- we better allow for many diverse scenarios. 
Accordingly I will emphasize general principles for designing search strategies for New Physics over specific and detailed examples. As Prof. Sanda set as an 
essential goal for this school: we want to help you prepare yourself for a future leadership role;  
that requires that you do your own thinking rather than `out-source' it. 

Let me add a personal comment: the lecture room at Villa Monastero -- not surprisingly -- used to be a chapel. This 
should present no problem for experimentalists; after all they talk about empirical facts. Yet for a 
theorist it constitutes a more dangerous environment.  I was thus greatly relieved when Prof. Manelli assured me he does not entertain the illusion 
that a theorist can speak the truth all the time; speaking in good faith is all he expects from a theorist. 
This I can promise. 

The outline of my five lectures is as follows: 
\begin{itemize}
\item 
{\bf Lecture I:  Introduction of the SM$^*$ -- Renormalizability, Neutral Currents, Mass Generation, 
GIM mechanism, \cp~violation a la CKM} 
\item 
{\bf Lecture II: CKM Phenomenology} 
\item 
{\bf Lecture III: \cp~Violation in $B$ Decays -- the `Expected' Triumph of a Peculiar Theory}
\item 
{\bf Lecture IV: Adding High Accuracy to High Sensitivity}
\item 
{\bf Lecture V: Searching for a New Paradigm 2005 \& Beyond Following Sam Beckett's Dictum}

\end{itemize}

\vspace{0.3cm}


\section{Lecture I: Introduction of the SM$^*$ -- Renormalizability, Neutral Currents, Mass Generation, 
GIM mechanism, \cp~violation a la CKM}
\label{LECT1}

\subsection{Introduction}
\label{Intro1}

A famous American Football coach once declared:"Winning is not the greatest thing -- 
it is the only thing!" This quote provides some useful criteria for sketching the status of the 
different components of the Standard Model (SM). It can be characterized by the carriers of its strong and electroweak forces described by {\em gauge} dynamics and the 
{\em mass matrices} for its quarks and leptons 
as follows: 
\beq 
{\rm SM}^* = SU(3)_C \times SU(2)_L \oplus {\rm `CKM'} (\oplus {\rm `PMNS'})  
\eeq
I have attached the asteriks to `SM' to emphasize the SM contains a very peculiar pattern of fermion 
mass parameters that is not illuminated at all by its gauge structure. 
I will address the status of these components in my first lecture. 

\subsection{QCD -- the `Only' Thing}
\label{QCD}

\subsubsection{`Derivation' of QCD}
\label{DERIVATION}

While it is important to subject QCD again and again to quantitative tests as the theory for the strong 
interactions, one should note that these serve more as tests of our theoretical control over 
QCD dynamics than of QCD itself. For its features can be inferred from a few general requirements 
and basic observations. A simplified list reads as follows:  
\begin{itemize}
\item 
Our understanding of chiral symmetry as a {\em spontaneously} realized one -- which allows treating 
pions as Goldstone bosons implying various soft pion theorems -- requires vector couplings for the gluons. 
\item 
The rates for $e^+ e^- \to {\rm had.}$, $\pi ^0 \to \gamma \gamma $ etc. etc. point to the need for three 
colours. 
\item 
Colour has to be implemented as an {\em un}broken symmetry. Local gauge theories are the only 
known way to couple {\em massless spin-one} fields in a {\em Lorentz invariant} way. The basic challenge is easily stated: $4 \neq 2$; i.e., while Lorentz covariance requires four component to describe a spin-one field, the latter contains only two physical degrees of freedom for massless fields. 
\item 
Combining confinement with asymptotic freedom requires a {\em non}-abelian gauge theory. 

\end{itemize}
In summary: for describing the strong interactions QCD is the {\em unique} choice among 
{\em local} quantum field theories. A true failure of QCD would thus create a geneuine paradigm 
shift, for one had to adopt an {\em intrinsically non-}local description. It should be remembered that string theory was first put forward for providing a framework for describing the strong interactions. 

\subsubsection{`Fly-in-the-Ointment': the Strong \cp~Problem of QCD}
\label{FLY}

A theoretical problem arises for QCD from an unexpected quarter that is however very relevant 
for our context here: QCD does {\em not automatically} conserve \op, \ot~ and \cp. To reflect the 
nontrivial topological structure of QCD's ground state one employs an 
{\em effective} Lagrangian containing an additional term to the usual QCD Lagrangian \cite{CPBOOK}: 
\beq 
{\cal L}_{eff} = {\cal L}_{QCD} + \theta \frac{g_S^2}{32\pi^2} G_{\mu \nu} 
\tilde G^{\mu \nu} \; ,  \; \tilde G_{\mu \nu} = \frac{i}{2} \epsilon _{\mu \nu \rho \sigma}G^{\rho \sigma} 
\label{LQCDEFF}
\eeq
Since $G_{\mu \nu} \tilde G^{\mu \nu}$ is a gauge invariant operator, its appearance in general cannot be forbidden, and what is not forbidden has to be considered allowed in a quantum field theory. It represents a total divergence, yet in QCD -- unlike in QED -- it cannot be ignored due 
to the topological structure of the ground state. 

Since under parity $\op$ and time reversal $\ot$ one has 
\beq 
G_{\mu \nu} \tilde G^{\mu \nu} \stackrel{\op, \ot}{\Longrightarrow} - G_{\mu \nu} \tilde G^{\mu \nu} 
\; ;  
\eeq 
i.e., the last term in Eq.(\ref{LQCDEFF}) violates $\op$ as well as $\ot$! Since 
$G_{\mu \nu} \tilde G^{\mu \nu}$ is flavour-{\em diagonal}, it generates an electric dipole moment 
(EDM) for the neutron \cite{SANDALECT}.  From the upper bound on the latter one infers 
 \cite{RAMSEYLECT}
\beq 
\theta < 10^{-9} \; . 
\label{THETABOUND} 
\eeq
Being the coefficient of a dimension-four operator $\theta$ can be renormalized to any value, even zero. Yet the modern view of renomalization is more demanding: requiring the renormalized value to be smaller than its `natural' one by {\em orders of magnitude} is frowned upon, since it requires 
{\em finetuning} between the loop corrections and the counterterms. This is what happens here. For  
purely within QCD the only intrinsically `natural' scale for $\theta$ is unity. If $\theta \sim 0.1$ 
or even $0.01$ were found, one would not be overly concerned.  Yet the bound of 
Eq.(\ref{THETABOUND}) is viewed with great alarm as very {\em unnatural} -- unless a symmetry 
can be called upon. If any quark were massless -- most likely the $u$ quark -- chiral rotations representing symmetry transformations in that case could be employed to remove 
$\theta$ contributions. Yet a considerable phenomenological body rules against such a scenario. 

A much more attractive solution would be provided by transforming $\theta$ from a fixed parameter into the manifestation of a {\em dynamical} field -- as is done 
for gauge and fermion masses through the Higgs-Kibble mechanism, see below -- and imposing a 
Peccei-Quinn symmetry would lead {\em naturally} to $\theta \ll {\cal O}(10^{-9})$. Alas -- this attractive solution does not come `for free': it requires the existence of axions. Those have not been observed despite great efforts to find them. 

This is a purely theoretical problem. Yet I consider the fact that it remains unresolved a significant chink 
in the SM$^*$'s armour. I still have not given up hope that `victory can be snatched from the jaws of defeat': establishing a Peccei-Quinn-type solution would be a major triumph for theory.

\subsubsection{Theoretical Technologies for QCD}
\label{TECH}

True theorists tend to think that by writing down, say, a Lagrangian one has defined a theory. 
Yet to make contact with experiment one needs theoretical technologies to infer observable quantities 
from the Lagrangian. That is the task that engineers and plumbers like me have set for themselves. 
Examples for such technologies are: 
\begin{itemize}
\item 
perturbation theory; 
\item 
chiral perturbation; 
\item 
QCD sum rules; 
\item 
heavy quark expansions (which will be described in some detail in Lecture IV). 

\end{itemize}
Except for the first one they incorporate the treatment of nonperturbative effects. 

None of these can claim universal validity; i.e., they are all `protestant' in nature. There is only 
one `catholic' technology, namely lattice gauge theory 
\footnote{I hasten to add that lattice gauge theory -- while catholic in substance -- exhibits a different sociology: it has not developed an inquisition and deals with heretics in a rather gentle way.}: 
\begin{itemize}
\item 
It can be applied to nonperturbative dynamics in all domains (with the possible {\em practical} 
limitation concerning strong final state interactions). 
\item 
Its theoretical uncertainties can be reduced in a {\em systematic} way. 
\end{itemize}   
Chiral perturbation theory {\em is} QCD at low energies describing the dynamics of soft pions and kaons. The heavy quark expansions treating the nonperturbative effects in heavy flavour decays through an expansion in inverse powers of the heavy quark mass are tailor made  for describing 
$B$ decays; to which degree their application can be extended down to the charm scale is a more iffy 
question, to which I will return in Lecture IV. Different formulations of lattice QCD can approach the 
nonperturbative dynamics at the charm scale from below as well as from above. The degree to which they yield the same results for charm provides an essential cross check on their numerical reliability. 
In that sense the study of charm decays serves as an important bridge between our understanding of nonperturbative effects in heavy and light flavours.

\subsection{$SU(2)_L\times U(1)$ -- not even the Greatest Thing}
\label{SU(2)}

\subsubsection{Prehistory}
\label{PRE}

It was recognized from early on that the four-fermion-coupling of Fermi's theory for the weak forces  
yields an {\em effective} description only that cannot be extended to very high energies. The lowest order 
contribution violates unitarity around 250 GeV. Higher order contributions cannot be called upon to 
remedy the situation, since due to the theory being non-renormalizable those come with more 
and more untamable infinities. Introducing massive charged vector bosons softens the problem, yet does not solve it. Consider the propagator for a massive spin-one 
boson carrying momentum $k$: 
\beq 
\frac{- g_{\mu \nu} + \frac{k_{\mu}k_{\nu}}{M_W^2}}{k^2 - M_W^2}
\eeq
The second term in the numerator has great potential to cause trouble. For it can act like 
a coupling term with dimension $1/({\rm mass})^2$; this is quite analogous to the original ansatz of Fermi's theory and amounts to a non-renormalizable coupling. It is actually the {\em longitudinal} 
component of the vector boson that is at the bottom of this problem. 

This potential problem is neutralized, if these massive vector bosons couple to conserved currents. 
To guarantee the latter property, one needs a non-abelian gauge theory, which implies the 
existence of neutral weak currents.

\subsubsection{Strong Points}
\label{STRONG}

The requirements of unitarity, which is nonnegotiable, and of renormalazibility, which is to some degree, severely restrict  possible theories of the electroweak interactions. It makes the generation of mass a highly nontrivial one, as sketched below. There are other strong points as well: 

\noindent 
$\oplus$ Since there is a {\em single} $SU(2)_L$ group, there is a single set of gauge bosons. 
Their {\em self}-coupling controls also, how they couple to the fermion fields. As explain later in more detail, this implies the property of `weak universality'. 

\noindent $\oplus$ The SM truly {\em pre}dicted the existence of neutral currents characterized by one 
parameter, the weak angle $\theta_W$ and the masses of the $W$ and $Z$ bosons. 

\noindent 
$\oplus$: Most remarkably the $SU(2)_L\times U(1)$ gauge theory combines QED 
with a pure parity conserving vector coupling to a massless neutral force field with the weak interactions, 
where the charged currents exhibit {\em maximal} parity violation due to their $V-A$ coupling and a 
very short range due to $M_Z > M_W \gg m_{\pi}$.

\subsubsection{Generating Mass}
\label{MASSGEN}

A massive spin-one field with momentum $k_{\mu}$ and spin $s_{\mu}$ 
has four (Lorentz) components. Going into its rest frame one realizes that the Lorentz invariant constraint $k\cdot s = 0$ can be imposed, which leaves three independent components, as it has to be. 

A massless spin-one field is still described by four components, yet has only two physical degrees 
of freedom. It needs another physical degree of freedom to transmogrify itself into a massive field. 
This is achieved by having the gauge symmetry {\em realized spontaneously}. For the case at hand this 
is implemented through  an ansatz that should be -- although rarely is -- referred to as 
Higgs-Brout-Englert-Guralnik-Hagen-Kibble mechanism (HBEGHK). Consider as simplest case a complex 
scalar field $\phi$ with a potential invariant under $\phi (x) \to e^{i\alpha (x)}\phi (x)$: 
\beq 
V(\phi) = \lambda |\phi|^4  - \frac{m^2}{2} |\phi|^2 
\eeq
Its minimum is obviously not at $|\phi| = 0$, but at $\sqrt{m^2/4\lambda}$. Thus rather than having 
a {\em unique} ground state with $|\phi|=0$ one has an 
{\em infinity of different, yet equivalent} ground states with $|\phi| = \sqrt{m^2/4\lambda}$.  To understand the physical content of such a scenario, one 
considers oscillations of the field around the minimum of the potential: oscillations in the radial direction 
of the field $\phi$ represent a scalar particle with mass; in the polar direction (i.e. the phase of 
$\phi$) the potential is at its minimum, i.e. flat, and 
the corresponding field component constitutes a {\em massless} field. 

It turns out that this massless scalar field can be combined with the two transverse components 
of a $M=0$ spin-one gauge field to take on the role of the latter's longitudinal component leading to the 
emergence of a {\em massive} spin-one field. Its mass is thus controlled by the 
nonperturbative quantity $\langle 0|\phi|0\rangle$. 

Applying this generic construction to the SM one finds that a priori both $SU(2)_L$ doublet and triplet 
Higgs fields could generate masses for the weak vector bosons. The ratio {\em observed} for the 
$W$ and $Z$ masses is fully consistent with only doublets contributing. Intriguingly enough such 
doublet fields can eo ipso generate fermion masses as well. 

In the SM one adds a single complex scalar doublet field to the mix of vector boson and fermion 
fields. Three of its four components slip into the role of the longitudinal components of 
$W^{\pm}$ and $Z^0$; the fourth one emerges as an independent physical field -- `the' Higgs 
field. Fermion masses are then given by the product of the single vacuum expectation value 
(VEV) $\langle 0|\phi|0\rangle$ and their Yukawa couplings -- a point we will return to soon.

\subsubsection{Triangle or ABJ Anomaly}
\label{ABJ}

The diagram with an internal loop of only fermion lines, to which three external axial vector 
(or one axial vector and two vector) lines are attached, generates a `quantum anomaly' 
\footnote{It is referred to as `triangle' anomaly due to the form of the underlying diagram or 
A(dler)B(ell)J(ackiw) anomaly due to the authors that identified it \cite{ABJ}.}: it 
removes a {\em classical} symmetry as expressed through the existence of a conserved current. 
In this specific case it affects the conservation of the axialvector current $J_{\mu}^5$. Classically 
we have $\partial ^{\mu}J^5_{\mu} = 0$ for {\em massless} fermions; yet the triangle anomaly leads to 
\beq 
\partial ^{\mu}J^5_{\mu} =  \frac{g_S^2}{16\pi ^2} G \cdot \tilde G  \neq 0
\label{TRIANOM}
\eeq
even for massless fermions; $G$ and $\tilde G$ denote the gluonic field strength tensor and its 
dual, respectively, as introduced in Eq.(\ref{LQCDEFF}). 

While by itself it yields a finite result on the right hand side of Eq.(\ref{TRIANOM}), it destroys the renormalizability of the theory. It cannot be 
`renormalized away' (since in four dimensions it cannot be regularized in a gauge invariant way). 
Instead it has to be neutralized by requiring that adding up this contribution from all types of fermions in the theory yields a vanishing result. 

For the SM this requirement can be expressed very concisely that all electric charges of the fermions 
of a given family have to add up to zero.  This imposes a connection between the charges of quarks 
and leptons, yet does not explain it.

\subsubsection{Theoretical Deficiencies}
\label{DEFEC}

With all the impressive, even  amazing successes of the SM, it is natural to ask why is the community 
not happy with it. There are several drawbacks: 

\noindent 
$\ominus$ Since the gauge group is $SU(2)_L \times U(1)$, only partial unification has been 
achieved. 

\noindent $\ominus$ 
The HBEGHK mechanism is viewed as providing merely an `engineering' solution, in particular since the physical Higgs field has not been observed yet. Even if or when it is, theorists in particular will 
not feel relieved, since scalar dynamics induce {\em quadratic} mass renormalization and are viewed 
as highly `unnatural', as exemplified through the gauge hierarchy problem. This concern has 
lead to the conjecture of New Physics entering around the TeV scale, which has 
provided the justification for the LHC and the motivation for the ILC. 

\noindent $\ominus$
{\em maximal} violation of parity is implemented for the charged weak currents `par ordre du mufti'
\footnote{A French saying describing a situation, where a decision is imposed on someone with no 
explanation and no right of appeal.}, i.e. 
based on the data with no deeper understanding. 

\noindent $\ominus$
Likewise neutrino masses had been set to zero  `par ordre du mufti'.

\noindent $\ominus$
The observed quantization of electric charge is easily implemented and is instrumental in neutralizing 
the triangle anomaly  -- yet there is no understanding of it. 

One might say these deficiencies are merely `warts' that hardly detract from the beauty of the SM.  
Alas -- there is the whole issue of family replication!

\subsection{CKM -- an `Accidental' Miracle}
\label{MIRACLE}

The twelve known quarks and leptons are arranged into three families. Those families possess identical gauge couplings and are distinguished only by their mass terms, i.e. their Yukawa couplings. 
We do not understand this family replication or why there are three families. It is not even clear 
whether the number of families represents a fundamental quantity  or is due to the more or less accidental interplay of complex forces as one encounters when analyzing the structure of nuclei. 
The only hope for a theoretical understanding we can spot on the horizon is superstring or 
M theory -- which is merely a euphemistic way of saying we have no clue. 

Yet the circumstantial evidence that we miss completely a central element of Nature's `Grand Design' is even stronger. 

\subsubsection{Quark Masses, the GIM Mechanism \& \cp}
\label{GIM}

Let us consider the mass terms for the up- and down-type quarks as expressed through 
matrices ${\cal M}_{U/D}$ and vectors of quark fields $U^F= (u,c,t)^F$ and $D^F=(d,s,b)^F$ 
in terms of the {\em flavour} eigenstates denoted 
by the superscript $F$: 
\beq 
{\cal L}_M \propto \bar U_L^F{\cal M}_U U_R^F + \bar D_L^F{\cal M}_D D_R^F  \; . 
\label{MASSLAG}
\eeq
 A priori there is no reason 
why the matrices ${\cal M}_{U/D}$ should be diagonal. Yet applying bi-unitary rotations 
${\cal J}_{U/D,L}$ will 
allow to diagonalize them
\beq 
{\cal M}_{U/D}^{\rm diag} =     {\cal J}_{U/D,L} {\cal M}_{U,D}{\cal J}_{U/D,R}^{\dagger}
\eeq
and obtain the {\em mass} eigenstates of the quark fields: 
\beq 
U^m_{L/R} = {\cal J}_{U,L/R} U_{L/R}^F \; , \; \; D^m_{L/R} = {\cal J}_{D,L/R} D_{L/R}^F
\eeq
The eigenvalues of ${\cal M}_{U/D}$ represent the masses of the quark fields. The measured 
values exhibit a very peculiar pattern that hardly appears to be arbitrary being so hierarchical 
for up- and down-type quarks, charged and neutral leptons. 

Yet again, there is much more to it. Consider the neutral current coupling
\beq 
{\cal L}_{NC}^{U[D]} \propto \bar g_Z \bar U^F[\bar D^F] \gamma _{\mu} U^F[D^F] Z^{\mu} 
\label{LAGNC}
\eeq
It keeps its form when expressed in terms of the mass eigenstates
\beq 
{\cal L}_{NC}^{U[D]} \propto \bar g_Z \bar U^m[\bar D^m] \gamma _{\mu} U^m[D^m] Z^{\mu} \; ; 
\label{LAGNC2} 
\eeq 
i.e., there are {\em no} flavour changing neutral currents. This important property is referred to 
as the `generalized' GIM mechanism \cite{GIM}. 

However for the charged currents the situation is quite different: 
\beq 
{\cal L}_{CC} \propto \bar g_W \bar U^F_L \gamma _{\mu} D^F W^{\mu} = 
\bar g_W \bar U^m_L \gamma _{\mu} V_{CKM}D^mW^{\mu}
\label{LAGCC}
\eeq
with 
\beq 
V_{CKM} = {\cal J}_{U,L}{\cal J}_{D,L}^{\dagger} 
\eeq
There is no reason, why the matrix $V_{CKM}$ should be the identity matrix or even 
diagonal \footnote{Even if some speculative dynamics were to enforce an alignment 
between the $U$ and $D$ quark fields at some high scales causing their mass matrices 
to get diagonalized by the same bi-unitary transformation, this alignment would be likely 
to get upset by renormalization down to the electroweak scales.}. It means the charged 
current couplings of the mass eigenstates will be modified in an observable way. In which way and by how much this happens requires further analysis since the phases of fermion fields are not necessarily 
observables. Such an analysis was first given by Kobayashi and Maskawa \cite{KM}. 

Consider $N$ families. $V_{CKM}$ then represents an $N \times N$ matrix that has to be unitary 
based on two facts: 
\begin{itemize}
\item 
The transformations ${\cal J}_{U/D,L/R}$ are unitary by construction. 
\item 
As long as the carriers of the weak force are described by a {\em single} local gauge group -- 
$SU(2)_L$ in this case -- they have to couple to all other fields in a way fixed by their 
{\em self}coupling. This was already implied by Eq.(\ref{LAGCC}), when writing the weak coupling 
$\bar g_W$ as an overall factor.  

\end{itemize} 
The unitarity of $V_{CKM}$ implies {\em weak universality}, as addressed later in more detail. 
There are actually $N$ such relations characterized by 
\beq 
\sum _j |V(ij)|^2 = 1\; , \; \; i = 1, ..., N 
\label{WU}
\eeq
These relations are important, yet insensitive to weak phases; thus they provide no {\em direct} 
information on \cp~violation.  

{\em Violations} of weak universality can be implemented by adding dynamical layers to the SM. 
So-called horizontal gauge interactions, which differentiate between families and induce 
flavour-changing neutral currents, will do it. Another admittedly ad-hoc possibility is to introduce 
a separate $SU(2)_L$ group for each quark family while allowing the gauge bosons from the 
different $SU(2)_L$ groups to mix with each other. This mixing can be set up in such a way that the lightest 
mass eigenstates couple to all fermions with approximately universal strength. Weak universality 
thus emerges as an approximate symmetry. Flavour changing neutral currents are again induced, and they can generate electric dipole moments. 

After this aside on weak universality let us return to  $V_{CKM}$. There are $N^2 - N$ 
{\em orthogonality} relations:
\beq 
\sum _j V^*(ij)V(jk) = 0 \; , \; \; i\neq k 
\eeq
Those are very sensitive to complex phases and tell us {\em directly} about \cp~violation. 

An $N \times N$ complex matrix contains $2N^2$ real parameters; the unitarity constraints 
reduce it to $N^2$ independent real parameters. Since the phases of quark fields like other 
{\em fermion} fields can be rotated freely, $2N-1$ phases can be removed from 
${\cal L}_{CC}$ (a {\em global} phase rotation of all quark fields has no impact on 
${\cal L}_{CC}$). Thus we have $(N-1)^2$ {\em independent physical} parameters. 
Since an $N\times N$ {\em orthogonal} matrix has $N(N-1)/2$ angles, we conclude that 
an $N\times N$ {\em unitary} matrix contains also $(N-1)(N-2)/2$ {\em physical phases}. This 
was the general argument given by Kobayashi and Maskawa. Accordingly: 
\begin{itemize}
\item 
For $N=2$ families we have one angle -- the Cabibbo angle -- and zero phases. 
\item 
For $N=3$ families we obtain three angles and one irreducible phase; i.e. a three family 
ansatz can support \cp~violation with a single source -- the `CKM phase'. 
PDG suggests a "canonical" parametrization for the $3\times 3$ CKM 
matrix: 
$$ 
{\bf V}_{CKM} = 
\left(  
\begin{array}{ccc} 
V(ud) & V(us) & V(ub) \\
V(cd) & V(cs) & V(cb) \\
V(td) & V(ts) & V(tb) 
\end{array} 
\right) 
$$
\beq 
= \left( 
\begin{array}{ccc} 
c_{12}c_{13} & s_{12}c_{13} & s_{13}e^{-i \delta _{13}}  \\
- s_{12}c_{23} - c_{12}s_{23}s_{13}e^{i \delta _{13}} &
c_{12}c_{23} - s_{12}s_{23}s_{13}e^{i \delta _{13}} & 
c_{13}s_{23} \\
s_{12}s_{23} - c_{12}c_{23}s_{13}e^{i \delta _{13}} &
- c_{12}s_{23} - s_{12}c_{23}s_{13}e^{i \delta _{13}} &
c_{13}c_{23} 
\end{array}
\right) 
\label{PDGKM} 
\eeq 
where 
\beq 
c_{ij} \equiv {\rm cos} \theta _{ij} \; \; , \; \;  
s_{ij} \equiv {\rm sin} \theta _{ij}
\eeq  
with $i,j = 1,2,3$ being generation labels. 

This is a completely general, yet not unique parametrisation: a 
different set of 
Euler angles could be chosen; the phases can be shifted around 
among the matrix elements 
by using a different phase convention. 
\item 
For even more families we encounter a proliferation of angles and phases, namely six angles 
and three phases for $N=4$. 

\end{itemize}
These results obtain by algebraic means can be visualized graphically: 
\begin{itemize}
\item 
For $N=2$ we have two weak universality conditions and two orthogonality relations: 
\bea 
\nonumber 
V^*(ud)V(us) + V^*(cd)V(cs) &=& 0 \\
V^*(us)V(ud) + V^*(cs)V(cd) &=& 0
\eea
While the CKM angles can be complex, there can be no nontrivial phase 
($\neq 0,\pi$) between their observable combinations; i.e., there can be no 
\cp~violation for two families in the SM. 
\item 
For three families  the orthogonality relations read 
\beq 
\sum _{j=1}^{j=3} V^*(ij)V(jk) = 0 \; , \; \; i\neq k
\eeq
There are six such relations, and they represent triangles in the complex plane with in general 
nontrivial relative angles. 
\item 
While these six triangles can and will have quite different shapes, as we will describe later 
in detail, they all have to possess the {\em same area}, namely 
\bea
\nonumber  
{\rm area(every\; triangle)} &=& \frac{1}{2}J \\
J =  {\rm Im}[V(ud)V(cs)V^*(us)V^*(cd)]   
\label{JARL}
\eea
{\em If} $J=0$, one has obviously no nontrivial angles, and there is no \cp~violation. The 
fact that all triangles have to possess the same area reflects the fact that for three families 
there is but a {\em single} CKM phase. 
\item 
Only the angles, i.e. the relative phases matter, but not the overall orientation of the triangles 
in the complex plane. That orientation merely reflects the phase convention for the quark fields. 

\end{itemize}
{\em If} any pair of up-type or down-type quarks were mass {\em degenerate}, then 
{\em any} linear combination of those two would be a mass eigenstate as well. Forming 
different linear combinations thus represents symmetry transformations, and with this 
{\em additional} symmetry one can further reduce the number of physical parameters. 
For $N=3$ it means \cp~violation could still not occur. 

The CKM implementation of \cp~violation depends on the form of the quark mass matrices 
${\cal M}_{U,D}$, not so much on how those are generated. Nevertheless something can be inferred 
about the latter: within the SM all fermion masses are driven by a {\em single} VEV; to obtain an 
irreducible relative phase between different quark couplings thus requires such a phase in 
quark Yukawa couplings; this means that in the SM \cp~violation arises in dimension-{\em four} 
couplings, i.e., is `hard'. 

\subsubsection{`Maximal' \cp~Violation?}
\label{MAXCPV}

As already mentioned charged current couplings with their $V-A$ structure break parity and 
charge conjugation {\em maximally}. Since due to \cpt~invariance \cp~violation is expressed 
through couplings with complex phases, one might say that maximal \cp~violation 
is characterized by complex phases of $90^o$. However this would be fallacious: for by changing 
the phase {\em convention} for the quark fields one can change the phase of a given 
CKM matrix element and even rotate it away; it will of course re-appear in other matrix elements. 
For example $|s\rangle \to e^{i\delta _s}|s\rangle$ leads to $V_{qs} \to  e^{i\delta _s}V_{qs}$ 
with $q=u,c,t$. In that sense the CKM phase is like the `Scarlet Pimpernel': "Sometimes here, 
sometimes there, sometimes everywhere." 

One can actually illustrate with a general argument, why there can be no straightforward definition for maximal \cp~violation. Consider neutrinos: Maximal \cp~violation means there are 
$\nu _L$ and $\bar \nu _R$, yet no $\nu _R$ or $\bar \nu _L$  
\footnote{To be more precise: $\nu_L$ and $\bar \nu _R$ couple to weak gauge bosons, 
$\nu _R$ or $\bar \nu _L$ do not.}. Likewise there are $\nu_L$ and $\bar \nu_R$, but 
not $\bar \nu _L$ or $\nu _R$. One might then suggest that maximal \cp~violation means 
that $\nu_L$ exists, but $\bar \nu _R$ does not. Alas -- \cpt~invariance already enforces the existence 
of both. 

Similarly -- and maybe more obviously -- it is not clear what maximal \ot~violation would mean although 
some formulations have entered daily language like the `no future generation'  and the `woman 
without a past'. 

\subsubsection{Some Historical Remarks}
\label{HISTREM}

\cp~violation was discovered in 1964 through the observation of $K_L \to \pi^+\pi^-$, yet it was not realized for a number of years that dynamics known at {\em that} time could {\em not} generate it. 
We should not be too harsh on our predecessors for that oversight: as long as one did not have a 
renormalizable theory for the weak interactions and thus had to worry about {\em infinities} 
in the calculated rates, one can be excused for ignoring a seemingly marginal rate with a branching 
ratio of $2\cdot 10^{-3}$. Yet even after the emergence of the renormalizable 
Glashow-Salam-Weinberg model its {\em phenomenological} incompleteness was not recognized right 
away. There is a short remark by Mohapatra in a 1972 paper invoking the need for right handed 
currents to induce \cp~violation. 

It was the 1973 paper by Kobayashi and Maskawa \cite{KM} that fully stated the inability of even a 
two-family SM to produce \cp~violation and that explained what had to be added to it: right-handed 
charged currents, extra Higgs doublets -- or (at least) a third quark family. Of the three options 
Kobayashi and Maskawa listed, their name has been attached only to the last one as the CKM 
description. 

As pointed out by Sanda being at Nagoya University at that time gave Kobayashi and Maskawa a 
`competitive edge' through `insider information'. Physicists at most other places analyzed nature 
in terms of three quarks only -- $u$, $d$ and $s$ -- (if at all), i.e. without the benefit of having two complete families; they also tended to view quarks as convenient mathematical entities rather than 
physical objects. It might be hard to grasp this in retrospect. For it appears straightforward now that 
charm quarks had to exist. Embedding weak charged currents with their Cabibbo couplings 
\bea 
\nonumber 
J^{(+)}_{\mu} &=& {\rm cos}\theta _C \bar d_L \gamma _{\mu}u_L + 
{\rm sin}\theta _C \bar s_L \gamma _{\mu}u_L \\
J^{(-)}_{\mu} &=& {\rm cos}\theta _C \bar u_L \gamma _{\mu}d_L + 
{\rm sin}\theta _C \bar u_L \gamma _{\mu}s_L
\label{CABCUR}
\eea   
into an $SU(2)$ gauge theory to arrive at a renormalizable theory requires neutral currents of a structure as obtained from the commutator of $J^{(+)}_{\mu}$ and $J^{(-)}_{\mu}$. Using for 
the latter the expressions of Eq.(\ref{CABCUR}) one arrives unequivocally at 
\beq 
J_{\mu}^{(0)} = ... + {\rm cos}\theta_C {\rm sin}\theta _C (\bar s_L \gamma _{\mu} d_L + 
\bar d_L \gamma _{\mu} s_L) \; , 
\label{SChNC}
\eeq 
i.e., strangeness changing neutral currents. Yet their Cabibbo suppression is not remotely sufficient 
to make them compatible with the observed super-tiny branching ratios for $K_L \to \mu^+\mu^-$, 
$2 \gamma$ etc. 
\footnote{The observed huge suppression of strangeness changing neutral currents actually 
led to some speculation that also flavour {\em conserving} neutral currents are greatly suppressed.}.  
The huge discrepancy between observed and expected branching ratios lead 
some daring spirits \cite{GIM} to postulate a fourth quark with quite specific properties to complete 
a second quark family in such a way that no strangeness changing neutral currents arise at 
{\em tree} level. Yet I remember there was great skepticism felt in the community maybe best expressed 
by the quote: "Nature is smarter than Shelley (Glashow) -- she can do without charm quarks." 
\footnote{The fact that nature needed charm after all does not prove the inverse of this quote, 
of course.} These remarks can indicate how profound a shift in paradigm were begun by the 
observation of scaling in deep inelastic lepton-nucleon scattering and completed by the discovery of 
the $J/\psi$ in 1974 and its immediate aftermatch.  

The `genius loci' of Nagoya University anticipated these developments: 
\begin{itemize}
\item 
Since it was the home of the Sakata school and the Sakata model of elementary particles quarks 
were viewed as physical degrees of freedom from the start. 
\item 
It was also the home of Prof. Niu who in 1971 had observed 
\cite{NIU} a candidate for a charm decay in emulsion 
exposed to cosmic rays and actually recognized it as such. The existence of charm, its association 
with strangeness and thus of two complete quark families were thus taken for granted at Nagoya. 

\end{itemize}

\subsection{Summary of Lecture I}
\label{SUM1}

The SM 
\beq  
{\rm SM}^* = SU(3)_C \times SU(2)_L \times U(1) \oplus `CKM' \oplus `PMNS'
\eeq
is based on two pillars: 
\begin{itemize}
\item 
The observed forces are described by gauge bosons of $SU(3)_C \times SU(2)_L \times U(1)$. 
\begin{itemize}
\item 
$SU(3)_C$ is the only possible solution for the strong interactions among {\em local} quantum 
field theories due to very general considerations. If it ever were to fail, one had to adopt an 
intrinsically {\em non-}local description. It is at least amusing to remember that string theory was initially 
conceived as a theory for the strong interactions.  
\item 
The $SU(2)_L \times U(1)$ gauge structure is inferred from general considerations like renormalizability and from the data with some `theoretical engineering' required for generating 
masses for the gauge bosons -- and at least a whiff of {\em incompleteness}.
\end{itemize}
\item 
Matter fields -- quarks (\& leptons) --  obtain their masses from Yukawa couplings to a 
$SU(2)$ doublet scalar field that generates masses also for the gauge bosons; their charged 
current couplings are described in terms of the CKM (\& PMNS) matrices, the origins of which lie 
in the quark (\& lepton) mass matrices, which are a priori non-diagonal. 
\begin{itemize}
\item 
Its status can be described by saying: "All it does, it works." I.e., it describes the diverse body of electroweak data 
amazingly well, as we will discuss in the next lecture and 
\item 
it achieves this phenomenological success for no understood reason, 
\item 
yet we have reason to believe such deeper reason has to exist. 

\end{itemize}
It is for this lack of understanding of some conjectured deeper reason that I attach an 
asteriks * to SM. 
\end{itemize}

 

\section{Lecture II: CKM Phenomenology}
\label{LECT2}

\subsection{The Phenomenological Landscape through 1998}
\label{PHENO99}

By 1998 we had observed a multifaceted phenomenological landscape in flavour dynamics 
\cite{SOZZI}: 
\begin{itemize}
\item 
The `$\theta - \tau$ puzzle' -- i.e., the observation that two particles decaying into final states of 
opposite parity: $\theta \to 2 \pi$, $\tau \to 3 \pi$, exhibited the same mass and lifetime -- lead to the realization that parity was violated in weak interactions, and actually to a maximal degree in charged 
currents. This can be implemented in a gauge theory through $V-A$ currents. 
\item 
The observation that the production rate of strange hadrons exceeded their decay rates by many 
orders of magnitude -- a feature that gave rise to the term `strangeness' -- was attributed to 
`associate production' meaning the strong and electromagnetic forces conserve this new 
quantum number `strangeness', while weak dynamics do not. Subsequently it gave rise to the notion of quark families. Another aspect of this notion was Cabibbo universality as already explained in 
Lecture I: $|V(ud)|^2 + |V(us)|^2 = 1$. 
\item 
The observation of $\Gamma (K^+ \to \pi^+\pi^0) \ll \Gamma (K_S \to \pi^+\pi^-)$ was attributed to a 
$\Delta I=1/2$ rule {\em postulating} that in strange decays the amplitude for $\Delta I = 1/2$ transitions are 
enhanced relative to those for $\Delta I = 3/2$ by a factor of about twenty. While various enhancements 
have been found, the full strength of this effect has not been derived from QCD yet. 
\item 
$K^0 - \bar K^0$ oscillations including \cp~violation had been analyzed and characterized by the 
observables $\Delta \Gamma _K$, $\Delta M_K$ and $\epsilon_K$. Likewise $B_d - \bar B_d$ 
oscillations had been discovered and $\Delta M(B_d)$ been measured with good accuracy. 
\item 
The existence of charm introduced specifically to produce the observed suppression of 
strangeness changing neutral currents in $K_L \to \mu^+\mu^-$, $\gamma \gamma$ 
and $K^0 - \bar K^0$ oscillations  had been established with all its predicted properties. 
\item 
Beauty, the existence of which had been telegraphed by the discovery of the $\tau$ as the third 
charged lepton 
was indeed observed exhibiting a surprising feature, namely its `long' lifetime of about 
$10^{-12}$ sec.  One gets a rough estimate for $\tau (B)$ by relating it to the muon lifetime: 
\beq 
\tau (B) \simeq \tau _b \sim \tau (\mu ) \left( \frac{m(\mu )}{m(b)}  \right) ^5 \frac{1}{9} 
\frac{1}{|V(cb)|^2} \simeq 3 \cdot 10^{-14} \left| \frac{{\rm sin}\theta_C}{V(cb)}  \right|^2 \; {\rm sec}
\eeq
One had expected $|V(cb)|$ to be suppressed, since it represents an out-of-family coupling. 
Yet  one had assumed without deeper reflection that $|V(cb)| \sim {\rm sin}\theta_C$ -- what 
else could it be? The measured value for $\tau (B)$ however pointed to  
$|V(cb)| \sim |{\rm sin}\theta_C|^2$.  

\item 
That top quarks are unusually heavy -- i.e. heavier than even the weak vector bosons -- was first 
inferred from the speed of $B_d - \bar B_d$ oscillations and later -- with a higher degree of 
accuracy -- from radiative corrections to $Z^0$ decays at LEP. In the early nineties they were 
finally observed directly. The most recent data yield 
\beq 
m_t = 172.7 \pm 2.9 \; {\rm GeV}
\eeq

\end{itemize}

\subsection{On the Theoretical Technologies}
\label{THTECH}

Before quantifying the preceding remarks I want to introduce some of 
the central theoretical technologies that are employed. 

\subsubsection{Electroweak Dynamics}
\label{EWD}

Electroweak forces can be dealt with perturbatively. Consider the $\Delta S =1$ 
four-fermion  transition operator: $(\bar u_L\gamma ^{\mu}s_L)(\bar d_L \gamma _{\mu}u_L)$. 
It constitutes a dimension-{\em six} operator. Yet placing such an operator -- or any other operator 
with dimension larger than four -- into the Lagrangian creates {\em non}renormalizable 
interactions. What happened is that we have started out from a renormalizable Lagrangian 
\beq 
{\cal L}_{CC} = g_W \bar q^{(i)}_L\gamma_{\mu}q_L^{(j)}W^{\mu} \; , 
\label{LCCEX}
\eeq
iterated it to second order in $g_W$ with $(q^{(i)},q^{(j)}) = (u,s) \& (u,d)$ 
and then `integrated out' the heavy field, namely in this case the vector boson field $W^{\mu }$. 
That way one arrives at an effective Lagrangian containing only light quarks as `active' fields. 

Such effective field theories have experienced a veritable renaissance in the last ten years. 
Constructing them in a self-consistent way is greatly helped by adopting a Wilsonian prescription: 
\begin{itemize}
\item 
First one defines a field theory ${\cal L}(\Lambda _{UV})$ at a high ultraviolet scale 
$\Lambda _{UV} \gg$ germane scales of theory like $M_W$, $m_Q$ etc. 
\item 
For applications characterized by physical scales $\Lambda_{phys}$ one renormalizes the theory from 
the cutoff $\Lambda _{UV}$ down to $\Lambda_{phys}$. In doing so one integrates out the 
{\em heavy} degrees of freedom, i.e. with masses exceeding $\Lambda_{phys}$ -- like 
$M_W$ -- to arrive at an 
{\em effective low energy} field theory using the operator product expansion (OPE) as a tool: 
\beq 
{\cal L}(\Lambda _{UV})  \Rightarrow {\cal L}(\Lambda_{phys} ) = 
\sum _i c_i(\Lambda_{phys} , \Lambda _{UV}, M_W, ...) 
{\cal O}_i (\Lambda_{phys} )
\eeq
\begin{itemize}
\item 
The {\em local} operators ${\cal O}_i (\Lambda_{phys} )$ contain the {\em active} dynamical 
fields, i.e. those with frequencies below ${\cal O}_i (\Lambda_{phys} )$.

\item 
Their c number coefficients $c_i(\Lambda_{phys} , \Lambda _{UV}, M_W, ...)$ provide the gateway 
for heavy degrees of freedom with frequencies exceeding ${\cal O}_i (\Lambda_{phys} )$ to enter. 
They are shaped by short-distance dynamics and therefore usually computed perturbatively.

\end{itemize}
\item 
Lowering the value of ${\cal O}_i (\Lambda_{phys} )$ in general changes the form of the Lagrangian: 
${\cal L}(\Lambda ^{(1)}_{phys}) \neq {\cal L}(\Lambda ^{(2)}_{phys}) $ for 
$\Lambda ^{(1)}_{phys} \neq \Lambda ^{(2)}_{phys}$. In particular integrating out heavy degrees of 
freedom will induce higher-dimensional operators to emerge in the Lagrangian.  In the example 
above integrating the $W$ field from the dimension-four term in 
Eq.(\ref{LCCEX}) produces dimension six four-quark operators. 
\item 
As a matter of principle observables cannot depend on the choice of $\Lambda_{phys}$; the latter
primarily provides just a demarkation line: 
\beq 
{\rm short \; distances} < 1/\Lambda_{phys} < {\rm long \; distances} 
\eeq
\end{itemize}
In practice, however, its value must be chosen judiciously due to limitations of our 
(present) computational abilities: on one hand we want to be able to calculate radiative 
corrections perturbatively and thus require $\alpha_S (\Lambda_{phys}) < 1$. Taken by itself it would 
suggest to choose $ \Lambda_{phys}$ as large as possible. Yet on the other hand 
we have to evaluate hadronic matrix elements; there $\Lambda_{phys}$ can provide an UV cutoff on 
the momenta of the hadronic constituents. Since the tails of hadronic wave functions cannot be 
obtained from, say, quark models in a reliable way, one wants to pick 
$\Lambda_{phys}$ as low as possible. More specifically for heavy flavour hadrons one can 
expand their matrix elements in powers of $\Lambda_{phys}/m_Q$. Thus one encounters 
a Scylla \& Charybdis situation. A reasonable middle course can be steered by picking 
$\Lambda_{phys} \sim 1$ GeV, and hence I will denote this quantity and this value by 
$\mu$. 

Some concrete examples might illuminate these remarks. Consider 
$K^0 - \bar K^0$ oscillations, which represent $\Delta S=2$ transitions. As explained in more detail in 
Prof. Sanda's lectures, those are driven by the off-diagonal elements of a `generalized mass matrix': 
\beq 
{\cal M}_{12} = M_{12} + \frac{i}{2}{ \Gamma}_{12} = 
\matel{K^0}{{\cal L}_{eff}(\Delta S =2)}{\bar K^0}
\eeq
The observables $\Delta M_K$ and $\epsilon_K$ are given in terms of Re$M_{12}$ and 
Im$M_{12}$, respectively. 
In the SM ${\cal L}_{eff} (\Delta S=2)$, which generates $M_{12}$, is produced by iterating 
two $\Delta S=1$ operators: 
\beq 
{\cal L}_{eff} (\Delta S=2) = {\cal L}(\Delta S=1) \otimes 
{\cal L}(\Delta S=1) 
\eeq  
This leads to the well known 
quark box diagrams, which generate a {\em local} $\Delta S=2$ operator. The contributions that do 
{\em not} depend on the mass of the internal quarks cancel against 
each other due to the GIM mechanism. Integrating over the internal 
fields, namely the $W$ bosons and the top and charm quarks 
\footnote{The up quarks act merely as a subtraction term here.} 
then yields a convergent result: 
$$  
{\cal L}_{eff}^{box}(\Delta S=2, \mu ) = 
\left( \frac{G_F}{4\pi }\right) ^2 \cdot 
$$ 
\beq  
\left[  \xi _c^2 E(x_c) \eta _{cc} + 
\xi _t^2 E(x_t) \eta _{tt} + 
2\xi _c \xi _t E(x_c, x_t) \eta _{ct}
 \right]  [\alpha _S(\mu ^2)]^{- \frac{6}{27}} 
\left( \bar d \gamma _{\mu}(1- \gamma _5) s\right) ^2 
+ h.c. 
\label{LAGDELTAS2}
\eeq  
with $\xi _i$ denoting combinations of KM parameters 
\beq 
\xi _i = V(is)V^*(id) \; , \; \; i=c,t \; ; 
\eeq  
$E(x_i)$ and $E(x_c,x_t)$ reflect the box loops with equal and 
different internal quarks, respectively \cite{INAMI}:  
\beq 
E(x_i) = x_i 
\left(   
\frac{1}{4} + \frac{9}{4(1- x_i)} - \frac{3}{2(1- x_i)^2} 
\right) 
- \frac{3}{2} \left( \frac{x_i}{1-x_i}\right) ^3 
{\rm log} x_i 
\label{TOPBOX}
\eeq  
$$  
E(x_c,x_t) = x_c x_t 
\left[ \left( 
\frac{1}{4} + \frac{3}{2(1- x_t)} - \frac{3}{4(1- x_t)^2} \right) 
\frac{{\rm log} x_t}{x_t - x_c} + (x_c \leftrightarrow x_t) - 
\right. 
$$ 
\beq 
\left. - \frac{3}{4} \frac{1}{(1-x_c)(1- x_t)} \right]  
\eeq 
\beq 
x_i = \frac{m_i^2}{M_W^2}  \; . 
\eeq  
The $\eta _{ij}$ represent the QCD radiative corrections from 
evolving the effective Lagrangian from $M_W$ down to 
the internal quark mass. 
The factor $[\alpha _S(\mu ^2)]^{-6/27}$ 
reflects the fact that a scale 
$\mu$ must be introduced at which the four-quark operator 
$\left( \bar s \gamma _{\mu}(1- \gamma _5) d\right) ^2 $ is 
defined. This dependance on the auxiliary variable 
$\mu$ drops out when one takes the matrix element of this 
operator (at least when one does it correctly).  
Including next-to-leading log 
corrections one finds (for $m_t \simeq 180$ GeV) \cite{BURAS}: 
\beq 
\eta _{cc} \simeq 1.38 \pm 0.20 \; , \; \; 
\eta _{tt} \simeq 0.57 \pm 0.01 \; , \; \; 
\eta _{cc} \simeq 0.47 \pm 0.04 
\eeq                       
\footnote{There is also a {\em non}-local $\Delta S=2$ 
operator generated from the iteration of ${\cal L}(\Delta S=1)$. 
While it  
presumably provides a major contribution to $\Delta m_K$, 
it is not sizeable for $\epsilon _K$ within the KM ansatz,  
as be inferred from the observation that 
$|\epsilon ^{\prime}/\epsilon _K|\ll 0.05$.} 
The dominant contributions  
for $\Delta M(K)$ and $\epsilon_K$ are produced, when (in addition to the $W^{\pm}$ pair) the {\em internal} quarks are charm and top, respectively. In either case the {\em internal} quarks are 
heavier than the {\em external} ones: $m_d, m_s \ll m_c, m_t$, and 
evaluating the Feynman diagrams indeed corresponds to integrating out the heavy fields. 

The situation is qualitatively very similar for $\Delta M(B^0)$, and in some sense even simpler: 
for within the SM by far the leading contribution is due to internal top quarks. Evaluating 
the quark box diagram with internal $W$ and top quark lines corresponds to integrating those 
heavy degrees of freedom out in a straightforward way leading to: 
\beq   
{\cal L}_{eff}^{box}(\Delta B=2, \mu ) \simeq 
\left( \frac{G_F}{4\pi }\right) ^2 M_W^2\cdot     
\xi _t^2 E(x_t) \eta _{tt} 
\left( \bar q \gamma _{\mu}(1- \gamma _5) b\right) ^2 
+ h.c. 
\label{LAGDELTAS2}
\eeq  
with $q = d,s$. 
\begin{center}
{\bf First dish of `Food for thought' a.k.a. Homework assignment \# 1} 
\end{center} 
\noindent 
When one calculates $\Delta M(B)$ as a function of the top mass employing the quark box 
diagram, one finds, see Eq.(\ref{TOPBOX}) 
\beq 
\Delta M(B) \propto \left(   \frac{m_t}{M_W}    \right) ^2  \; \; {\rm for} \; \; m_t \gg M_W
\label{HW1}
\eeq
The factor on the right hand side of Eq.(\ref{HW1}) for $m_t \ll M_W$ reflects the familiar GIM suppression; yet for $m_t \gg M_W$ it constitutes a (huge) enhancement! It means that a low 
energy observable, namely $\Delta M(B)$, is controlled more and more by a state or field at 
asymptotically high scales. Does this not violate decoupling theorems and even common sense? 
Does it violate decoupling -- and if so, why is it allowed to do so -- or not? 
\begin{center}
{\bf End of Homework \# 1} 
\end{center} 

While quark box diagrams contribute also to $\Gamma_{12}(\Delta S=2)$, it would be absurd to assume they are significant. For $\Gamma_K$ is dominated by the impact of hadronic phase 
space causing $\Gamma (K_{neut} \to 2 \pi) \gg \Gamma (K_{neut} \to 3 \pi)$. Yet even beyond that it is 
unlikely that such a computation would make much sense: to contribute to $\Delta \Gamma_K$ 
the internal quark lines in the quark box diagram have to be $u$ and $\bar u$ quarks, i.e. 
{\em lighter} than the external quarks $s$ and $\bar s$. That means calculating this 
Feynman diagram  does not correspond to integrating out the heavy degrees of freedom. 
For the same reason (and others as explained later in more detail) computing quark box diagrams 
tells us little of value concerning $D^0 - \bar D^0$ oscillations, since the internal quarks on the leading 
CKM level  -- $s$ and $\bar s$ -- are lighter than the external charm quarks. 

A new and more intriguing twist concerning quark box diagrams occurs when addressing $\Delta \Gamma$ for $B^0$ mesons. Those diagrams again do not generate a {\em local} operator, since the internal charm quarks carry less than half the mass of the external $b$ quarks. 
Nevertheless it can be conjectured that the on-shell $\Delta B=2$ transition operator generating 
$\Delta \Gamma _B$ is largely shaped by short distance dynamics. 

QCD radiative corrections affect the strength of these effective weak transition operators -- and create different types of such operators. Consider $\Delta S=1$ transitions. On the tree graph 
level there is one operator, namely $(\bar u_L \gamma _{\mu}s_L)(\bar d_L \gamma ^{\mu}u_L)$.  
Including one-loop diagrams where a gluon is exchanged between quark lines one obtains 
${\cal O}(\alpha_S)$ contributions to the original 
$(\bar u_L \gamma _{\mu}s_L)(\bar d_L \gamma ^{\mu}u_L)$ operator -- and to the new coupling 
$(\bar u_L \gamma _{\mu}t^is_L)(\bar d_L \gamma ^{\mu}t^iu_L)$, where the 
$t^i$ denote the generators of colour $SU(3)$. I.e., the two operators 
$O^{1\times 1}= (\bar u_L \gamma _{\mu}s_L)(\bar d_L \gamma ^{\mu}u_L)$ and 
$O^{8\times 8}= (\bar u_L \gamma _{\mu}t^is_L)(\bar d_L \gamma ^{\mu}t^iu_L)$, 
where the former [latter] represents the product of two colour-singlet[octet] currents, mix under 
QCD renormalization already on the one-loop level: 
\beq 
(\bar u_L \gamma _{\mu}s_L)(\bar d_L \gamma ^{\mu}u_L) \; \; \; 
\stackrel{{\rm QCD\, 1-loop\, renormalization}}\Longrightarrow \; \; \; 
c_{1\times 1}O^{1\times 1} + c_{8\times 8}O^{8\times 8}
\eeq
with $c_{1\times 1} = 1 +{\cal O}(\alpha_S)$, whereas $c_{8\times 8} = {\cal O}(\alpha_S)$. Since 
some of these $\alpha_S$ corrections are actually enhanced by numerically sizeable 
log$(M_W/\mu)$ factors, they are quite significant. Therefore one wants to identify the 
{\em multiplicatively} renormalized transition operators with 
\beq 
\tilde O \; \; \; \; \; 
\stackrel{{\rm QCD\, 1-loop\, renormalization}}\Longrightarrow 
\; \; \; \; \; \tilde c \; \tilde O
\eeq
This can be done even without brute-force computations by relying on isospin arguments: consider 
the weak scattering process between quarks 
\beq 
s_L + u_L \to u_L + d_L
\eeq 
proceeding in an S wave. 
It can be driven by two $\Delta S=1$ operators, namely 
\beq 
O_{\pm} = \frac{1}{2} \left[ (\bar u_L\gamma_{\mu} s_L)(\bar d_L \gamma^{\mu}u_L) \pm 
(\bar d_L\gamma_{\mu} s_L)(\bar u_L \gamma^{\mu}u_L) \right] 
\eeq
The operator $O_+ [O_-]$ produces an $ud$ pair in the final state that is [anti]symmetric in isospin and thus 
carries $I=1[I=0]$; since the initial $su$ pair carries $I=1/2$, $O_+[O_-]$ generates 
$\Delta I = 1/2\&3/2$ [only $\Delta I =1/2]$ transitions. 

With QCD conserving isospin, its radiative corrections cannot mix the operators $O_{\pm}$, 
which therefore are {\em multiplicatively} renormalized:  
\beq 
O_+\; [O_-] \; \; \; \; \; 
\stackrel{{\rm QCD\, 1-loop\, renormalization}}\Longrightarrow 
\; \; \; \; \;  c_+  \, O_+\; [c_- \, O_-]  \; . 
\eeq 
and therefore 
\beq 
{\cal L}_{eff}^{(0)}(\Delta S =1)= O_+ + O_- 
\; \; \;  
\stackrel{{\rm QCD\, 1-loop\, ren.}}\Longrightarrow 
\; \; \;   {\cal L}_{eff}(\Delta S =1)=c_+  \, O_+ +c_- \, O_-
\eeq
with $c_{\pm} = 1 + {\cal O}(\alpha _S)$. 

Integrating out those loops containing a $W$ line in addition to the gluon line and two 
quark lines yields terms $\propto \alpha_S {\rm log}(M_W^2/\mu ^2)$, which are not necessarily 
small. Using the renormalization group equation to sum those terms terms one finds on the leading 
log level 
\beq 
c_{\pm} = \left[\frac{ \alpha_S(M_W^2)}{\alpha_S(\mu ^2)}
\right]^{\gamma_{\pm}} \; , \; \; \gamma _+= \frac{6}{33-2N_F}=-\frac{1}{2}\gamma_- \; . 
\label{LEADLOG1}
\eeq
I.e., 
\footnote{The expressions of Eq.(\ref{LEADLOG1}) hold in the `leading log approximation'; 
including terms $\sim \alpha_S^{n+1}{\rm log}^n(M_W^2/\mu ^2)$ modifies them, yet 
$c_- > 1 > c_+$ and  $c_- c_+^2 \simeq 1$ still hold.} 
\beq 
c_- > 1 > c_+ \; , \; \; c_- c_+^2 = 1
\label{LEADLOG2} 
\eeq
That means that QCD radiative corrections provide a quite sizeable $\Delta I=1/2$ 
enhancement. Corresponding effects arise for ${\cal L}_{eff}(\Delta C/B=1)$. 

QCD radiative corrections create yet another effect, namely they lead to the emergence 
of `Penguin' operators. Without the gluon line their diagram would decompose into two 
{\em disconnected} parts and thus not contribute to a transition operator. These Penguin 
diagrams can drive only $\Delta I = 1/2$ modes. Furthermore in the loop all three quark families 
contribute; the diagram thus contains the irreducible CKM phase -- i.e. it generates 
{\em direct} \cp~violation in strange decays. Similar effects arise in beauty, but not necessarily in 
charm decays. 

The main message of these more technical considerations was to show that while QCD conserves 
flavour, it has a highly nontrivial impact on flavour transitions by not only affecting the strength of 
the bare weak operator, but also inducing new types of weak transition operators already on the 
perturbative level. In particular, QCD creates a source of {\em direct} \cp~violation in strange decays  naturally, albeit with a significantly reduced strength.   

\subsubsection{Nonperturbative Dynamics}
\label{NONPERTDYN}

Perturbative dynamics does of course not suffice to calculate decays rates for hadrons. 
Applying the OPE to the description of a 
transition $H \to f$ we need to evaluate on-shell hadronic matrix elements: 
\beq 
T(H \to f) \propto \matel{f}{{\cal L}_{eff}}{H} \propto 
\sum_i c_i(\mu) \matel{f}{O_i(\mu)}{H} 
\eeq
where $\mu$ denotes the demarkation line between long and short distance dynamics. 

We can call on several allies to take up the challenge of nonperturbative dynamics posed by 
$\matel{f}{O_i(\mu)}{H}$. 
\begin{itemize}
\item 
{\em Quark Models:} We can still get considerable mileage out of this `old war horse', if we do 
not overburden it. They are an excellent tool to train our intuition and arrive at {\em first} answers -- 
yet are unsatisfactory for {\em final} answers. 
\item 
{\em Chiral Perturbation Theory:} It represents QCD at low energies -- yet does not provide 
a fool proof algorithm. 
\item 
{\em Heavy Quark Theory:} It reflects QCD for heavy flavour hadrons and will be described in more detail later.  
\item 
{\em QCD Sum Rules:} They are genuinely based on QCD, yet there is typically a bound, below which 
their intrinsic uncertainties cannot be reduced. 
\item 
{\em Lattice QCD (LQCD):} the perceived panacea. 

\end{itemize}
Let me present just one example, albeit one of central interest. The quantities 
\beq 
\matel{K}{(\bar s_L \gamma_{\mu}d_L)(\bar s_L \gamma^{\mu}d_L)}{\bar K} \, , \, \, 
\matel{B}{(\bar b_L \gamma_{\mu}d_L)(\bar b_L \gamma^{\mu}d_L)}{\bar B}
\label{KKBARME}
\eeq
control $K^0 - \bar K^0$ and $B_d - \bar B_d$ oscillations, respectively, in the SM. One can 
write: 
\beq 
\matel{K}{(\bar s_L \gamma_{\mu}d_L)(\bar s_L \gamma^{\mu}d_L)}{\bar K} = 
\frac{4}{3} B_K f_K^2 M_K^2 
\eeq
with $f_K$ denoting the $K$ meson decay constant. This expression does not entail any 
loss of generality as long as the parameter $B_K$ is left open 
\footnote{$B_K$ is often called the kaon `bag factor'. This name goes back to the heydays of the 
MIT bag model used to calculate a host of hadronic matrix elements. I admit freely and without shame 
that I have used the MIT bag model myself.}. The parametrization is convenient since 
on general grounds one expects $B_K \sim {\cal O}(1)$ based on the following argument. 

The Hilbert space of all hadronic states can be spanned by the states with no hadrons -- $|0\rangle$ -- and with an increasing number of hadrons -- $|n_{had}\rangle$. The identity in this Hilbert space can then be 
written as follows:  
\beq 
{\bf 1} = |0\rangle \langle 0| + \sum _n |n_{had}\rangle \langle n_{had}| 
\eeq
Inserting it into Eq.(\ref{KKBARME}) one obtains 
$$ 
\matel{K}{(\bar s_L \gamma_{\mu}d_L)(\bar s_L \gamma^{\mu}d_L)}{\bar K} = 
\matel{K}{(\bar s_L \gamma_{\mu}d_L)}{0}\matel{0}{(\bar s_L \gamma^{\mu}d_L)}{\bar K} + 
$$
\beq 
\sum _n \matel{K}{(\bar s_L \gamma_{\mu}d_L)}{n_{had}}
\matel{n_{had}}{(\bar s_L \gamma^{\mu}d_L)}{\bar K}
\label{KKME}
\eeq
The `vacuum saturation' (VS) or `factorization' approximation consists of assuming that the overall 
contribution from all nontrivial hadronic intermediate states, i.e. the 
second line in Eq.(\ref{KKME}), is zero or at least small: 
\beq 
\matel{K}{(\bar s_L \gamma_{\mu}d_L)(\bar s_L \gamma^{\mu}d_L)}{\bar K}_{VA} =  
\matel{K}{(\bar s_L \gamma_{\mu}d_L)}{0}\matel{0}{(\bar s_L \gamma^{\mu}d_L)}{\bar K}
\eeq
Since $\matel{0}{(\bar s_L \gamma^{\mu}d_L)}{\bar K(p)} = i f_K p_{\mu}$ we get 
\beq 
\matel{K}{(\bar s_L \gamma_{\mu}d_L)(\bar s_L \gamma^{\mu}d_L)}{\bar K}_{VA} = 
\frac{4}{3} f_K^2 M_K^2  
\eeq
with the colour factor $\frac{4}{3}=\left( 1 + \frac{1}{N_C}\right)$ reflecting the two possible quark 
line contractions. The  
VS hypothesis $B_K \simeq 1$ itself does {\em not} assume contributions from individual hadronic intermediate states to 
be small, only that their sum is smallish, since different $|n_{had}\rangle$ contribute with alternating signs. This expectation was first supported by analyses based on $1/N_C$ arguments, QCD sum rules and later more quantitatively by LQCD: 
\beq 
B_K^{theor.} = 0.79 \pm 0.04 \pm 0.09
\eeq 
Some technical subtleties have to be treated properly to arrive at a well defined quantity. One has 
to keep in mind that when assuming VS, one has to specify at which scale VS is assumed. For 
contributions that are factorizable at one scale $\mu_1$ are in general not purely factorizable at a different scale $\mu_2$. This is particularly relevant for 
$\matel{B}{(\bar b_L \gamma_{\mu}d_L)(\bar b_L \gamma^{\mu}d_L)}{\bar B}$: does one 
assume VS at a high scale like $M_B$ and then evolve merely the factorized expression down to 
ordinary hadronic scales $\mu \sim 1$ GeV. Or does one evolve the full expression down to 
$\mu$ and assume factorization at this low scale? The two procedure actually yield quite different results; the former actually makes little sense. 

\subsection{The CKM Paradigm of Large \cp~Violation in $B$ Decays}
\label{CKMPARAD}

\subsubsection{Basics}

As explained in detail in Prof. Sanda's lectures here at the school, the 
decays $K_L \to \pi \pi$ are controlled by two types of 
{\em coherent} processes, namely 
\begin{itemize}
\item 
$\Delta S=2$ dynamics, which generates the two mass eigenstates $K_L$ and $K_S$; 
\item 
$\Delta S=1$ reactions that cause the decay of the kaon into a final state 
with{\em out} strangeness: 

\end{itemize}
\beq 
[K^0 \stackrel{\Delta S =2}{\longleftrightarrow} \bar K^0] \Rightarrow 
K_L \stackrel{\Delta S=1}{\longrightarrow} \pi \pi 
\eeq
Thus there are a priori two classes of gateways, through which \cp~violation can enter. This is made 
explicit in the usual notation: 
\bea 
\nonumber 
\eta_{+-,00} &=& \frac{T(K_L \to \pi^{+,0}\pi^{-,0})}{T(K_S \to \pi^{+,0}\pi^{-,0})} \\
\nonumber 
\eta_{+-} &=& \epsilon_K + \epsilon^{\prime} \\
\eta_{00} &=& \epsilon_K - 2\epsilon^{\prime} 
\label{ETADEF}
\eea
$\eta_{+-}$ or $\eta_{00}$ $\neq 0$ constitutes \cp~violation. More specifically 
we see that $\epsilon_K$ parametrizes \cp~violation common to both decay modes; it thus 
reflects \cp~violation in the composition of the decaying state -- $K_L$ -- as produced by 
$\Delta S=2$ forces; 
$\epsilon^{\prime}$ on the other hand differentiates between the two channels $\pi^+\pi^-$ and 
$\pi^0\pi^0$ and is thus the result of $\Delta S =1$ dynamics. With less than Shakespearean 
flourish one says $\epsilon_K$ and $\epsilon^{\prime}$ describe 
{\em indirect} and {\em direct} \cp~violation, respectively. 

This classification can be extended in a straightforward way to the decays of any hadron carrying flavour quantum number  $F$ like beauty and charm hadrons. \cp~violation in $\Delta F=1$ 
and in $\Delta F=2$ transitions are of the `direct' and  `indirect' variety, respectively. Indirect 
\cp~violation can thus arise only for neutral mesons  
\footnote{There is one subtlety to be noted: as explained 
later, 
the distinction between direct and indirect \cp~violation is not always unambiguous in the 
decays of neutral mesons.}. I want to stress the following: the processes $K_L \to \pi \pi$ or 
$B_d \to \psi K_S$ involve two {\em phenomenologically} distinct dynamics, namely 
$\Delta F=1\& 2$. It is important to deduce from the data, to which degree both contribute 
to the observed modes -- yet in the end an underlying theory has to explain both. One should 
also note that the `superweak' model is {\em not} a theory,  actually not even a model -- it is 
merely a convenient {\em classification} scheme. For a given theory one has to analyze, whether 
it is a dynamical implementation -- exact or approximate -- of a superweak scenario. As 
explained in detail later on, CKM theory is not of the superweak variety.

\subsubsection{Prelude: Before 1973}

The discovery of \op~violation in weak decays in 1957 caused great shock in the physics community -- 
yet even the theorists quickly recovered by arguing they had placed unnecessary demands 
on nature. Look at politics for example: `left' and `right' is often defined -- at least on the 
gut level -- in terms of `good' and `bad' or `positive' and `negative'. There is a slight problem, though. 
There is no universal consensus about who the good and bad guys are. Consider pion decay:  
$\pi \to e \nu$. Maximal \op~violation means that the emerging electrons are purely left-handed. This 
statement assumes implicitly one is considering the decays of negative pions: 
\beq 
\pi ^- \to e^-(L) \bar \nu
\eeq 
The decay of positive pions on the other hand produce purely right-handed positrons: 
\beq 
\pi ^+ \to e^+(R) \nu
\eeq
Those two transitions are related by a \cp~transformation; thus they are equivalent as long as 
\cp~is conserved. One can say that a pion of a given charge will produce leptons of only 
one helicity; yet what one means by `left' depends on the definition of negative charge and vice versa: 
\beq 
`L' = f(`-') 
\eeq
This is like saying: "The thumb is left on the right hand." -- a correct as well as useless statement since 
circular. Thus -- to address a question raised by Pauli after 1957 -- even maximal parity violation when 
coupled with \cp~symmetry does {\em not} signal an absolute preference of nature for `left' -- 
only a correlation of `left' and `right' with the sign of the electric charge. Even Landau, who was a late 
convert to parity violation made his peace with this situation. 

Alas, the `fall back' position was shattered in 1964 by the discovery of \cp~violation through 
$K_L \to \pi^+\pi^-$ and subsequently through 
\beq 
\Gamma (K_L \to l^+(R)\nu \pi^-) > \Gamma (K_L \to l^-(L)\nu \pi^+) \; ; 
\label{CPCLKL}
\eeq 
the latter allows to define the sign of electric charge based on data rather than a convention and likewise for the handedness of the charged lepton. 

How much this discovery shook the HEP community is best gauged by noting the efforts made to 
reconcile the observation of $K_L \to \pi^+\pi^-$ with \cp~invariance: 
\begin{itemize}
\item 
To infer that $K_L \to \pi \pi$ implies \cp~violation one has to invoke the superposition principle of 
quantum mechanics. One can introduce \cite{ROOS} {\em non}linear terms into the Schr\" odinger 
equation in such a way as to allow $K_L \to \pi^+\pi^-$ with \cp~invariant dynamics. While completely ad hoc, it is possible in principle. Such efforts were ruled out by further data, most decisively by 
$\Gamma (K^0(t) \to \pi^+\pi^-) \neq \Gamma (\bar K^0(t) \to \pi^+\pi^-)$. 
\item 
One can try to emulate the success of Pauli's neutrino hypothesis. An apparent violation of 
energy-momentum conservation had been observed in $\beta$ decay $n \to p e^-$, since 
the electron exhibited a {\em continuous} momentum spectrum. Pauli postulated that the reaction actually was 
\beq 
n \to p e^- \bar \nu 
\eeq  
with $\bar \nu$ a neutral and light particle that had escaped direct observation, yet let to a continuous 
spectrum for the electron. I.e., Pauli postulated a new particle -- and a most whimsical one at that -- to save a symmetry, namely the one under  translations in space and time responsible for the 
conservation of energy and momentum. Likewise it was suggested that the real reaction was 
\beq 
K_L \to \pi^+\pi^- U 
\label{UPART}
\eeq 
with $U$ a neutral and light particle with {\em odd} intrinsic \cp~parity. 
I.e., a hitherto unseen particle with presumably whimsical properties was introduced to save a symmetry.  
This attempt at evasion was also soon rejected 
experimentally (see Homework \# 2). This represents an example of the ancient Roman saying: 

\begin{center} 
"Quod licet Jovi, non licet bovi." \\
"What is allowed Jupiter, is not allowed a bull."
\end{center} 
I.e., we mere mortals cannot get away with speculations like `Jupiter' Pauli. 

\end{itemize} 
\begin{center}
{\bf Homework assignment \# 2} 
\end{center} 
\noindent 
What was the conclusive argument to rule out the reaction of Eq.(\ref{UPART}) taking place even for 
a very tiny  $U$ mass? 
\begin{center}
{\bf End of Homework \# 2} 
\end{center}

There are more special features for \cp~violation not shared by \op~violation: 
\begin{itemize}
\item 
Due to \cpt~invariance -- an almost 
inescapable consequence of {\em local} relativistic quantum field theories -- \cp~violation 
implies a commensurate violation of microscopic time reversal invariance. 
\item 
As explained above \cp~violation allows to define `positive' vs. `negative', `left-' vs. 
`right-'handed, `matter' vs. `antimatter' in a {\em convention independent} way. 
\item 
If one wants to understand the observed baryon number of the Universe not as an 
arbitrary {\em initial condition}, but as a {\em dynamically generated} quantity -- similar to the success one has 
achieved in understanding the abondances of light nuclei in the Universe -- three ingredients 
are needed as pointed out by Sakharov just a year after the discovery of \cp~violation 
\cite{SAKH}: 
\begin{enumerate}
\item 
Dynamics that can change baryon number;   
\item 
\cp~violation, since otherwise baryon production or destruction is matched by the corresponding 
processes for antibaryons; 
\item 
the Universe has to be out of thermal equilibrium, since otherwise \cpt~invariance 
acts globally like \cp~symmetry. 

\end{enumerate}
How these ingredients work together and how they can be implemented in specific models is 
discussed in Prof. Dolgov's lectures at this school. 
\item 
It is the smallest observed violation of a symmetry as characterized by the off-diagonal element in the 
generalized $K^0 - \bar K^0$ mass matrix: 
\beq 
{\rm Im} M_{12}^K \simeq 1.1 \cdot 10^{-8} \; eV 
\footnote{This number is often expressed through the dimensionless ratio 
Im$M_{12}^K/M_K \simeq 2.2 \cdot 10^{-17}$; however using the kaon mass as a yardstick is arbitrary, 
since  it is mainly produced by the strong interactions rather than the (super)weak forces behind 
$M_{12}$.}
\eeq
Since a world with \cp~symmetry is fundamentally different from one without it, 
such a `near miss' seems peculiar when contrasted with maximal \op~violation.

\end{itemize}
The phenomenology of \cp~violation was quickly developed. Yet, as outlined in Lecture I, the lack 
of a theory was not realized for a number of years even after the renormalizibility of the 
$SU(2)_L\times U(1)$ electroweak SM was recognized. The socalled `superweak' model put forward 
by Wolfenstein already in 1965 is not a theory; it can hardly qualify even as a model -- it is 
basically a classification scheme.  

\subsubsection{`Growing up': 1973 -- 1994}

The early 1970's marked a turning point culminating in the `October revolution of 1974', the discovery of 
the $J/\psi$ and $\psi ^{\prime}$. Like any true revolution, it had several events leading up to it, chiefly 
among them the observation of Bjorken scaling and approximate scale invariance in deep inelastic 
lepton nucleon scattering and the large cross section for $e^+e^-  \to had$. Crucial further developments were the realization of QCD being asymptotically free, the discovery of 
the $\tau$ lepton and in 1976 of the $\Upsilon$ resonance. The latter was for most authors, who had 
lived through the charmonium revolution, `deja vue all over again'. It was readily accepted 
that hadrons with the new quantum number beauty had to exist with masses 
$\sim M(\Upsilon )/2$; that they together with $\tau$ leptons presumably are part of a third 
quark-lepton family and that they decay weakly preferably to charm hadrons with a reduced CKM 
parameter $V(cb)$ with a typical guestimate $|V(cb)| \sim |V(us)|$. 

In 1979 it was predicted that the channels $B_d \to K^+\pi^-$ should reveal  sizable {\em direct} 
\cp~violation \cite{SONI79}and in 1980 that a host of $B$ decays should exhibit sizable or even large 
\cp~asymmetries involving quantum mechanical state  mixing and oscillations \cite{CARTER,BS80}
\footnote{By `state mixing' I mean the fact that different states can contribute coherently and by 
`oscillations' that one pure state can evolve into another pure state in time and evolve back again. The 
superposition principle of quantum mechanics is central to both effects, yet you can have the former without the latter.}, 
in particular in the 
mode $B_d \to \psi K_S$ \cite{BS80}. For proper perspective -- and not merely for establishing bragging rights for theorists -- one should note that these predictions were made, before a single $B$ decay mode had been identified and before their lifetime was measured. 

The first indication that the $B$ lifetime is significantly longer and thus $|V(cb)|$ smaller than anticipated came in 1982. It was then confirmed that $B$ mesons live about 1 psec. This pointed to  
$|V(cb)| \sim {\cal O}(\lambda ^2)$ with $\lambda = {\rm sin}\theta_C$. Together with the 
expected observation $|V(ub)| \ll |V(cb)|$ and coupled with the assumption 
of three-family unitarity this allows to expand the CKM matrix in powers of $\lambda$, which 
yields the following most intriguing result through order $\lambda ^5$, as first recognized by Wolfenstein: 
\beq 
{\bf V}_{CKM} = 
\left( 
\begin{array}{ccc} 
1 - \frac{1}{2} \lambda ^2 & \lambda & 
A \lambda ^3 (\rho - i \eta + \frac{i}{2} \eta \lambda ^2) \\
- \lambda & 1 - \frac{1}{2} \lambda ^2 - i \eta A^2 \lambda ^4 & 
A\lambda ^2 (1 + i\eta \lambda ^2 ) \\ 
A \lambda ^3 (1 - \rho - i \eta ) \\
& - A\lambda ^2 & 1 
\end{array}
\right) 
\label{WOLFKM}
\eeq  
The three Euler angles and one complex phase of the representation given in Eq.(\ref{PDGKM}) 
is taken over by the four real quantities $\lambda$, $A$, $\rho$ and $\eta$; 
$\lambda$ is the expansion parameter with $\lambda \ll 1$, whereas $A$, $\rho$ and $\eta$ 
are a priori of order unity, as will be discussed in some detail later on. I.e., the `long' 
lifetime of beauty hadrons of around 1 psec together with beauty's affinity to transform itself into charm 
and the assumption of only three quark families tell us 
that the CKM matrix exhibits a very peculiar hierarchical pattern in powers of $\lambda$: 
\beq 
V_{CKM} = 
\left( 
\begin{array}{ccc} 
1 & {\cal O}(\lambda ) & {\cal O}(\lambda ^3) \\ 
{\cal O}(\lambda ) & 1 & {\cal O}(\lambda ^2) \\ 
{\cal O}(\lambda ^3) & {\cal O}(\lambda ^2) & 1 
\end {array} 
\right) 
\; \; \; , \; \; \; \lambda = {\rm sin}\theta _C 
\eeq 
As explained in Lecture I, we know this matrix has to be unitary. Yet in addition it is almost 
the identity matrix, almost symmetric and the moduli of its elements shrink with the distance from the diagonal. 
It has to contain a message from nature -- albeit in a highly encoded form. 

My view of the situation is 
best described by a poem by the German poet Joseph von Eichendorff from the late romantic 
period \footnote{I have been told that early romantic writers would have used the term `symmetry' 
instead of `song'.}: 

\vspace{3mm}

\begin{tabular} {ll}
Schl\"aft ein Lied in allen Dingen, &  There sleeps a song in all things \\
die da tr\"aumen fort und fort, & that dream on and on, \\
und die Welt hebt an zu singen, & and the world will start to sing, \\
findst Du nur das Zauberwort. & if you find the magic word.
\end{tabular}
 
\vspace{3mm}

\noindent 
The six triangle relations obtained from the unitarity condition fall into three categories:  
\begin{enumerate}
\item 
$K^0$ triangle: 
\beq 
\begin{array}{ccc} 
V^*(ud)V(us) + &V^*(cd)V(cs) + &V^*(td) V(ts) = 
\delta _{ds}= 0 \\
{\cal O}(\lambda ) & {\cal O}(\lambda ) & {\cal O}(\lambda ^5) 
\end{array} 
\label{TRI1} 
\eeq 
$D^0$ triangle:
\beq 
\begin{array}{ccc} 
V^*(ud)V(cd) + &V^*(us)V(cs) + &V^*(ub) V(cb) = 
\delta _{uc}= 0 \\
{\cal O}(\lambda ) & {\cal O}(\lambda ) & {\cal O}(\lambda ^5) \; , 
\end{array} 
\label{TRI2} 
\eeq 
where below each product of matrix elements I have noted 
their size in powers of $\lambda $. 
These two triangles are extremely `squashed': two sides are 
of order $\lambda $, the third one of order $\lambda ^5$ and their 
ratio of order $\lambda ^4 \simeq 2.3 \cdot 10^{-3}$; 
Eq.(\ref{TRI1}) and Eq.(\ref{TRI2}) control the situation in 
strange and charm decays; the relevant weak phases there 
are obviously tiny.

\item 
$B_s$ triangle: 
\beq 
\begin{array}{ccc} 
V^*(us)V(ub) + &V^*(cs)V(cb) + &V^*(ts) V(tb) = 
\delta _{sb}= 0 \\
{\cal O}(\lambda ^4) & {\cal O}(\lambda ^2) & {\cal O}(\lambda ^2) 
\end{array} 
\label{TRI3} 
\eeq 
$tc$ triangle: 
\beq 
\begin{array}{ccc} 
V^*(td)V(cd) + &V^*(ts)V(cs) + &V^*(tb) V(cb) = 
\delta _{ct}=0 \\
{\cal O}(\lambda ^4) & {\cal O}(\lambda ^2) & {\cal O}(\lambda ^2) 
\end{array} 
\label{TRI4} 
\eeq 
The third and fourth triangles are still rather squashed, yet less so: 
two sides are of order $\lambda ^2$ and the third one of order 
$\lambda ^4$. 

\item 
$B_d$ triangle: 
\beq 
\begin{array}{ccc} 
V^*(ud)V(ub) + &V^*(cd)V(cb) + &V^*(td) V(tb) = 
\delta _{db}=0 \\
{\cal O}(\lambda ^3) & {\cal O}(\lambda ^3) & {\cal O}(\lambda ^3) 
\end{array} 
\label{TRI5} 
\eeq 
$ut$ triangle: 
\beq 
\begin{array}{ccc} 
V^*(td)V(ud) + &V^*(ts)V(us) + &V^*(tb) V(ub) = 
\delta _{ut}=0 \\
{\cal O}(\lambda ^3) & {\cal O}(\lambda ^3) & {\cal O}(\lambda ^3) 
\end{array}
\label{TRI6}  
\eeq 
The last two triangles have sides that are all of the same 
order, namely $\lambda ^3$. All their angles are therefore 
naturally large, i.e. $\sim$ several $\times$ $10$ degrees! Since to 
leading order in $\lambda$ one has 
\beq 
V(ud) \simeq V(tb) \; , \; V(cd) \simeq - V(us) \; , \; 
V(ts) \simeq - V(cb) 
\eeq 
we see that the triangles of Eqs.(\ref{TRI5}, \ref{TRI6}) 
actually coincide to that order.  

The sides of this triangle having naturally large angles are 
given by $\lambda \cdot V(cb)$, $V(ub)$ and 
$V^*(td)$; these are all quantities that control important 
aspects of $B$ decays, namely CKM favoured and disfavoured 
$B_{u,d}$ decays and $B_d - \bar B_d$ oscillations. The $B_d$ triangle of Eq.(\ref{TRI5}) 
is usually referred to as {\em `the' CKM unitarity triangle}.  

\end{enumerate} 
Let the reader be reminded that all six triangles, despite their very different shapes, have 
the same area, see Eq.(\ref{JARL}), reflecting the single CKM phase for three families.  

Some comments on notation might not be completely useless. The BABAR collaboration 
and its followers refer to the three angles of the CKM unitarity triangle as 
$\alpha$, $\beta$ and $\gamma$; the BELLE collaboration instead has adopted the 
notation $\phi_1$, $\phi_2$ and $\phi_3$. While it poses no problem to be 
conversant in both languages, the latter has not only historical priority on its side 
\cite{BJSANDA}, but is 
also more rational. For the angles $\phi_i$ in the 
`$bd$' triangle of Eq.(\ref{TRI5}) are always opposite the side defined by 
$V^*(id)V(ib)$.   Furthermore this classification scheme can readily be generalized to all six 
unitarity triangles;  those triangles can be labeled by $kl$ with 
$k \neq l= d,s,b$ or $k\neq l = u,c,t$, see Eqs.(\ref{TRI1}) -- (\ref{TRI6}). Its 18 angles 
can then be unambiguously denoted by $\phi_i^{kl}$: it is the angle in triangle $kl$ opposite 
the side $V^*(ik)V(il)$ or $V^*(ki)V(li)$, respectively. Therefore I view the notation 
$\phi_i^{(kl)}$ as the only truly Cartesian one. 

For complex phases to become observable,  we need two different, yet coherent 
amplitudes to contribute to the same process. The best and most spectacular implementation 
of this requirement is provided by $B^0 - \bar B^0$ oscillations. Such oscillations for $B_d$ mesons were discovered by the ARGUS collaboration \cite{ARGUSOSC}in 1986 with 
\footnote{ARGUS' discovery came as a surprise, since the size of its signal exceeded most theoretical 
expectations. In fairness it should be remembered that the discovery of top quarks had been claimed 
with $m_t \sim 40$ GeV. Since the observable signal, the ratio of `wrong' to `right' sign leptons in 
semileptonic $B$ decays, is given by $x^2/(2+x^2)$ and $x(B_d)$ very roughly 
scales like $m_t^2$ this ratio is enhanced by two orders of magnitude when going from 
$m_t = 40$ GeV to $160$ GeV.}  
\beq 
x(B_d) = \frac{\Delta M(B_d)}{\Gamma (B_d)} \simeq 0.75 \; ; 
\eeq  
i.e., the oscillation rate $\Delta M(B_d)$ and decay rate $\Gamma (B_d)$ are very close to each other, 
which is optimal. This observation was the first experimental hint (albeit an indirect one) that top quarks had to be 
super-heavy, namely $m_t > 100$ GeV. This is very similar to though less precise than later LEP I 
findings on $m_t$. 
The discovery of $B_d - \bar B_d$ oscillations {\em defined} the 
`CKM Paradigm of Large \cp~Violation in $B$ Decays' that had been {\em anticipated} in 1980: 
\begin{itemize}
\item 
A host of nonleptonic $B$ channels has to exhibit sizable \cp~asymmetries. 
\item 
For $B_d$ decays to flavour-nonspecific final states (like \cp~eigenstates) the \cp~asymmetries 
depend on the time of decay in a very characteristic manner; their size should typically be measured 
in units of 10\% rather than 0.1\%. 
\item 
{\em There is no plausible deniability for the CKM description, if such asymmetries are 
not found.}  
\item 
For $m_t \geq 150$ GeV the SM prediction for $\epsilon_K$ is dominated by the top quark 
contribution like $\Delta M(B_d)$. It thus drops out from their ratio, and sin$2\phi_1$ can be 
predicted within the SM irrespective of the (superheavy) top quark mass. In the early 1990's, i.e., 
before the direct discovery of top quarks, it was predicted \cite{BEFORETOP} 
\beq 
\frac{\epsilon_K}{\Delta M(B_d)} \propto {\rm sin}2\phi_1 \sim 0.6 - 0.7
\eeq  
with values for $B_Bf_B^2$ inserted as now estimated by LQCD. 
\item 
The \cp~asymmetry in the Cabibbo favoured channels $B_s \to \psi \phi/\psi \eta$ is Cabibbo suppressed, i.e. below 4\%, for reasons very specific to CKM theory, as pointed out already in 
1980 \cite{BS80}. 

\end{itemize}

In 1974 finally top quarks were observed directly with a mass fully consistent 
with the indirect estimates given above; the most recent analyses from CDF \& D0 list  
\beq 
m_t = 172.7 \pm 2.9 \; {\rm GeV}
\eeq

\subsubsection{Data in 1998}
\label{DATA(*}

\cp~violation had been observed only in the decays of neutral kaons, and all its 
manifestations -- $K_L \to \pi^+\pi^-$, $\pi^0\pi^0$, 
$K^0 \to \pi^+\pi^-$ vs. $\bar K^0 \to \pi^+\pi^-$, $K_L\to l^+\nu \pi^-$ 
vs. $K_L \to l^- \bar \nu \pi^+$ -- could be described for 35 years with a {\em single real} number, 
namely $|\eta_{+-}|$ or $\Phi (\Delta S=2) = {\rm arg}(M_{12}/\Gamma_{12})$.  

There was intriguing, though not conclusive evidence for {\em direct} \cp~violation: 
\beq 
\frac{\epsilon ^{\prime}}{\epsilon_K} = 
\left\{ 
\begin{array}{ll} 
(2.30 \pm 0.65)\cdot 10^{-3} & {\rm NA31} \\
(0.74 \pm 0.59)\cdot 10^{-3} & {\rm E731}
\end{array}
\right.
\eeq
These measurements were made in the 1980's and had been launched by theory 
guestimates  
suggesting values for $\epsilon^{\prime}$ that would be within the reach of these experiments. 
Theory, however, had `moved on' favouring values $\leq 10^{-3}$ -- or so it was claimed.

\subsubsection{The Completion of a Heroic Era}

{\em Direct} CP violation 
has been unequivocally established in 1999. The present world average dominated 
by the data from NA48 and KTeV reads as follows \cite{SOZZI}: 
\beq 
\langle \epsilon ^{\prime}/\epsilon _K \rangle = 
(1.63 \pm 0.22) \cdot 10^{-3} 
\eeq 
Quoting the result in this way does not do justice to the experimental 
achievement, since $\epsilon _K$ is a very small number itself. 
The sensitivity achieved becomes more obvious when quoted in terms of actual 
widths \cite{SOZZI}:
\beq 
\frac{\Gamma (K^0 \to \pi ^+ \pi ^-) - 
\Gamma (\bar K^0 \to \pi ^+ \pi ^-)}
{\Gamma (K^0 \to \pi ^+ \pi ^-) + 
\Gamma (\bar K^0 \to \pi ^+ \pi ^-)} = 
(5.04 \pm 0.82) \cdot 10^{-6} \; !
\label{DIRECTK}
\eeq
This represents a discovery of the very first rank 
\footnote{As a consequence of Eq.(\ref{DIRECTK}) I am not impressed by \cpt~tests falling short 
of the $10^{-6}$ level.}. Its significance does not depend on whether the  
SM can reproduce it or not -- which is the most concise confirmation of 
how important it is. 
The HEP community can take pride 
in this achievement; the tale behind it is a most fascinating one about imagination and perseverance. The two groups and their predecessors deserve our respect; they 
have certainly earned my admiration. 

The experimental findings are consistent with CKM theory on the qualitative level, since the latter 
does not represent a superweak scenario even for strange decays due to the existence of Penguin 
operators. It is not inconsistent with it even quantitatively. One should keep in mind that within the 
SM $\epsilon^{\prime}/\epsilon_K$ has to be considerably suppressed.  
$\epsilon^{\prime}$ requires interference between $\Delta I = 1/2\,  \& \, 3/2$ amplitudes and is thus 
reduced by the `$\Delta I = 1/2$ rule': $|T(\Delta I = 3/2)/T(\Delta I = 1/2)| \sim 1/20$. Furthermore 
$\epsilon^{\prime}$ is generated by loop diagrams -- as is $\epsilon_K$; yet the top quark mass 
enhances $\epsilon_K$ powerlike -- $|\epsilon_K| \propto m_t^2/M_W^2$ -- whereas 
$\epsilon^{\prime}$ only logarithmically. When there is only one weak phase -- as is the case 
for CKM theory -- one has $|\epsilon^{\prime}/\epsilon_K| \propto {\rm log}m_t^2/m_t^2$, i.e. 
greatly reduced again for superheavy top quarks (revisit HW \# 1). 

CKM theory can go beyond such semiquantitative statements, but one should not expect a 
{\em precise} prediction from it in the near future. For the problem of uncertainties in the evaluation 
of hadronic matrix elements  is compounded by the fact that the two main contributions to 
$\epsilon^{\prime}$ are similar in magnitude, yet opposite in sign \cite{EPSPRIMETH}.

\subsubsection{CKM Theory at the End of the 2nd Millenium}

\begin{figure}[ht]
\begin{center}
\epsfig{
height=5truecm, width=10truecm,
        figure=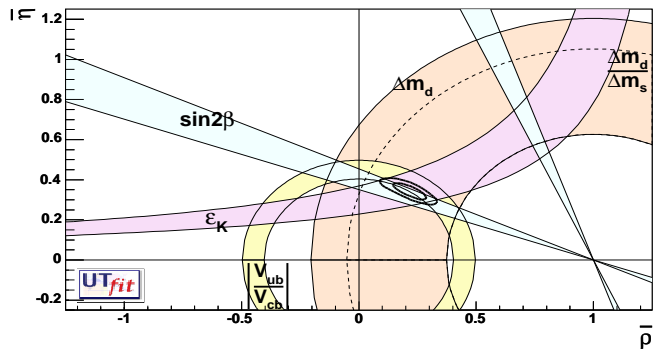}
\caption{The CKM Unitarity Triangle fit.
\label{CKMTRIANGLEFIT}  
}
\end{center}
\end{figure}

It is indeed true that large fractions of the observed values for $\Delta M_K$, $\epsilon_K$ and 
$\Delta M_B$ and even most of $\epsilon^{\prime}$ could be due to New Physics given 
the limitations in our theoretical control over hadronic matrix elements. Equivalently constraints from 
these and other data translate into `broad' bands in plots of the unitarity triangle, see 
Fig.\ref{CKMTRIANGLEFIT}. 
The problem with this statement is that it is not even wrong -- it misses the real point. Let me illustrate 
it by a local example first. If you plot the whereabouts of the students at this school on a local map of Varenna, you would find a seemingly broad band; however when you look at the `big' picture -- say a 
map of Europe -- you realize these students are very closely bunched together 
in one tiny spot on the map. 
This cannot be by accident, there has to be a good reason for it, which, I hope, is obvious in this specific case of students 
being concentrated in Varenna. Likewise for the problem at hand: Observables like 
$\Gamma (B \to l \nu X_{c,u})$, $\Gamma (K \to l \nu \pi)$, $\Delta M_K$, $\Delta M_B$, 
$\epsilon _K$ and sin$2\phi_1$ etc. represent very different dynamical regimes that proceed on 
time scales hat span several orders of magnitude.  The very fact that CKM theory 
can accommodate such diverse observables always within a factor two or better and 
relate them in such a manner that its parameters can be plotted as meaningful constraints on a 
triangle is highly nontrivial and -- in my view -- must reflect some underlying, yet unknown 
dynamical layer. Furthermore the CKM parameters exhibit an unusual hierarchical pattern -- 
$|V(ud)| \sim |V(cs)| \sim |V(tb)| \sim 1$, $|V(us)| \simeq |V(cd)| \simeq \lambda$, 
$|V(cb)| \sim |V(ts)| \sim {\cal O}(\lambda ^2)$, $|V(ub)| \sim |V(td)| \sim {\cal O}(\lambda ^3)$ -- 
as do the quark masses culminating in $m_t \simeq 175$ GeV. Picking such values for these parameters would have been seen as frivolous at best -- had they not been forced upon us by 
(independent) data. Thus I view it already as a big success for CKM theory that the experimental constraints on its parameters can be represented through triangle plots in a meaningful way. 

\begin{center}
{\bf Interlude: Singing the Praise of Hadronization}
\end{center}

\noindent 
Hadronization and nonperturbative dynamics in general are usually viewed as unwelcome 
complication, if not outright nuisances. A case in point was already mentioned: while 
I view the CKM predictions for $\Delta M_K$, $\Delta M_B$, $\epsilon_K$ to be in 
remarkable agreement with the data, significant contributions from New Physics could 
be hiding there behind the theoretical uncertainties due to lack of computational control 
over hadronization. Yet {\em without} hadronization bound states of quarks and antiquarks will not form; without 
the existence of kaons $K^0 - \bar K^0$ oscillations  obviously cannot occur. 
It is hadronization that provides the `cooling' of the (anti)quark degrees of freedom, which 
allows subtle quantum mechanical effects to add up coherently over macroscopic distances. 
Otherwise  
one would not have access to a super-tiny energy difference Im${\cal M}_{12} \sim 10^{-8}$ eV, 
which is very sensitive to different layers of dynamics, 
and indirect \cp~violation could not manifest itself. The same would hold for $B$ mesons and 
$B^0 - \bar B^0$ oscillations. 

\noindent 
To express it in a more down to earth way:  
\begin{itemize}
\item 
Hadronization leads to the formation of kaons and pions with masses exceeding 
greatly (current) quark masses.  
It is the {\em hadronic} phase space that suppresses the \cp~{\em conserving} rate for 
$K_L \to 3 \pi$ by a factor $\sim 500$, since the $K_L$ barely resides above the three pion threshold. 
\item 
It awards `patience'; i.e. one can `wait' for a pure $K_L$ beam to emerge after starting out with a 
beam consisting of $K^0$ and $\bar K^0$. 
\item 
It enables \cp~violation to emerge in the {\em existence} of a reaction, namely 
$K_L \to 2 \pi$ rather than an asymmetry; this greatly facilitates its observation. 
\end{itemize}
For these reasons alone we should praise hadronization as the hero in the tale of \cp~violation 
rather than the villain it is all too often portrayed. 

\begin{center}
{\bf End of Interlude}
\end{center}
 
Looking at the present CKM triangle fit shown in Fig. \ref{CKMTRIANGLEFIT} one realizes another 
triumph of CKM theory appears imminent: if one removed at present the constraints from $\epsilon_K$ 
and sin$2\phi_1$, i.e. \cp~constraints,  then a `flat' CKM `triangle' is barely 
compatible with the constraints on $|V(td)|$ from $\Delta M(B_d)$ and on 
$|V(ub)/V(cb)|$ from semileptonic $B$ decays -- processes not sensitive to \cp~violation per se.  However  
a measurement of 
$\Delta M(B_s)$ through resolving $B_s - \bar B_s$ oscillations in the near future would 
definitely require 
this triangle to be nontrivial: `\cp~insensitive observables would imply \cp~violation'! 

The first unequivocal manifestation of a Penguin contribution surfaced in radiative $B$ decays, 
first the exclusive channel $B \to \gamma K^*$ and subsequently the inclusive one 
$B\to \gamma X_s$ \cite{LANCERI}. These transitions represent flavour changing neutral currents and as such 
represent a one-loop, i.e. quantum process. 

By the end of the second millenium a rich and diverse body of data on flavour dynamics had been 
accumulated, and CKM theory provided a surprisingly successful description of it. This prompted some daring spirits to perform detailed fits of the CKM triangle to infer a rather accurate prediction for 
the \cp~asymmetry in $B_d \to \psi K_S$: 
\beq 
{\rm sin}2\phi_1 = 0.72 \pm 0.07
\eeq

\subsubsection{CKM `Exotica'}
\label{EXOT}

Here I list three classes of effects that are sensitive to \cp~violation.  
 
{\em Electric dipole moments:} 

\noindent 
The energy shift of a system placed inside a weak 
electric field can be expressed through an expansion in terms of the components of that 
field $\vec E$: 
\beq 
\Delta {\cal E} = d_i E_i + d_{ij}E_iE_j + {\cal O}(E^3) 
\eeq
The coefficients $d_i$ of the term linear in the electric field form a vector $\vec d$, 
called an electric dipole moment (EDM). For a {\em non-}degenerate system -- it does not have to be elementary -- one infers from symmetry considerations that this vector has to be proportional to 
that systems spin: 
\beq 
\vec d \propto \vec s
\eeq
Yet, since 
\beq
E_i \stackrel{\ot} \to E_i \; \; , \; \; s_i \stackrel{\ot} \to - s_i 
\eeq
under time reversal \ot , a non-vanishing EDM constitutes \ot~violation. 

No EDM has been observed yet; the upper bounds of the neutron and electron EDM read 
as follows \cite{RAMSEYLECT}: 
\bea 
d_N &<& 5 \cdot 10^{-26} \; {\rm e\, cm} \; \; \;  {\rm [from \; ultracold\; neutrons]}\\
d_e &<& 1.5 \cdot 10^{-27} \; {\rm e\, cm} \; \; \; {\rm [from\; atomic \; EDM]}
\eea
The experimental sensitivity achieved can be illustrated as follows: (i) An 
neutron EDM of $5\cdot 10^{-26}$ e cm of an object with a radius 
$r_N \sim 10^{-13}$ cm scales to a displacement of about 7 micron, i.e. less than 
the width of human hair, for an object of the size of the earth. (ii) Expressing the uncertainty 
in the measurement of the electron's magnetic dipole moment -- 
$\delta ((g-2)/2) \sim 10^{-11}$ in analogy to its EDM, one finds a sensitivity level of 
$\delta(F_2(0)/2m_e) \sim 2 \cdot 10^{-22}$ e cm compared to 
$d_e< 2\cdot 10^{-26}$ e cm. 

Despite the tremendous sensitivity reached -- the tour de force required  is nicely described in 
Prof. Ramsey's lectures \cite{RAMSEYLECT} -- these numbers are still several orders of magnitude above what is expected in CKM theory: 
\bea 
d_N^{CKM} &\leq &  10^{-30} \; {\rm e\, cm}  \\
d_e^{CKM} &\leq &  10^{-36} \; {\rm e\, cm} \; , 
\eea
where in $d_N^{CKM}$ I have ignored any contribution from the strong \cp~problem. 
These numbers are so tiny for reasons very specific to CKM theory, namely its chirality structure 
and the pattern in the quark and lepton masses. Yet New Physics scenarios with right-handed 
currents, flavour changing neutral currents, a non-minimal Higgs sector, heavy neutrinos etc. are 
likely to generate considerably larger numbers: $10^{-28} - 10^{-26}$ e cm represents a 
very possible range there quite irrespective of whether these new forces contribute to 
$\epsilon_K$ or not. This range appears to be within reach in the foreseeable future. 
It requires tremendous efforts -- yet the potential insights to be gained by finding a nonzero 
EDM somewhere are tremendous. 

{\em Pol$_{\perp}(\mu)$ in $K_{\mu 3}$ decays:} 

\noindent 
The correlation 
\beq 
{\rm Pol}_{\perp}(\mu) \equiv \frac{\langle \vec s(\mu) \cdot (\vec p(\mu) \times \vec p(\pi))\rangle}
{|\vec p(\mu) \times \vec p(\pi)|} 
\eeq
between momenta $\vec p$ and spin $\vec s$ 
for the mode $K \to \mu^+ \nu \pi$ 
is called \ot~{\em odd} or a \ot~{\em odd moment}, since it changes sign under time reversal 
\ot:
\beq 
\vec s \to - \vec s \; , \; \vec p \to - \vec p \; \; \; \Longrightarrow \; \; \; 
{\rm Pol}_{\perp}(\mu) \to - {\rm Pol}_{\perp}(\mu)
\eeq
It should be noted that ${\rm Pol}_{\perp}(\mu) \neq 0$ does not necessarily imply \ot~violation. 
In $K_L \to \mu^+ \nu \pi^-$ one expects 
${\rm Pol}_{\perp}(\mu) \leq {\cal O}(\alpha_S/\pi) \sim 10^{-3}$ even when \ot~ is conserved due 
to Coulomb interaction between the two charged particles in the final state. 

The reason that a parity odd moment implies \op~violation, whereas a non-zero \ot~odd moment 
can arise even with \ot~invariant dynamics, is due to the fact that the \op~operator is linear, while 
\ot~is {\em anti}linear: 
\beq 
\ot (\alpha |a\rangle ) = \alpha ^* \ot |a \rangle 
\eeq
This property of \ot~is enforced by the commutation relation $[X,P] = i \hbar$, since 
\bea 
\ot ^{-1} [X,P]\ot &=& - [X,P] \\
\ot ^{-1} i\hbar \ot  &=& - i \hbar 
\eea
This antilinearity comes into play when the transition amplitude is described {\em through second} 
(or even higher) order in the effective interaction, i.e. when final state interactions are 
included denoted symbolically by 
\beq 
\ot ^{-1} ({\cal L}_{eff}\Delta t + \frac{i}{2}({\cal L}_{eff}\Delta t)^2 + ...)\ot = 
{\cal L}_{eff}\Delta t -  \frac{i}{2}({\cal L}_{eff}\Delta t)^2 + ... \neq 
{\cal L}_{eff}\Delta t +  \frac{i}{2}({\cal L}_{eff}\Delta t)^2 + ... 
\eeq  
even for $[\ot, {\cal L}_{eff}]=0$. 

For $K^+ \to \mu^+ \nu \pi^0$ on the other hand there are no strong and only highly 
suppressed electromagnetic 
final state interactions; ${\rm Pol}_{\perp}(\mu) \geq 10^{-5}$ represents a genuine \ot~violation 
that has to be matched by a commensurate (direct) \cp~violation. Data show 
\beq 
{\rm Pol}_{\perp}(\mu) (K^+ \to \mu^+ \nu \pi^0) = (-1.7 \pm 2.3 \pm 1.1)\cdot 10^{-3} \; . 
\eeq 
CKM dynamics can produce merely 
${\rm Pol}_{\perp}^{CKM}(\mu) (K^+ \to \mu^+ \nu \pi^0) \sim 10^{-7}$. The effect  
is so tiny, since interference between helicity changing and conserving amplitudes is 
needed to induce ${\rm Pol}_{\perp}(\mu)$; such a contribution is highly suppressed in the SM 
with its purely left chiral charged currents. 

Charged Higgs exchange on the other hand can naturally produce ${\rm Pol}_{\perp}(\mu)$ 
through interference with $W$ exchange. 

{\em $\tau$ Decays:} 

\noindent 
They provide a very intriguing laboratory to search for 
\cp~violation. That would be caused by New Physics except for 
$\tau ^{\pm} \to \nu \pi^{\pm}K_S$, where $K_S$' slight tilt towards antimatter creates a 
\cp~asymmetry of $3.27 \cdot 10^{-3}$, as explained later \cite{BSCPTAU}. 

\subsection{Summary of Lecture II}
\label{SUM2}

Just before the turn of the millenium the situation could be characterized as follows: 
\begin{itemize}
\item 
\cp~violation had been established in 1964 through the observation of  
\beq 
{\rm BR}(K_L \ra \pi ^+ \pi ^-) = 2.3 \cdot 10^{-3} \neq 0 
\eeq
\item 
$K^- - \bar K^0$ oscillations exhibit a commensurate \ot~violation. Its most direct, although 
not most significant manifestation is given by the `Kabir Test' performed by 
the CPLEAR collaboration \cite{CPLEAR}: 
\beq 
A_T= \frac{\Gamma (K^0 \Rightarrow \bar K^0) - 
\Gamma (\bar K^0 \Rightarrow K^0)} 
{\Gamma (K^0 \Rightarrow \bar K^0) + 
\Gamma (\bar K^0 \Rightarrow K^0)} = 
(6.3 \pm 2.1 \pm 1.8) \; , 
\cdot 10^{-3} \; \; \; 
{\rm CPLEAR} 
\label{CPLEAR}
\eeq 
which is fully consistent with the expectation 
$A_T = 4 {\rm Re}\epsilon _K = 6.48 \cdot 10^{-3}$. 
\item 
Other manifestations had been found: 
\begin{itemize}
\item 
\beq 
\frac{{\rm BR}(K_L \ra l^+ \nu \pi ^-)}
{{\rm BR}(K_L \ra l^- \nu \pi ^+)} \simeq 1.00654 \neq 1 
\eeq
again consistent with $1+4{\rm Re}\epsilon_K = 1.00648$. 
\item 
A large \ot~odd moment was found in the rare $K_L$ mode --
BR$(K_L \ra \pi ^+ \pi ^- e^+ e^-) = 
(3.32 \pm 0.14 \pm 0.28 ) \cdot 10^{-7}$:    
With $\phi$ defined as the angle between the planes 
spanned by 
the two pions and the two leptons in the $K_L$ 
restframe:  
$$   
\phi \equiv \angle ( \vec n_l, \vec n_{\pi})
$$ 
\beq  
\vec n_l = \vec p_{e ^+}\times \vec p_{e ^-}/
|\vec p_{e ^+}\times \vec p_{e ^-}| \; , \;  
\vec n_{\pi} = \vec p_{\pi ^+}\times \vec p_{\pi ^-}/ 
|\vec p_{\pi ^+}\times \vec p_{\pi ^-}|
\label{PHISEHGAL}
\eeq    
one analyzes 
the decay rate as a function of $\phi$: 
\beq 
\frac{d\Gamma}{d\phi} = \Gamma _1 {\rm cos}^2\phi + 
\Gamma _2 {\rm sin}^2\phi + 
\Gamma _3 {\rm cos}\phi \, {\rm sin} \phi 
\eeq 
Since  
\beq 
{\rm cos}\phi \, {\rm sin} \phi = 
(\vec n_l \times \vec n_{\pi}) \cdot 
(\vec p_{\pi ^+} + \vec p_{\pi ^-}) 
(\vec n_l \cdot \vec n_{\pi})/
|\vec p_{\pi ^+} + \vec p_{\pi ^-}| 
\eeq
one notes that 
\beq 
{\rm cos}\phi \, {\rm sin} \phi \; \; \; 
\stackrel{{\bf T},{\bf CP}}{\longrightarrow} \; \; \; 
- \; {\rm cos}\phi \, {\rm sin} \phi 
\eeq    
under both \ot~ and \cp~transformations; i.e. the observable  
$\Gamma _3$ represents a \ot~- and \cp~-odd correlation. 
It can be projected out by comparing the $\phi$ 
distribution integrated over two quadrants: 
\beq 
A = 
\frac{\int _0^{\pi/2} d\phi \frac{d\Gamma}{d\phi} - 
\int _{\pi /2}^{\pi} d\phi \frac{d\Gamma}{d\phi}}
{\int _0^{\pi} d\phi \frac{d\Gamma}{d\phi}} = 
\frac{2\Gamma _3}{\pi (\Gamma _1 + \Gamma _2)} 
\eeq
It was first measured by KTEV and then confirmed by NA48:  
\beq 
A = (13.8 \pm 2.2)\% \, .
\label{KTEVSEHGAL2}
\eeq 
$A\neq 0$ is induced by $\epsilon_K$, the \cp~violation in the $K^0 - \bar K^0$ mass matrix, 
leading to the prediction \cite{SEGHALKL}
\beq 
A = (14.3 \pm 1.3)\% \, .
\eeq
The observed value for the \ot~odd moment $A$ is fully consistent with \ot~violation. Yet 
$A\neq 0$ {\em by itself} does not establish 
\ot~violation \cite{BSTODD}. 
\end{itemize} 
\end{itemize}
\begin{itemize}
\item 
Very sensitive searches for {\em direct} \cp~violation had been 
undertaken with the following measurements: 
\beq 
{\rm Re} \frac{\epsilon ^{\prime}}{\epsilon _K} = 
\left\{ 
\begin{array}{l} 
(2.3 \pm 0.7) \cdot 10^{-3} \; \; NA\, 31 \\ 
(0.6 \pm 0.58 \pm 0.32 \pm 0.18) \cdot 10^{-3} \; \; 
E\, 731 
\end{array}  
\right. 
\eeq
While the NA 31 number represented strong evidence for direct \cp~violation, the 
E 731 did not. The experimental situation was thus not conclusive.  
\item 
An impressive amount of experimental ingenuity, acumen and 
commitment 
went into producing this list. We had known since 1964 that \cp~violation 
unequivocally exists in nature. After 34 years of dedicated experimentation 
all its direct manifestations (i.e. ignoring the baryon number of the Universe) could 
still be  
characterized by a {\em single} non-vanishing quantity: 
\beq 
{\rm Im} M_{12} \simeq 1.1 \cdot 10^{-8} \; eV \neq 0 \; \; \hat = \; \; 
\Phi (\Delta S = 2) \equiv {\rm arg}\frac{M_{12}}{\Gamma_{12}} = (6.54 \pm 0.24)\cdot 10^{-3} 
\eeq  
\item 
The KM ansatz allows us to incorporate \cp~violation into the 
Standard Model. Yet it does not regale us with an understanding. 
Instead it relates the origins of \cp~violation to central 
mysteries of the Standard Model: Why are there families? 
Why are there three of those?  What is underlying 
the observed pattern in the fermion masses? 
\end{itemize}
Just before the turn of the millenium, in 1999,  the situation was clarified, direct  
\cp~violation was established in $K_L$ decays on both sides of the Atlantic, at CERN as well as 
at FNAL with the world average
\beq 
\left. {\rm Re} \frac{\epsilon ^{\prime}}{\epsilon _K}\right|_{WA} =  (1.63 \pm 0.22)\cdot 10^{-3} \; . 
\eeq

Also other manifestations of \cp~ and \ot~violation have been pursued with great vigour:    
\beq
{\rm Pol}_{\perp}^{K^+}(\mu ) = (-1.85\pm 3.60) \cdot 10^{-3} 
\eeq
\beq 
d_N < 12 \cdot 10^{-26} \; \; e\, cm 
\eeq   
\beq 
d_{Tl} = (1.6 \pm 5.0) \cdot 10^{-24} \; \; e\, cm 
\stackrel{theor.}{\Longrightarrow} 
d_e = (-2.7 \pm 8.3) \cdot 10^{-27} \; \; e\, cm
\eeq 
No effect has been observed yet -- and CKM theory predicts that none could realistically be ever 
observed in these cases. 

The situation at the beginning of the new millenium could then be sketched as follows: 
\begin{itemize}
\item 
The KM ansatz succeeds in {\em accommodating} the data in an 
unforced way: $\epsilon _K$ emerges to be naturally 
small, $\epsilon ^{\prime}$ naturally tiny (once the huge 
top mass is accounted for), the EDM's for neutrons [electrons] 
naturally (tiny)$^2$ [(tiny)$^3$] etc. 
\item 
\ot~violation manifesting itself in 
${\rm Pol}_{\perp}^{K^+}(\mu ) \neq 0$ requires dynamics involving both chirality 
conserving and violating weak couplings combined with a relative phase between 
them. $W$ exchange from the SM provides the former, while exchange of a 
(charged) Higgs can generate the latter. Searching for ${\rm Pol}_{\perp}^{K^+}(\mu )$ 
thus represents a sensitive probe for nonminimal Higgs dynamics and therefore should be 
encouraged. 
\item 
CKM theory can be characterized as follows: One takes a model with a set of mass related 
basic quantities -- fermion masses, CKM parameters -- and assigns them values that any 
sober person would view as frivolous, were those not forced upon us by data, in particular 
since we have no deeper understanding of mass generation, especially for fermions. One 
would have little reason to expect success in describing flavour dynamics proceeding in 
diverse environments on vastly different scales. Yet it did seem to work, at least on a semi-quantitative 
level. 
\item 
Cynics might feel that reproducing a single number after the fact does not pose a stiff  
challenge to a determined theorist. Yet CKM theory did more. Due to the unexpectedly `long' 
$B$ meson lifetimes and the huge top quark mass it predicted that a phenomenon that had been 
tiny in kaon decays, namely \cp~violation, had to be close to maximal, i.e 100 \%, in certain 
$B$ deccay modes, in particular, but not exclusively, in $B_d \to \psi K_S$. There was, to steal a phrase 
from the US political scene in the1980's `no plausible deniability' 
or following an older tradition: "Hic Rhodos, hic salta!" for CKM theory. 

\item 
The fact that this `CKM paradigm of large \cp~violation' exists and can be probed 
experimentally is due to several favourable factors, whose confluence must be seen as 
a gift from nature, who had 
\begin{enumerate}
\item 
arranged for a huge top quark mass, 
\item 
a `long' $B$ lifetime, 
\item 
the $\Upsilon (4S)$ resonance being above $B \bar B$, yet below $B\bar B^*$ thresholds and 
\item
had presented us previously with charm hadrons, which prompted the development of 
microvertex detectors with an effective resolution that was needed for $B$ decays. 
\end{enumerate} 
\item 
One of the central theoretical tools are effective field theories: after writing down the 
Lagrangian of the`true' theory at some high energy scale $\Lambda_{UV}$, 
one evolves it down to lower 
energy scales $\mu$ by integrating out all `heavy' fields, i.e. those 
with frequencies higher than $\mu$ and retaining only the `light' fields: 
\beq 
{\cal L}(\Lambda_{UV} \longrightarrow {\cal L}_{eff}(\mu) = 
\sum_i c_i (\mu, \Lambda_{UV}) O_i(\mu) \; , 
\eeq
where the local operators $O_i$ are built exclusively from the light fields. The c number coefficients 
$c_i$ are shaped by the heavy degrees of freedom; 
{\em thus they provide the gateway for New Physics}. 
						
\end{itemize}
Some of us had concluded that while the observed pattern of the CKM theory strongly points to 
the existence of 
a deeper level of dynamics underlying it, its phenomenological sucesses suggest that one cannot 
{\em count} on New Physics intervening in $B $ decays in a {\em numerically} massive manner.

\section{Lecture III: : \cp~Violation in $B$ Decays -- the `Expected' Triumph of a 
Very Peculiar Theory}
\label{LECTIII}

As explained in the previous lecture, within CKM theory one is unequivocally lead to a paradigm 
of large \cp~violation in $B$ decays. This realization became so widely accepted that 
two $B$ factories 
employing $e^+e^- \to \Upsilon (4S) \to B \bar B$ were constructed -- one at KEK in Japan and one 
in Stanford in the US -- together with specialized detectors, around which two collaborations 
gathered, the BELLE and BABAR collaborations, respectively. Details can be find in the talks 
by Aihara, Lanceri, Giorgi and Hitlin at this school.

\subsection{Establishing the CKM Description as a Theory -- 
\cp~Violation in $B$ Decays}

\begin{figure}[ht]
\begin{center}
\epsfig{
height=5truecm, width=10truecm,
        figure=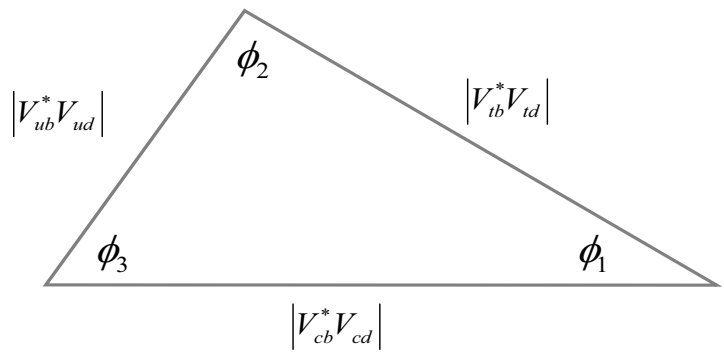}
\caption{The CKM Unitarity Triangle 
\label{CKMTRIANGLENOT}  
}
\end{center}
\end{figure}

The three angles $\phi_{1,2,3}$ in the CKM unitarity triangle 
(see Fig.\ref{CKMTRIANGLENOT}  for notation) can be determined -- or at least 
probed -- through \cp~asymmetries in the three modes $B_d(t) \to \psi K_S, \pi^+\pi^-$ 
and $B_d \to K^+ \pi^-$. Those will be addressed in three acts plus two interludes. 

\subsubsection{Act 1: $B_d(t) \to \psi K_S$ and $\phi_1$ (a.k.a. $\beta$) }
\label{ACT1}

The first published result on the \cp~asymmetry in $B_d \to \psi K_S$ was actually obtained by 
the OPAL collaboration at LEP I \cite{OPAL}: 
\beq 
{\rm sin}2\phi_1 = 3.2 ^{+1.8}_{-2.0} \pm 0.5 \; , 
\eeq
where the `unphysical' value of sin$2\phi_1$ is made possible, since a large background subtraction 
has to be performed. 
The first value inside the physical range was obtained by CDF \cite{CDFPHI1} : 
\beq 
{\rm sin}2\phi_1 =0.79 \pm 0.44 
\eeq
In 2000 the two $B$ factory collaborations BABAR and BELLE presented their first measurements 
\cite{LANCERI}: 
\beq 
{\rm sin}2\phi_1 =  
\left\{ 
\begin{array}{l} 
0.12 \pm 0.37 \pm 0.09 \; \; {\rm BABAR\;  '00} \\ 
0.45 \pm 0.44 \pm 0.09 \; \; 
{\rm BELLE\;  '00}
\end{array}  
\right. 
\eeq
Already one year later these inconclusive numbers turned into conclusive ones, and 
the first \cp~violation outside the $K^0 - \bar K^0$ complex was established: 
\beq 
{\rm sin}2\phi_1 =  
\left\{ 
\begin{array}{l} 
0.59 \pm 0.14 \pm 0.05 \; \; {\rm BABAR\; '01} \\ 
0.99 \pm 0.14 \pm 0.06 \; \; 
{\rm BELLE\;  '01}
\end{array}  
\right. 
\eeq
By 2003 the numbers from the two experiments had well converged  
\beq 
{\rm sin}2\phi_1 =  
\left\{ 
\begin{array}{l} 
0.741 \pm 0.067 \pm 0.03 \; \; {\rm BABAR\; '03} \\ 
0.733 \pm 0.057 \pm 0.028 \; \; 
{\rm BELLE\; '03}
\end{array}  
\right. 
\eeq
allowing one to state just the world averages, which is actually a BABAR/BELLE average: 
\beq 
{\rm sin}2\phi_1 =  
\left\{ 
\begin{array}{l} 
0.726 \pm 0.037  \; \; {\rm WA\;  '04} \\ 
0.685 \pm 0.032  \; \; 
{\rm WA \; '05}
\end{array}  
\right. 
\label{WA0405} 
\eeq
{\bf The \cp~asymmetry in $B_d \to \psi K_S$ is there, is huge and as expected even quantitatively.}  
For CKM fits based on constraints from $|V(ub)/V(cb)|$, $B^0 - \bar B^0$ oscillations and -- as 
the only \cp~sensitive observable -- $\epsilon_K$ yield 
\beq 
{\rm sin}2\phi_1|_{\rm CKM}  = 0.725 \pm 0.065 \; , 
\label{PHI1CKM}
\eeq
which is in impressive agreement with the data. This is illustrated by 
Fig.\ref{CKMTRIANGLEFIT} showing these constraints. 

Before turning to Acts Two and Three, I will present an Interlude. 

\subsubsection{Interlude 1: "Praise the Gods Twice for EPR Correlations"} 
\label{INTER1} 

 The BABAR and BELLE analyses are based on a glorious application of quantum mechanics and in 
 particular EPR correlations\cite{EPR}.  The \cp~asymmetry in $B_d \to \psi K_S$ had been predicted to 
 exhibit a peculiar dependence on the time of decay, since it involves $B_d - \bar B_d$ oscillations 
 in an essential way: 
\beq 
{\rm rate} (B_d(t)[\bar B_d(t)] \to \psi K_S) \propto e^{-t/\tau _B} (1- [+] A {\rm sin}\Delta m_B t) \; , 
\label{ASYM}
\eeq  
At first it would seem that an asymmetry of the form given in Eq.(\ref{ASYM}) could not be measured for practical reasons. For in the reaction
\beq 
 e^+e^- \to \Upsilon (4S) \to B_d \bar B_d
\label{UPS4S}
\eeq
the point where the $B$ meson pair is produced is ill determined due to the finite size of the electron and positron beam spots: the latter amounts to about 1 mm in the longitudinal direction, while a $B$ meson typically travels only about a quarter of that distance before it decays. 
It would then seem that the length of the flight path of the $B$ mesons is poorly known and that averaging over this ignorance would greatly dilute or even eliminate the signal. 
 
It is here where the existence of a EPR correlation comes to the rescue. While the two $B$ mesons in the reaction of Eq.(\ref{UPS4S}) oscillate back and forth between a $B_d$ and $\bar B_d$, they change their flavour identity in a {\em completely correlated} way.  For the $B \bar B$ pair forms a \oc~{\em odd} state; Bose statistics then tells us that there cannot be two identical flavour hadrons in the final state: 
\beq 
 e^+e^- \to \Upsilon (4S) \to B_d \bar B_d \not \to B_d B_d, \; \bar B_d \bar B_d
 \label{NOTID}
 \eeq
 Once one of the $B$ mesons decays through a flavour specific mode, say $B_d \to l^+\nu X$ 
 [$\bar B_d \to l^- \bar \nu X$], then we know unequivocally that the other $B$ meson was a 
 $\bar B_d$ [$B_d$] at {\em that} time. The time evolution of $\bar B_d(t) [B_d(t)] \to \psi K_S$ as described by 
 Eq.(\ref{ASYM}) starts at {\em that} time as well; i.e., the relevant time parameter is the {\em interval between} 
 the two times of decay, not those times themselves. That time interval is related to -- and thus can be inferred from -- 
 the distance between the two decay vertices, which is well defined and can be measured. 
 
\begin{figure}[t]
 \begin{center}
  \mbox{\psfig{file=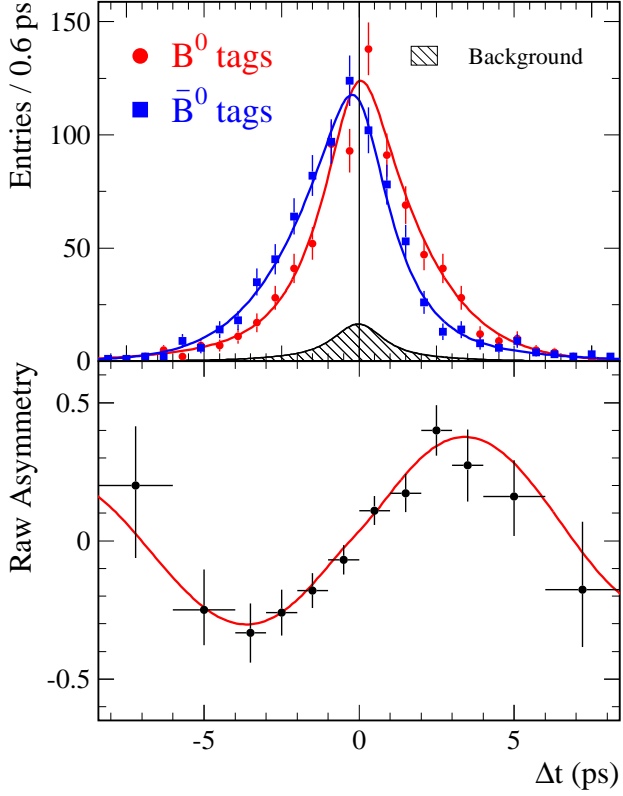,width=6in}} 
  \caption{The observed decay time distributions for $B^0$ (red) and
  $\bar B^0$ (blue) decays.  \label{timeasymmfig}}
 \end{center}
\end{figure}
 
 
 The great {\em practical} value of the EPR correlation is instrumental for another consideration as well, namely how to 
 see directly from the data that \cp~violation is matched by T violation. Fig.\ref{timeasymmfig} shows two distributions, one for the 
 interval  $\Delta t$ between the times of decays $B_d \to l^{+}X$ and $\bar B_d \to \psi K_S$ and the other one for the 
\cp~conjugate process $\bar B_d \to l^{-}X$ and $B_d \to \psi K_S$. They are clearly different proving that \cp~is broken. Yet they show more: the shape of the two distributions is actually the same 
(within experimental uncertainties) the only difference being that the average of $\Delta t$ is 
{\em positive} for $(l^-X)_{\bar B} (\psi K_S)$ and 
 {\em negative} for $(l^+X)_{B} (\psi K_S)$ events. I.e., there is a (slight) preference for 
 $B_d \to \psi K_S$ 
 [$\bar B_d \to \psi K_S$] to occur {\em after} [{\em before}] and thus more [less] slowly (rather than just more rarely) than $\bar B \to l^-X$ [$B \to l^+ X$]. Invoking \cpt~invariance merely for semileptonic $B$ decays -- yet not for nonleptonic 
 transitions -- synchronizes the starting point of the $B$ and $\bar B$ decay `clocks', and the 
 EPR correlation keeps them synchronized. We thus see that \cp~and 
\ot~violation are `just' 
 different sides of the same coin. 
 As explained above, EPR correlations are essential for 
 this argument! 
 
 The reader can be forgiven for feeling that this argument is of academic interest only, since 
 \cpt~invariance of all 
 processes is based on very general arguments. Yet the main point to be noted is that EPR correlations, which 
 represent some of quantum mechanics' most puzzling features, serve as an essential precision tool, which is routinely used in these measurements. I feel it is thus inappropriate to refer to EPR correlations as a paradox.

\subsubsection{Act 2: $B_d(t) \to \pi^+ \pi^-$ and $\phi_2$ (a.k.a. $\alpha$)}
\label{ACT2}

The situation is theoretically more complex than for $B_d (t) \to \psi K_S$ due to two 
reasons: 
\begin{itemize}
\item 
While both final states $\pi \pi$ and $\psi K_S$ are \cp~eigenstates, the former unlike the latter 
is not reached through an isoscalar transition. The two pions can form an $I=0$ or $I=2$ 
configuration (similar to $K\to 2\pi$), which in general will be affected differently by the strong interactions. 
\item 
For all practical purposes $B_d \to \psi K_S$ is described by two tree diagrams 
representing the two effective operators 
$(\bar c_L\gamma _{\mu}b_L)(\bar s_L\gamma ^{\mu}c_L)$ and 
$(\bar c_L\gamma _{\mu}\lambda _i b_L)(\bar s_L\gamma ^{\mu}\lambda _ic_L)$ with the 
$\lambda_i$ representing the $SU(3)_C$ matrices. Yet for $B\to \pi \pi$ we have 
effective operators 
$(\bar d_L\gamma _{\mu}\lambda _i b_L)(\bar q\gamma ^{\mu}\lambda _i q)$ generated by 
the Cabibbo suppressed Penguin loop diagrams in addition to the two tree 
operators $(\bar u_L\gamma _{\mu}b_L)(\bar d_L\gamma ^{\mu}u_L)$ and 
$(\bar u_L\gamma _{\mu}\lambda _i b_L)(\bar d_L\gamma ^{\mu}\lambda _iu_L)$. 
\end{itemize} 
This greater complexity manifests itself already in the phenomenological description of 
the time dependent \cp~asymmetry: 
\beq 
\frac{R_+(\Delta t) - R_-(\Delta t)}{R_+(\Delta t) + R_-(\Delta t)} = S {\rm sin}(\Delta M_d \Delta t) - 
C {\rm cos}(\Delta M_d \Delta t) \; , \;  S^2 + C^2 \leq 1 
\eeq
where $R_{+[-]}(\Delta t)$ denotes the rate for 
$B^{tag}(t)\bar B_d(t+\Delta)[\bar B^{tag}(t)B_d(t+\Delta)]$ and 
\beq 
S = \frac{2{\rm Im}\frac{q}{p}\bar \rho_{\pi^+\pi^-}}{1+\left|\frac{q}{p}\bar \rho_{\pi^+\pi^-}\right| ^2} \; , \; \; 
C = \frac{1 - \left|\frac{q}{p}\bar \rho_{\pi^+\pi^-}\right| ^2}
{1+\left|\frac{q}{p}\bar \rho_{\pi^+\pi^-}\right| ^2} 
\eeq
As before, due to the EPR correlation between the two neutral $B$ mesons, it is the 
{\em relative} time interval $\Delta t$ between the two $B$ decays that matters, not their 
lifetime. The new feature is that one has also a cosine dependence on $\Delta t$. 

$C\neq 0$ obviously represents {\em direct} \cp~violation. Yet what is often overlooked is that 
also $S$ can reveal such \cp~violation. For if one studies $B_d$ decays into two \cp~eigenstates 
$f_a$ and $f_b$ and finds   
\beq 
S(f_a) \neq \eta(f_a) \eta(f_b) S(f_b) 
\eeq
with $\eta_i$ denoting the \cp~parities of $f_i$, then one has established 
{\em direct} \cp~violation. For the case under study that means even if $C(\pi \pi )= 0$, 
yet $S(\pi^+\pi^-) \neq - S(\psi K_S$, one has observed unequivocally {\em direct} 
\cp~violation. One should note that such direct \cp~violation might not necessarily induce 
$C\neq 0$. For the latter requires, as explained in Prof. Sanda's lectures and below in 
Sect. \ref{INT2} (see Eq.(\ref{PARTIAL})), that two different amplitudes 
contribute coherently to $B_d \to f_b$ with non-zero relative weak as well as strong phases. 
$S(f_a) \neq \eta(f_a) \eta(f_b) S(f_b)$ on the other hand only requires that the two 
overall amplitudes for $B_d \to f_a$ and $B_d \to f_b$ possess a relative phase. This can be 
illustrated with a familiar example from CKM dynamics: If there were no Penguin operators for 
$B_d \to \pi^+\pi^-$ (or it could be ignored quantitatively), one would have 
$C(\pi^+\pi^-) = 0$, yet at the same time $S(\psi K_S) = {\rm sin}(2\phi_1)$ together with 
$S(\pi^+\pi^-)  = {\rm sin}(2\phi_2)$.  

BELLE finds  
\beq 
S = - 0.67 \pm 0.16 \pm 0.06 \; , \; \; C =  0.56 \pm 0.12 \pm 0.06 \; , 
\eeq
which amounts to a 5.2 $\sigma$ CP asymmetry and {\em direct} CP violation 
with 3.3 $\sigma$ significance. For  {\em without} direct CP violation one would have to 
find $C=0$ and $S = - {\rm sin}2\phi_1$ \cite{ELEFANT}. However this signal has so far not been confirmed 
by BABAR, which has found $S=-0.30 \pm 0.17 \pm 0.03$, $C=- 0.09 \pm 0.15 \pm 0.04$. Future data  will hopefully clarify this unsettled situation. 

Once this has been settled, one can take up the challenge of extracting a value for 
$\phi_2$ from the data 
\footnote{The complications due to the presence of the Penguin contribution are all too often referred to as `Penguin pollution'. Personally I find it quite unfair to blame our lack of theoretical control on water fowls rather than on the guilty party, namely us. }. This can be done in a model independent way by 
analyzing $B_d(t) \to \pi^+\pi^-$, $\pi^0\pi^0$ and $B^{\pm} \to \pi^{\pm}\pi^0$ transitions 
and performing an isospin decomposition. For the Penguin contribution cannot affect 
$B_d(t) \to [\pi \pi]_{I=2}$ modes. Unfortunately there is a seriouus experimental bottle neck, 
namely to study $B_d (t) \to \pi^0\pi^0$ with sufficient accuracy \cite{LANCERI}. Therefore alternative 
decays have been suggested, in particular $B \to \rho \pi$ and $\rho \rho$. I will comment on them 
later.

\subsubsection{Act 3: $B_d \to K^+\pi^-$}
\label{ACT3}
It was pointed out in a seminal paper \cite{SONIBKPI} that (rare) transitions like 
$\bar B_d \to K^- + \pi$'s have the ingredients for sizeable direct \cp~asymmetries: 
\begin{itemize}
\item 
Two different amplitudes can contribute coherently, namely the highly CKM 
suppressed tree diagram with $b \to u \bar u s$ and the Penguin diagram 
with $b \to s \bar qq$. 
\item 
The tree diagram contains a large weak phase from $V(ub)$. 
\item 
The Penguin diagram with an internal charm quark loop exhibits an imaginary part, which can be 
viewed -- at least qualitatively -- as a strong phase generated by the production and subsequent 
annihilation of a $c \bar c$ pair (the diagram with an internal $u$ quark loop acts merely as a 
subtraction point allowing a proper definition of the operator). 
\item 
While the Penguin diagram with an internal top quark loop is actually not essential, the 
corresponding effective operator can 
be calculated quite reliably, since integrating out first the top quarks and then the $W$ boson leads to a truly local operator. Determining its matrix elements is however another matter. 

\end{itemize}
To translate these features into accurate numbers represents a formidable task, we have not mastered yet. In Ref. \cite{CECILIABOOK} an early and detailed effort was made to treat $\bar B_d \to K^-\pi^+$ theoretically with the following results: 
\beq 
{\rm BR}(\bar B_d \to K^- \pi ^+) \sim 10^{-5} \; , \; \; 
A_{\cp} \sim  - 0.10
\label{BKPIPRED}
\eeq 
Those numbers turn out to be rather prescient, since they are in gratifying agreement with the data 
$$
{\rm BR}(\bar B_d \to K^- \pi ^+) = (1.85 \pm 0.11) \cdot 10^{-5} 
$$
\beq 
A_{\cp} =
\left \{  
\begin{array}{l} 
 - 0.133 \pm 0.030 \pm 0.009  \; \; \; BABAR \\
- 0.113 \pm 0.021   \; \; \; BELLE
\end{array}  
\right. 
\label{BKPIDATA}
\eeq 
Cynics might point out that the authors in  \cite{CECILIABOOK} did not give a specific estimate of the 
theoretical uncertainties in Eq.(\ref{BKPIPRED}). More recent authors have been more ambitious -- 
with somewhat mixed success. I list the predictions inferred from pQCD \cite{pQCD} and 
QCD Factorization \cite{QCDFACT} and the data for the three modes 
$\bar B_d \to K^-\pi^+$ and $B^- \to K^- \pi^0, \, \bar K^0 \pi^-$: 
\beq 
A_{\cp} (B_d \to K^-\pi^+)=
\left \{  
\begin{array}{l} 
 - 0.133 \pm 0.030 \pm 0.009 \; \; \; {\rm BABAR} \\
- 0.113 \pm 0.021 \; \; \; {\rm BELLE}\\
-0.09 ^{+0.05+0.04}_{-0.08-0.06} \; \; \; {\rm pQCD} \\
+ 0.05 \pm 0.09 \; \; \; {\rm QCD\; Fact.} 
\end{array}  
\right. 
\label{K-PI+}
\eeq 
\beq 
A_{\cp} (B^- \to K^-\pi^0)=
\left \{  
\begin{array}{l} 
 + 0.06 \pm 0.06 \pm 0.01 \; \; \; BABAR \\
+ 0.04 \pm 0.04 \pm 0.02 \; \; \; BELLE\\
-0.01 ^{+0.03+0.03}_{-0.05-0.05}\; \; \;  {\rm pQCD}\\
+ 0.07 \pm 0.09 \; \; \; {\rm QCD\; Fact.}  
\end{array}  
\right. 
\label{K-PI0}
\eeq 
\beq 
A_{\cp} (B^- \to \bar K^0\pi^-)=
\left \{  
\begin{array}{l} 
 - 0.09 \pm 0.05 \pm 0.01  \; \; \; BABAR \\
+ 0.05 \pm 0.05 \pm 0.01   \; \; \; BELLE\\
+ 0.00 \; \; \; {\rm pQCD}\\
+ 0.01 \pm 0.01\; \; \; {\rm QCD\; Fact.} 
\end{array}  
\right. 
\label{K0PI-}
\eeq

\subsubsection{Interlude 2: On Final State Interactions and \cpt~Invariance}
\label{INT2}

Due to \cpt~invariance  
\cp~violation can be implemented only through a  
complex phase in some effective couplings.  For it to become  
observable two different, yet coherent amplitudes have to  
contribute to an observable. There are two types of scenarios for 
implementing this requirement: 
\begin{enumerate}
\item 
When studying a final state $f$ that can be reached by a $\Delta B = 1$ transition 
from $B^0$ as well as $\bar B^0$, then $B^0-\bar B^0$ oscillations driven by $\Delta B =2$ dynamics 
provide the second amplitude, the weight of 
which varies with time. 
\item 
Two different $\Delta B =1$ amplitudes ${\cal M}_{a,b}$ of fixed ratio --
distinguished by,  say, their isospin content -- exist leading 
{\em coherently} to the same final state: 
\beq  
T(B \to f) = \lambda_a {\cal M}_a  + \lambda_b {\cal M}_b   
\label{T2M}
\eeq 
I have factored out the {\em weak} couplings $\lambda_{a,b}$ while allowing 
the amplitudes ${\cal M}_{a,b}$ to be still complex due to strong or electromagnetic FSI.   
For the \cp~conjugate reaction one has  
\beq  
T(\bar B \to \bar f) = \lambda_a^* {\cal M}_a + 
\lambda_b^* {\cal M}_b  
\eeq 
It is important to note that the reduced amplitudes 
${\cal M}_{a,b}$ remain 
unchanged, since strong and electromagnetic forces conserve \cp .  
Therefore we find   
\beq  
\Gamma (\bar B \to \bar f) - \Gamma (B \to  f) = 
\frac{2{\rm Im}\lambda_a \lambda_b^* \cdot {\rm Im}{\cal M}_a{\cal M}_b^*}
{|\lambda_a|^2|{\cal M}_a|^2 + |\lambda_b|^2|{\cal M}_b|^2 
+ 2{\rm Re}\lambda_a \lambda_b^* \cdot {\rm Re}{\cal M}_a{\cal M}_b^*} 
\label{PARTIAL} 
\eeq 
i.e. for a \cp~asymmetry to become observable, two  
conditions have to satisfied simultaneously irrespective of the underlying dynamics: 
\begin{itemize}
\item 
Im $\lambda_a \lambda_b^* \neq 0$, i.e.  there has to be a relative phase between the weak 
coulings $\lambda_{a,b}$. 
\item 
Im${\cal M}_a{\cal M}_b^* \neq 0$, i.e. final state interactions (FSI) have to induce a phase shift 
between ${\cal M}_{a,b}$. 
\end{itemize}

\end{enumerate} 
While this is well known, it is often not fully appreciated that \cpt~invariance places constraints on 
the phases of the ${\cal M}_{a,b}$. For it implies much more than equality of masses and lifetimes of particles and antiparticles. It tells us that the 
widths for {\em sub}classes of transitions for particles and 
antiparticles have to coincide already, either identically or 
at least practically. Just writing down strong phases in an equation 
like Eq.(\ref{T2M}) does {\em not automatically} satisfy \cpt~constraints.

I will illustrate this feature first with two 
simple examples and then express it in more general terms. 
\begin{itemize}
\item 
\cpt~invariance already implies $\Gamma (K^- \to \pi ^- \pi ^0) = 
\Gamma (K^+ \to \pi ^+ \pi ^0)$ up to small electromagnetic 
corrections, since in that case there are no other channels it 
can rescatter with. 
\item 
While 
$\Gamma (K^0 \to \pi^+\pi^-) \neq \Gamma (\bar K^0 \to \pi^+\pi^-)$ 
and $\Gamma (K^0 \to \pi^0\pi^0) \neq \Gamma (\bar K^0 \to \pi^0\pi^0)$ 
one has 
$\Gamma (K^0 \to \pi^+\pi^- + \pi^0\pi^0) = 
\Gamma (\bar K^0 \to \pi^+\pi^- + \pi^0\pi^0)$. 
\item 
Let us now consider a scenario where a particle $P$ and its antiparticle 
$\bar P$ can each decay into two final states only, namely $a,b$ and 
$\bar a, \bar b$, respectively 
\cite{WOLFFSI,URIFSI}. Let us further assume that strong (and
electromagnetic) forces drive transitions among $a$ and $b$ -- and 
likewise for $\bar a$ and $\bar b$ -- as described by an S matrix 
${\cal S}$. The latter can then be decomposed into two parts  
\beq 
{\cal S} = {\cal S}^{diag} + {\cal S}^{off-diag} \; , 
\eeq
where ${\cal S}^{diag}$ contains the diagonal transitions 
$a \Rightarrow a$, $b \Rightarrow b$ 
\beq 
{\cal S}^{diag}_{ss} = e^{2i\delta _s} \; , s=a,b 
\eeq
and 
${\cal S}^{off-diag}$ the off-diagonal ones 
$a \Rightarrow b$, $b \Rightarrow a$:
\beq 
{\cal S}^{off-diag}_{ab} = 
2i{\cal T}^{resc}_{ab} e^{i(\delta _a + \delta _b)} 
\eeq 
with 
\beq 
{\cal T}^{resc}_{ab} = {\cal T}^{resc}_{ba} = 
({\cal T}^{resc}_{ab})^* \; , 
\eeq
since the strong and electromagnetic forces driving the 
rescattering 
conserve \cp~and \ot . The resulting S matrix is unitary to first 
order in ${\cal T}^{resc}_{ab}$. 
\cpt~invariance implies the following relation between the 
weak decay amplitude of $\bar P$ and $P$: 
\bea 
T(P \to a) &=& e^{i\delta _a}\left[ T_a + T_b i{\cal T}^{resc}_{ab}\right] \\
T(\bar P \to \bar a) &=& e^{i\delta _a}\left[ T^*_a +T^*_bi{\cal T}^{resc}_{ab}\right]
\eea
and thus 
\beq 
\Delta \gamma (a) \equiv |T(\bar P \to \bar a)|^2 - 
|T(P \to  a)|^2 = 4 {\cal T}^{resc}_{ab} {\rm Im}T^*_a T_b \; ; 
\label{DELTAA}
\eeq
likewise
\beq 
\Delta \gamma (b) \equiv |T(\bar P \to \bar b)|^2 - 
|T(P \to  b)|^2 = 4 {\cal T}^{resc}_{ab} {\rm Im}T^*_b T_a 
\eeq
and therefore as expected 
\beq 
\Delta \gamma (b) = - \Delta \gamma (b)
\eeq
Some further features can be read off from Eq.(\ref{DELTAA}):
\begin{enumerate} 
\item 
If the 
two channels that rescatter have comparable widths -- 
$\Gamma (P \to a) \sim \Gamma (P \to b)$ -- one would like the
rescattering 
$b \leftrightarrow a$ to proceed via the usual strong forces; for
otherwise  the asymmetry $\Delta \Gamma $ is suppressed relative to these 
widths by the
electromagnetic coupling. 
\item 
If on the other hand the channels 
command very different widths -- say 
$\Gamma (P \to a) \gg \Gamma (P \to b)$ -- then a large {\em relative} 
asymmetry in $P \to b$ is accompagnied by a tiny one in 
$P \to a$. 
\end{enumerate} 
This simple scenario can easily be extended to two sets $A$ and $B$ of 
final states s.t. for all states $a$ in set $A$ the transition 
amplitudes have the same weak coupling and likewise for states 
$b$ in set $B$. One then finds 
\beq 
\Delta \gamma (a) = 4 \sum _{b\, \in \, B}{\cal T}^{resc}_{ab}{\rm Im}T_a^*T_b 
\eeq
The sum over all CP asymmetries for states $a \, \in \, A$ cancels the 
correponding sum over $b\, \in \, B$: 
\beq 
\sum _{a\, \in \, A} \Delta \gamma (a) 
= 4 \sum _{b\, \in \, B}{\cal T}^{resc}_{ab}{\rm Im}T_a^*T_b = 
- \sum _{b\, \in \, B} \Delta \gamma (b)
\eeq
\end{itemize}
These considerations tell us that the \cp~asymmetry averaged over certain 
classes of channels defined by their quantum numbers has to 
vanish. Yet these channels can still be very heterogenous, 
namely consisting of two- and quasi-two-body modes, 
three-body channels and other multi-body decays. 
Hence we can conclude: 
\begin{itemize}
\item 
If one finds a direct \cp~asymmetry in one channel, 
one can infer -- based on rather general grounds -- which other channels
have to exhibit the compensating  asymmetry as required by \cpt~invariance.
Observing them would enhance  the significance of the measurements very
considerably.  
\item 
Typically there can be several classes of rescattering channels. 
The SM weak dynamics select a subclass of those where the 
compensating asymmetries have to emerge. QCD frameworks like 
generalized factorization can be invoked to estimate the 
relative weight of the asymmetries in the different classes. 
Analyzing them can teach us important lessons about the 
inner workings of QCD. 
\item 
If New Physics generates the required weak phases (or at least 
contributes significantly to them), it can induce 
rescattering with novel classes of channels. The pattern in the 
compensating asymmetries then can tell us something about the 
features of the New Physics 
involved.  
\end{itemize}

I want to end this Interlude by adding that Penguins (diagrams) are rather smart beings: they know about these \cpt~constraints. For when one considers the imaginary parts of the Penguin diagrams, 
which are obtained by cutting the internal quark lines, namely the up and charm quarks (top quarks do not contribute there, since they cannot reach their mass shell in $b$ decays), one realizes that 
\cp~asymmetries in $B \to K +\pi$'s are compensated by those in $B\to D \bar D_s + \pi$'s ,...

\subsection{On Measuring other \cp~Observables in $B$ Decays}
\label{OTHEROB}

An obvious, yet still useful criterion for \cp~observables is that they must be 
`re-phasing' invariant under $|\bar B^0\rangle \to e^{-i\xi}|\bar B^0\rangle$. There are three 
classes of such observables:
\begin{enumerate}
\item 
$|T(B\to f)| \neq |T(\bar B \to \bar f)|$; 

\noindent it reflects pure $\Delta B = 1$ dynamics and thus amounts 
to direct \cp~violation. 
\item 
$|q| \neq |p|$; 

\noindent it requires \cp~violation in $\Delta B =2$ dynamics. 
\item 
Im$\frac{q}{p}\bar \rho (f) \neq 0$, $\bar \rho (f) = \frac{T(\bar B \to f)}{T(B\to f)}$  
\footnote{This condition is formulated for the simplest case of $f$ being a \cp~eigenstate.}; 

\noindent  
such an effect requires the interplay of $\Delta B=1$ \& $\Delta B =2$ forces. 

\end{enumerate}
Decays of beauty baryons and of charged $B$ mesons can realize only the first scenario, 
whereas semileptonic transitions $\bar B^0 \to l^+X$ vs. $B^0 \to l^-X$ only the second one due 
to the SM's $\Delta B = \Delta Q_l$ selection rule. 

The third scenario with its interplay of direct and indirect \cp~violation requires a nonleptonic mode. 
As already stated for the example $B \to \pi^+\pi^-$ one has in general 
\beq 
A_{\cp} (t) = S {\rm sin}(\Delta M_B t) + C {\rm cos}(\Delta M_B t) \; . 
\eeq 
with the following features: 
\begin{itemize}
\item 
\beq 
C = \frac{1- |(q/p)\bar \rho (f_{\cp})|^2}{1+ |(q/p)\bar \rho (f_{\cp})|^2} \neq 0
\eeq
unambiguously reflects {\em direct} \cp~violation. 
\item 
On the other hand the interpretation of 
\beq
S = \frac{2{\rm Im}(q/p)\bar \rho(f_{\cp})}{ 1+ |(q/p)\bar \rho (f_{\cp})|^2} \neq 0 
\eeq
requires some subtlety: 
\begin{itemize}
\item 
As long as it is observed in a single channel only, there is no true distinction between 
direct and indirect \cp~violation. For changing the phase convention of quark fields allows to 
shift the observed phase from the $\Delta B=2$ to the $\Delta B =1$ sector and viceversa. 
\item 
Once it is seen in at least two channels $f_a$ and $f_b$ of the same $B^0$, the verdict becomes clearer. If one finds 
\beq 
S(f_a) \neq \eta (f_a) \eta (f_b) S(f_b) 
\eeq
with $\eta_i$ denoting the \cp~parity of the final states $f_i$, then one has established 
the presence also of {\em direct} \cp~violation. {\em One should that such a manifestation 
of direct \cp~violation might not surface through $C(f_i) \neq 0$. For the latter requires, as explained 
above, the presence of two different, yet still coherent transitions in $B\to f_i$; this might not be 
the case.}

\end{itemize}

\end{itemize}

\subsubsection{$\phi_2$ from \cp~Violation in $B_d \to$ Multi-Pions}
\label{PIONS}

In Sect. \ref{ACT2} I have already mentioned the principal as well as practical 
complications in determining $\phi_2$ from $B \to \pi \pi$. An isospin decomposition can be undertaken 
also in $B \to 3\pi$ and $4\pi$ to disentangle the Penguin contribution. 
\begin{itemize}  
\item 
\beq 
B_{d,u} \to \rho \pi
\eeq
These channels are less challenging {\em experimentally} than $B_{d,u} \to 2\pi$, yet they pose some complex  
{\em theoretical} problems. For going from the experimental starting point 
$B \to 3 \pi$ to $B \to \pi \rho$ configurations is quite nontrivial. There are other contributions to the three-pion final state like $\sigma \pi$, and cutting on the dipion mass provides a rather imperfect filter 
due to the large $\rho$ width. It hardly matters in this context whether the $\sigma$ is a bona fide resonance or some 
other dynamical enhancement. This actually leads to a further complication, namely that the $\sigma$ structure cannot be 
described adequately by a Breit-Wigner shape. As analyzed first in \cite{ANTONELLO} and then 
in more detail in \cite{ULF} ignoring such complications can induce a systematic uncertainty in the extracted value of $\phi_2$.
\item 
\beq 
B_{d,u} \to \rho \rho 
\eeq
These transitions while providing better rates experimentally, contain even more theoretical 
complexities, since they have to be extracted from $B \to 4 \pi$ final states, where one has to 
allow for $\sigma \rho$, $2\sigma $, $\rho 2\pi$ etc. in addition to $2\rho$.  

\end{itemize}
My point here is one of caution rather than of agnosticism. The concerns sketched above might 
well be more academic than practical with the present statistics. My main conclusions are the 
following: (i) I remain unpersuaded that {\em averaging} over the values for 
$\phi_2$ obtained {\em so far} from the three methods listed above provides a reliable value, since I do 
not think that the systematic uncertainties have sufficiently been analyzed. (ii) It will be 
mandatory to study those in a comprehensive way, before we  can make full use of the even larger data sets that will become available in the next few years. As I have emphasized repeatedly, our aim has 
to be to reduce the uncertainty down to at least the 5\% level in a way that can be {\em defended}. 
(iii) In the end we will need 
\begin{itemize}
\item
to perform time {\em dependent} Dalitz plot analyses (and their generalizations) and 
\item 
involve the expertise that already exists or can obtained concerning low-energy 
hadronization processes like final state interactions among low energy pions and kaons; 
valuable information can be gained on those issues from $D_{(s)} \to \pi$'s, kaons etc. as well 
as $D_{(s)} \to l \nu K\pi/\pi \pi /KK$, in particular when analyzed with state-of-the-art tools of  
chiral dynamics. 
\end{itemize}

\subsubsection{$\phi_3$ from $B^+ \to D_{neut}K^+$ vs. $B^- \to D_{neut} K^-$}
\label{DNEUTK}

As first mentioned in 1980 \cite{CS80}, then explained in more detail in 1985 \cite{BS85} 
and further developed in \cite{GRONWYL},  
the modes $B^{\pm} \to D_{neut}K^{\pm}$ should exhibit direct \cp~violation driven by the 
angle $\phi_3$, if the neutral $D$ mesons decay to final states that are {\em common} to 
$D^0$ and $\bar D^0$. Based on simplicity the original idea was to rely on two-body modes like 
$K_S\pi^0$, $K^+K^-$, $\pi^+\pi^-$, $K^{\pm}\pi^{\mp}$. One drawback of that method are the small 
branching ratios and low efficiencies. 

A new method was pioneered by BELLE and then implemented also by BABAR, namely to employ 
$D_{neut} \to K_S \pi^+\pi^-$ and perform a full Dalitz plot analysis. This requires a very 
considerable analysis effort -- yet once this initial investment has been made, it will pay handsome profit in the long run. For obtaining at least a decent description of the full Dalitz plot population 
provides  considerable cross checks concerning systematic uncertainties and thus a high degree of 
confidence in the results. BELLE and BABAR find \cite{LANCERI}:  
\beq 
\phi_3 = 
\left\{
\begin{array}{ll} 68^o  \pm 15^o (stat) \pm 13^o(syst) \pm 11^o ({\rm model}) & {\rm BELLE}\\
70^o  \pm 31^o (stat) \pm 12^o(syst) \pm 14^o ({\rm model}) & {\rm BABAR}
\end{array}
\right.
\eeq 
I view it still a pilot study, yet a most promising one. It exemplifies how the complexities of 
hadronization can be harnessed to establish confidence in the accuracy of our results. I consider 
this to be the way of the future. 

\subsubsection{$\phi_1$ from \cp~Violation in $B_d \to$ 3 Kaons}
\label{KAONS}

Analysing \cp~violation in $B_d \to \phi K_S$ decays is a most promising way to search 
for New Physics. For the underlying quark-level transition $b \to s \bar s s$ represents a pure 
loop-effect in the SM, it is described by a {\em single} $\Delta B=1$\& $\Delta I=0$ operator (a `Penguin'), a reliable 
SM prediction exists for it \cite{GROSS} -- 
sin$2\phi_1(B_d \to \psi K_S) \simeq {\rm sin}2\phi_1(B_d \to \phi K_S)$ -- 
and the $\phi$ meson represents a {\em narrow} resonance.  

Great excitement was created when BELLE reported a large discrepancy between the predicted and 
observed \cp~asymmetry in $B_d \to \phi K_S$ in the summer of 2003:  
\beq 
{\rm sin}2\phi_1(B_d \to \phi K_S) = 
\left\{
\begin{array}{ll} - 0.96 \pm 0.5 \pm 0.10 & {\rm BELLE}\\
0.45 \pm 0.43 \pm 0.07 & {\rm BABAR}
\end{array}
\right. \;  ; 
\eeq
Based on more data taken, this discrepancy has 
shrunk considerably: the BABAR/BELLE average for 2005 yields \cite{LANCERI}
\beq 
{\rm sin}2\phi_1(B_d \to \psi K_S) = 0.685 \pm 0.032
\eeq
compared to 
\beq 
{\rm sin}2\phi_1(B_d \to \phi K_S) = 
\left\{
\begin{array}{ll} 0.50 \pm 0.25 ^{+0.07}_{-0.04} & {\rm BABAR}\\
0.44 \pm 0.27 \pm 0.05 & {\rm BELLE}
\end{array}
\right. \;  ; 
\eeq
BABAR's as well as BELLE's numbers are below the prediction, albeit by one sigma only. It 
is ironic that such a smaller deviation, although not significant, is actually more believable as 
signaling an incompleteness of the SM than 
the large one originally reported by BELLE. 

This issue has to be pursued with vigour, since this reaction provides such a natural portal to 
New Physics. One complication has to be studied, though, in particular if the observed 
value of sin$2\phi_1(B_d \to \phi K_S)$ falls below the predicted one by a moderate amount only. 
For one is actually observing $B_d \to K^+K^-K_S$. If there is a single weak phase like in the SM one finds 
\beq 
{\rm sin}2\phi_1(B_d \to \phi K_S) = - {\rm sin}2\phi_1(B_d \to `f_0(980)' K_S) \; , 
\eeq 
where $`f_0(980)'$ denotes any {\em scalar} $K^+K^-$ configuration with a mass close to that of the 
$\phi$, be it a resonance or not. A smallish pollution by such a $`f_0(980)' K_S$ -- by, say, 
10\% {\em in amplitude} --  
can thus reduce the asymmetry assigned to $B_d \to \phi K_S$ significantly -- by 20\% in this example. 

In the end it is therefore mandatory to perform a {\em full time dependent Dalitz plot analysis} 
for $B_d \to K^+K^-K_S$ and compare it with that for $B_d \to 3 K_S$ and 
$B^+ \to K^+K^-K^+, \, K^+K_SK_S$ and also with $D \to 3K$. This is a very challenging task, but 
in my view essential. There is no `royal' way to fundamental insights. 
\footnote{The ruler of a Greek city in southern Italy once approached the resident sage with the request 
to be educated in mathematics, but in a `royal way', since he was very busy with many 
obligations. Whereupon the sage replied with admirable candor: 
"There is no royal way to mathematics."} 

An important intermediate step in this direction is given by one application of 
{\em Bianco's Razor} \cite{RIO}, namely to analyze the \cp~asymmetry in $B_d \to [K^+K^-]_MK_S$ as a 
function of the cut $M$ on the $K^+K^-$ mass.

\subsection{Loop Induced Rare $B_{u,d}$ Transitions}
\label{LOOPDEC}

Processes that require a loop diagram to proceed -- i.e. are classically forbidden -- provide 
a particularly intriguing stage to probe fundamental dynamics. 

It marked a tremendous  success for the SM, when radiative $B$ decays were measured, first in the 
exclusive mode $B \to \gamma K^*$ and subsequently also inclusively: $B \to \gamma X$. 
Both the rate and the photon spectrum are in remarkable agreement with SM prediction; they have 
been harnessed to extracting heavy quark parameters, as explained in Lecture IV. 

More recently the next, i.e. even rarer level has been reached with transitions to final states 
containing a pair of charged leptons:   
\beq 
{\rm BR}(B \to l^+l^- X) = 
\left\{
\begin{array}{ll} (6.2 \pm 1.1\pm 1.5)\cdot 10^{-6} & {\rm BABAR/BELLE}\\
(4.7 \pm 0.7)\cdot 10^{-6} & {\rm SM}
\end{array}
\right. \;   . 
\eeq
Again the data are consistent with the SM prediction \cite{ALIETAL1}, yet the present experimental uncertainties 
are very sizable. We are just at the beginning of studying $B \to l^+l^-X$, and it has to be 
pursued in a dedicated and comprehensive manner for the following reasons: 
\begin{itemize}
\item 
With the final state being more complex than for $B \to \gamma X$, it is described by a larger number 
of observables: rates, spectra of the lepton pair masses and the lepton energies, their forward-backward asymmetries and \cp~asymmetries. 
\item 
These observables provide independent information, since there is a larger number of effective transition operators than for $B \to \gamma X$. By the same token there is a much wider window 
to find New Physics and even diagnose its salient features. 
\item 
It will take the statistics of a Super-B factory to mine this wealth of information on New Physics. 
\item 
Essential insights can be gained also by analyzing the exclusive channel $B\to l^+l^-K^*$ at hadronic 
colliders like the LHC, in particular the position of the zero in the lepton forward-backward asymmetry. 
For the latter appears to be quite insensitive to hadronization effects in this exclusive mode 
\cite{HILLER1}. 
\end{itemize}

Let me present a brief case study that can illustrate how various analyses get intertwined. 
While the electromagnetic Penguin operator drives $B \to \gamma X_s $, its strong counterpart 
generates $B_d \to s \bar s s \bar d \Rightarrow \phi K_S$. The observed rate for the former, 
which is very consistent with the SM prediction, will therefore constrain deviations from the SM in the latter. There is a loop hole in the argument, though: if New Physics produces a photon in 
$B \to \gamma X$ that carries a helicity {\em opposite} to that predicted in the SM, then it will contribute 
{\em quadratically} to $\Gamma (B \to \gamma X_s)$, yet {\em linearly} to the \cp~asymmetry in 
$B_d \to \phi K_S$. Thus it could well hide in the noise level of the former while having a substantial impact on the latter. 

\noindent 
How could one check such a scenario? One could attempt to determine the photon 
polarization in the radiative $B$ decays. This can be done by measuring angular correlations in 
$B \to \gamma K^{**} \to \gamma (K\pi\pi)$ modes \cite{PIRJOL}. The same operator will contribute 
also to 
$B \to l^+l^- X$, and its contribution can be disentangled there. This can even be achieved 
by analyzing angular correlations in the exclusive channel $B \to l^+l^- K^* \to l^+l^- (K\pi)$ 
\cite{MELIKHOV}. 

The analogous decays with a $\nu \bar \nu$ instead of the $l^+l^-$ pair is irresistibly attractive to theorists -- although quite resistibly so to experimentalists:
\beq 
{\rm BR}(B \to \nu \bar \nu X) 
\left\{
\begin{array}{ll} \leq 7.7 \cdot 10^{-4} & {\rm ALEPH}\\
= 3.5 \cdot 10^{-5} & {\rm SM}
\end{array}
\right. \;   ; 
\eeq
\beq 
{\rm BR}(B \to \nu \bar \nu K) 
\left\{
\begin{array}{ll} \leq 7.0 \cdot 10^{-5} & {\rm BABAR}\\
= (3.8^{+1.2}_{-0.6}) \cdot 10^{-6} & {\rm SM} 
\end{array}
\right. \;   , 
\eeq
where the SM predictions are taken from Refs.\cite{BUBU} and  \cite{BUHIISI}, respectively. 
To measure such a rare transition with no striking experimental signature requires a 
most hermetic detector. The justification for taking on this challenge lies in the fact that 
$B \to l^+l^-X$ and $B\to \nu \bar \nu X$ are complementary for diagnosing New Physics 
couplings with $l^{\pm}$ and $\nu$ representing the `down' and `up' members, respectively, 
of an $SU(2)$ doublet \cite{SIMULA}.

\subsection{Other Rare Decays}
\label{OTHERRARE}

There are some relatively rare $B$ decays that could conceivably reveal New Physics, although they 
proceed already on the tree level. One well known example is $B^+ \to \tau \nu$ that is sensitive 
to charged Higgs fields. This applies also to semileptonic $B$ decays. As will be described in 
Lecture IV, the Heavy Quark Expansion (HQE) has provided a sturdy and accurate description of 
$B \to l \nu X_c$ that allowed to extract $|V(cb)|$ with less than 2\% uncertainty. With it and other heavy quark parameters determined with considerable accuracy one can predict 
$\Gamma (B\to \tau \nu X_c)$  within the SM and compare it with the data. 
A discrepancy can be attributed to New Physics, presumably in the form of a {\em charged} 
Higgs field. 
Measuring also its hadronic mass moments can serve as a valuable cross check. Such studies will probably require the statistics 
of a Super-B factory. 

This is true also for studying the exclusive channel $B \to \tau \nu D$. As pointed out in Ref.\cite{MIKI}, 
one could find that the ratio $\Gamma (B \to \tau \nu D)/\Gamma (B \to \mu \nu D)$ deviates from its 
SM value due to the exchange of a charged Higgs boson with a mass of even several hundred GeV. 
This is the case in particular for `large tg$\beta$ scenarios' of two-Higgs-doublet models.  
There is a complication, though. Contrary to the suggestion in the literature the hadronic form factors 
do {\em not} drop out from this ratio. One should keep in mind that (i) the contribution from the second form factor $f_-$, which is proportional to the square of the lepton mass, cannot be ignored for 
$B \to \tau \nu D$ and (ii) the form factors  are not taken at the same momentum transfer in 
the two modes. 

These complications can be overcome by Uraltsev's BPS approximation \cite{BPS}.  
Relying on it one can extract $|V(cb)|$ from $B \to e/\mu \nu D$ and compare it with the 
`true' value obtained from $\Gamma _{SL}(B)$. {\em If} this comparison is successful and our 
theoretical control over $B \to l \nu D$ thus validated, one can apply the BPS approximation 
to $B \to \tau \nu D$. Since, as mentioned above, the second form factor $f_-$ can be measured there, 
one has another cross check.

\subsection{$B_s$ Decays -- an Independent Chapter in Nature's Book}
\label{BSDEC}

When the program for the $B$ factories was planned, it was thought that studying $B_s$ transitions 
will be required to construct the CKM triangle, namely to determine one of its sides and the angle 
$\phi_3$. As discussed above a powerful method has been developed to extract 
$\phi_3$ from $B^{\pm} \to D^{neut}K^{\pm}$, and the effort has started to obtain 
$|V(td)/V(ts)|$ from $\Gamma (B \to \gamma \rho/\omega)/\Gamma (B \to \gamma K^*)$. None of this, however, reduces the importance of a 
comprehensive program to study $B_s$ decays -- on the contrary! With the basic CKM parameters 
fixed or to be fixed in $B_{u,d}$ decays, $B_s$ transitions can be harnessed as powerful probes 
for New Physics and its features. 

In this context it is essential to think `outside the box' -- pun intended. The point here is that several 
relations that hold in the SM (as implemented through quark box and other loop diagrams) are 
unlikely to extend beyond minimal extensions of the SM. In that sense $B_{u,d}$ and $B_s$ decays 
constitute two different and complementary chapters in Nature's book on fundamental dynamics. 

\subsubsection{$B_s - \bar B_s$ Oscillations}
\label{BSOSC}

There is hope, even optimism, that we are about to establish $B_s - \bar B_s$ oscillations as 
characterized by the two observables $\Delta M(B_s)$ and $\Delta \Gamma (B_s)$. 

$\Delta M(B_s)$: The experimental sensitivity has reached a domain, where the SM most likely puts it: 
\beq 
\Delta M(B_s)  
\left\{
\begin{array}{ll} > 14.5 \; {\rm psec}^{-1} & {\rm exp.}\\
\sim 15 - 30  \; {\rm psec}^{-1} & {\rm SM-CKM}
\end{array}
\right. \;   ; 
\eeq
the numerical input for the SM prediction comes from the observed rate of 
$B_d - \bar B_d$ oscillations via the expression: 
\beq 
\frac{\Delta M(B_s)}{\Delta M(B_d)} \simeq \frac{B_sf^2(B_s)}{B_df^2(B_d)}\frac{|V(ts)|^2}{|V(td)|^2}\; . 
\eeq
This relation also exhibits the phenomenological interest in measuring $\Delta M(B_s)$, namely to 
obtain an accurate value for $|V(td)|$. Lattice QCD is usually invoked to gain theoretical control over the first ratio of hadronic 
quantities. I fully subscribe to the importance of this measurement, 
irrespective of the validity of the SM. 

$\Delta \Gamma (B_s)$: In June 2004 the CDF collaboration first presented an intriguing analysis exhibiting two 
surprisingly large lifetimes controlling $B_s \to \psi \phi|_{CP =\pm}$ \cite{CDF}:  
\bea
\tau [CP = +] = (1.05 ^{+0.16}_{-0.13} \pm 0.02) ps \; \; \; \; vs. \; \; \; \; 
\tau [CP = -] = (2.07 ^{+0.58}_{-0.46} \pm 0.03) ps \\
\frac{\Delta \Gamma _s}{\bar \Gamma_s} \equiv 
\frac{\Gamma (B_s[CP=+]) - \Gamma (B_s[CP=-])}
{\frac{1}{2}(\Gamma (B_s[CP=+]) + \Gamma (B_s[CP=-]) )}
 = (65^{+25}_{-33} \pm 1)\% 
 \label{CDFDELTA}
\eea
More recently D0 has presented a similar analysis \cite{D0}: 
\beq
\frac{\Delta \Gamma _s}{\bar \Gamma_s} =  25^{+14}_{-15} \% 
\label{D0DELTA}
\eeq
My heart wishes that $\frac{\Delta \Gamma _s}{\bar \Gamma_s}$ were indeed as large 
as 0.5 or even larger. For it would open up a whole new realm of \cp~studies in 
$B_s$ decays with a great potential to identify New Physics. Yet my head tells me that values 
exceeding 0.25 or so are very unlikely; it would point at a severe 
limitation in our theoretical understanding of $B$ lifetimes. For only on-shell intermediate 
states $f$ in $B^0 \to f \to \bar B^0$ can contribute to $\Delta \Gamma (B)$, and for 
$B^0 = B_s$ these are predominantly driven by $b\to c \bar c s$. Let $R(b\to c \bar cs)$ denote their fraction of all $B_s$ decays. 
If these transitions contribute only 
to $\Gamma (B_s(CP=+))$ one has $\Delta \Gamma_s/\bar \Gamma_s = 2R(b\to c \bar cs)$. 
Of course this is actually an upper bound quite unlikely to be even remotely saturated. 
With the estimate $R(b\to c \bar cs) \simeq 25\% $, which is consistent with the data on 
the charm content of $B_{u,d}$ decays this upper bound reads 50\%. More realistic calculations 
have yielded considerably smaller predictions: 
\beq 
\frac{\Delta \Gamma_s}{\bar \Gamma_s} = \left\{
\begin{array}{l} 22\% \cdot \left(\frac{f(B_s)}{220\, \MeV}\right)^2 \\
12 \pm 5\% 
\end{array}
\right. \;  ; 
\label{DELTAPRED}
\eeq 
where the two predictions are taken from Refs.\cite{AZIMOV} and \cite{LENZ}, respectively.  
A value as high as $20 - 25$ \% is thus not out of the question theoretically, and Eq.(\ref{CDFDELTA}) is still consistent with it. One should note that invoking New Physics would actually `backfire' since it leads to a lower prediction.
If, however, a value exceeding 25\% were established experimentally, we had to draw at least one of the following conclusions: (i) $R(b\to c \bar cs)$ actually exceeds the estimate of 25\% significantly. This would imply 
at the very least that the charm content is higher in $B_s$ than $B_{u,d}$ decays by a 
commensurate amount and the $B_s$ semileptonic branching ratio lower. 
(ii) Such an enhancement of  $R(b\to c \bar cs)$ would presumably -- though not necessarily -- imply 
that the average $B_s$ width exceeds the $B_d$ width by more than the predicted 1-2\% level. 
That means in analyzing $B_s$ lifetimes one should allow $\bar \tau (B_s)$ to float 
{\em freely}. (iii) If in the end one found the charm content of $B_s$ and $B$ decays to be quite 
similar and 
$\bar \tau (B_s)$ $\simeq$ $\tau (B_d)$, yet  
$\Delta \Gamma_s/\bar \Gamma_s$ to exceed 0.25, we had to concede a loss of theoretical control 
over $\Delta \Gamma$. This would be disappointing, yet not inconceivable: the a priori reasonable 
ansatz of evaluating both  
$\Delta \Gamma_B$ and $\Delta M_B$ from quark box diagrams -- with the only 
manifest difference being that the internal quarks are charm in the former and top in the 
latter case -- obscures the fact that the dynamical situation is actually different. In the latter case the 
effective transition operator is a local one involving a considerable amount of averaging over 
off-shell transitions; the former is shaped by on-shell channels with a relatively small amount 
of phase space: for the $B_s$ resides barely 1.5 GeV above the $D_s \bar D_s$ threshold. 
To say it differently: the observable $\Delta \Gamma_s$ is more vulnerable to limitations 
of quark-hadron duality than $\Delta M_s$ and even beauty lifetimes 
\footnote{These are all dominated by nonleptonic transitions, where duality violations can be 
significantly larger than for semileptonic modes.} . 

In summary: establishing $\Delta \Gamma_s \neq 0$ amounts to important qualitative progress in our 
knowledge of beauty hadrons; it can be of great practical help in providing us with novel probes of CP 
violations in $B_s$ decays, and it can provide us theorists with a reality check concerning the reliability 
of our theoretical tools for nonleptonic $B$ decays.

\subsubsection{\cp~Violation in Nonleptonic $B_s$ Modes}
\label{CPBSNL}

One class of nonleptonic $B_s$ transitions does not follow the paradigm of large \cp~violation in 
$B$ decays \cite{BS80}: 
$$ 
A_{\cp}(B_s(t) \to [\psi \phi]_{l=0}/\psi \eta ) = {\rm sin}2\phi (B_s) {\rm sin}\Delta M(B_s)t 
$$
\beq
{\rm sin}2\phi (B_s) =
{\rm Im}\left[ \frac{(V^*(tb)V(ts))^2}{|V^*(tb)V(ts)|^2}\frac{(V(cb)V^*(cs))^2}{(V(cb)V^*(cs))^2}  \right]
\simeq 2\lambda ^2 \eta   \sim 0.02  \; . 
\label{BSPSISM}
\eeq
This is easily understood: on the leading CKM level only quarks of the second and third families 
contribute to $B_s$ oscillations and $B_s \to \psi \phi$ or $\psi \eta$; therefore on that level there can be no \cp~violation making the \cp~asymmetry Cabibbo suppressed. {\em Yet New Physics of various ilks can quite conceivably generate sin$2\phi (B_s) \sim $ several $\times$ 10 \%.} 

Analyzing the decay rate evolution in proper time of 
\beq 
B_s(t) \to \phi \phi 
\label{BSPHIPHI} 
\eeq
with its direct as well as indirect \cp~violation is much more than a repetition of the 
$B_d(t) \to \phi K_S$ saga: 
\begin{itemize}
\item 
${\cal M}_{12}(B_s)$ and ${\cal M}_{12}(B_d)$ -- the off-diagonal elements in the mass matrices for 
$B_s$ and $B_d$ mesons, respectively -- provide in principle independent pieces of information 
on $\Delta B=2$ dynamics. 
\item
While the final state $\phi K_S$ is described by a single partial wave, namely $l=1$, there are 
three partial waves in $\phi \phi$, namely $l= 0,1,2$. Disentangling the three partial rates and their 
\cp~asymmetries -- or at least separating $l$ = even and odd contributions -- provides a new 
diagnostics about the underlying dynamics. 
\end{itemize}

\subsubsection{Leptonic, Semileptonic and Radiative Modes}
\label{BSLEPTSL}

The decays into a lepton pair and to `wrong-sign' leptons should be studied also for $B_d$ mesons; however here I discuss only $B_s$ decays, where one can expect more dramatic effects.

$\bullet$ The mode $B_s \to \mu^+\mu^-$ is necessarily very rare since it suffers from helicity 
suppression $\propto (m(\mu)/M(B_s))^2$ and `wave function suppression' 
$\propto (f_B/M(B_s))^2$, which reflects the practically zero range of the weak interactions. 
These tiny factors can be partially compensated in some 
large tg$\beta$ SUSY scenarios, where an  enhancement factor of tg$^6\beta$ arises. 
Upper bounds on the branching ratio established at FNAL already cut into the a priori 
allowed parameter space. 

$\bullet$ Due to the rapid $B_s$ oscillations those mesons have a practically equal probability 
to decay into `wrong' and `right' sign leptons. One can then search for an asymmetry in the 
wrong sign rate for mesons that initially were $B_s$ and $\bar B_s$: 
\beq 
a_{SL}(B_s) \equiv \frac{\Gamma (\bar B_s \to l^+X) - \Gamma (B_s \to l^-X)}
{\Gamma (\bar B_s \to l^+X) + \Gamma (B_s \to l^-X)}
\eeq
This observable is necessarily small; among other things it is proportional to 
$\frac{\Delta \Gamma (B_s)}{\Delta M(B_s)} \ll 1$. yet within CKM theory it is truly tiny:  
\beq 
a_{SL}(B_s) \sim {\cal O}(10^{-4}) 
\eeq
due to a feature quite specific to CKM; analogous to $B_s \to \psi \phi$ 
quarks of only the second and third family participate on the leading CKM level, which therefore 
cannot exhibit \cp~violation. Yet again, New Physics can enhance $a_{SL}(B_s)$, this time  
by two orders of magnitude to the 1\% level. 

$\bullet$ As already emphasized $B_s \to \gamma X$ and $B_s \to l^+l^-X$ should be studied in a 
comprehensive manner.

\subsection{Summary of Lecture III}
\label{SUM3}

Not only has the first \cp~asymmetry outside the $K^0 - \bar K^0$ complex been observed -- 
the CKM paradigm has become a {\em tested} theory. The term `ansatz' with its patronizing flavour is no longer adequate. 
The predicted `Paradigm  of large \cp~Violation in $B$ Decays' has been established now in qualitative as well as quantitative agreement with CKM theory in three quite distinct $B_d$ channels:  
$B_d \to \psi K_S$ 
\footnote{The evolution of the experimental measurements of sin$2\phi_1$ as sketched above should 
illustrate for theorists that when experimentalists quote uncertainties those mean something 
very real. Shifts by more than one sigma are quite possible.}, $\pi^+\pi^-$ and $K^+\pi^-$. The observed effects in the first two transitions that 
involve $B_d - \bar B_d$ oscillations are fully commensurate with \ot~violation, and the last two modes 
exhibit large {\em direct} \cp~violation. 

My referring to it as the `expected triumph of a very peculiar theory'  should not be seen as 
downgrading these observations. On the contrary -- the discovery of large \cp~asymmetries in 
$B$ decays is a very momentous one validating an a priori unlikely theory, namely CKM dynamics, 
and demonstrating a more general point, namely that \cp~violating phases are not intrinsically small. 
I maintain that this novel success of CKM theory should not have come as a shock after the previous successes. Yet even for a theorist like me the 
strongest fascination of physics lies in the fact that data are the final arbiter. Verifying something empirically thus always adds a higher quality. In a similar vein I would argue that the discovery 
of the weak bosons marks one of the major intellectual human triumphs of the second half of the 
twentieth century; the fact that  those states had been predicted by the SM does {\em not} 
detract from it.   

It is more than a bon mot to say "$K$ and $B$ decays are exactly the same, only different'. 
Table \ref{tab:OSCPARAM} sketches the oscillation parameters for neutral mesons carrying 
strangeness, charm and beauty. There are at least qualitative similarities in the patterns. Furthermore there are actually domains in $K_L$ decays that exhibit truly large \cp~asymmetries, namely 
the angular correlation between the di-pion and the di-electron planes in 
$K_L \to \pi^+\pi^-e^+e^-$, see Eq.(\ref{KTEVSEHGAL2}), and the difference between 
$K^0 \to \pi^+\pi^-$ and 
$\bar K^0 \to \pi^+\pi^-$ for intermediate times of decay. 

Nevertheless the statement that "\cp~violation in $B$ decays is much larger than in 
$K$ decays" is an empirical fact. Suffice it to say that the mass eigenstates of neutral kaons are 
very well approximated by \cp~eigenstates as well -- in marked contrast to the situation in 
$B$ decays 
\footnote{CKM theory explains why \cp~invariance is a `near miss' in $K_L$ decays in the 
following sense: the first and second families are almost decoupled there from the third.}.

\begin{table}
\begin{center}
\begin{tabular}{llll}
\hline
$K^0$      & $D^0$   & $B_d$ & $B_s$    \\
\hline
      $\Delta M_K \simeq \bar \Gamma _K$  & $\Delta M_D \ll \bar \Gamma _D$    
      & $\Delta M (B_d) \sim \bar \Gamma (B_d)$ & $\Delta M (B_s) \gg \bar \Gamma (B_s)$\\
      $\Delta \Gamma _K \simeq 2\bar \Gamma _K$  & $\Delta \Gamma_D \ll \bar \Gamma _D$    
      & $\Delta \Gamma (B_d) \ll \bar \Gamma (B_d)$ & 
      $\Delta \Gamma (B_s) \sim {\cal O}(\bar \Gamma (B_s))$ \\
     $\Delta \Gamma _K \sim \Delta M_K$  & $\Delta \Gamma _D \sim \Delta M_D$    
      & $\Delta \Gamma (B_d) \ll  \Delta M (B_d)$ & 
      $\Delta \Gamma (B_s) \ll  \Delta M (B_s)$ \\
    \hline
\end{tabular}  
\caption{Comparing oscillation parameters for neutral $K$, $D$ and $B$ mesons}
\label{tab:OSCPARAM}
\end{center} 
\end{table}

Even beyond the success of the specific CKM theory a `demystification' of \cp~violation has occurred: if the dynamics are 
sufficiently multi-layered to support \cp~violation -- for the case under study it means there are 
at least three families --, the latter can be large, i.e. the observable phases 
can be of order unity. Having been raised by my mother in the spirit of the Enlightenment I welcome this 
development wholeheartedly. This demystification will be completed, if \cp~violation is found anywhere in leptodynamics -- a point I will return to in Lecture V. 

There is no longer a need for -- and at present little sense in -- searching for an {\em alternative} theory for the observed \cp~violation. Yet the decisive success of the SM description does {\em not at all} weaken the need to search for New Physics in $B$ decays as {\em corrections} to SM effects. 
\begin{itemize}
\item
It makes the peculiar features of the SM in general and the 
CKM dynamics in particular even more mysterious thus suggesting an even stronger need for a more 
fundamental explanation. 
\item
We know, as explained in Dolgov's lectures \cite{DOLGOV} that CKM dynamics are irrelevant for baryogenesis in 
the Universe. I see that actually as a rather positive statement in the sense that the assumed 
baryogenesis -- i.e., the conjecture that the observed baryon number is a 
{\em dynamically generated}  
quantity rather than an {\em arbitrary initial value} -- implies the existence of New Physics. This 
is further strengthened by the aforementioned demystification telling us there is no general intrinsic 
reason why \cp~phases should be small. 
\item 
On a more pragmatic level, \cp~studies provide highly sensitive probes of the underlying dynamics. 
A comprehensive program of 
precision studies will be essential in diagnosing the New Physics anticipated for the 
TeV scale. I will return to this point in more detail in Lecture V. In this context it is essential to make the utmost efforts to bring hadronization effects under quantitative control \cite{RIO}. 
\item 
In most cases -- with notable exceptions like $B_s(t) \to \psi \phi$ -- we cannot {\em count} on 
New Physics inducing large deviations from SM expectations. Accordingly it is only now that we 
are reaching `territory', where significant discrepancies with SM  predictions can `realistically' 
be hoped for. 
\item 
Studying $B_s$ transitions in a dedicated fashion will allow us to `read' an independent chapter in 
Nature's book on fundamental dynamics. 
\end{itemize}

\section{Lecture IV: Adding High Accuracy to High Sensitivity}
\label{LECTIV}

None of the impressive successes of the SM weaken the case for it being incomplete, i.e that 
New Physics has to exist, quite conceivably already at the TeV scale. Apart from the general 
observation that the CKM structure looks very peculiar thus suggesting an underlying layer of 
hitherto unknown dynamics, we have some more specific theoretical arguments and 
data of mostly celestial origin: 
\begin{itemize}
\item 
theoretical short comings: 
\begin{itemize}
\item 
the UV instability of Higgs dynamics and the gauge hierarchy problem; 
\item 
the Strong \cp~Problem, see Sect. \ref{FLY}. 
\end{itemize}
\item 
`Heavenly pointers': 
\begin{itemize}
\item 
the baryon number of the Universe; 
\item 
neutrino oscillations: the solar neutrino deficit and 
the `atmospheric anomaly' together with the earthly KAMLAND and 
K2K data \cite{CADENAS};  
\item 
the evidence for ubiquitous dark matter; 
\item 
the most intriguing evidence for a mysterious dark energy. 

\end{itemize}

\end{itemize} 
I harbour few doubts that New Physics will be discovered at the LHC `directly', i.e. through the 
observation of  the production of new quanta. Yet our goal has to be to identify also the nature 
of this anticipated New Physics. I am convinced that it will be essential in this context to search 
for its indirect impact on low energy processes. Among such indirect searches for New Physics one usually distinguishes between two classes, namely those based 
\begin{itemize}
\item 
on high {\em accuracy} with the most celebrated example that of the $g-2$ of muons and 
\item 
on high {\em sensitivity} as in $\epsilon_K$, $\Delta M(B^0)$ and EDM's. 
\end{itemize}
The present successes of the SM suggest that manifestations of New Physics on heavy flavour 
transitions might be subtle rather than numerically massive. Therefore I formulate the 
\begin{center}
{\bf New Paradigm of Hevay Flavour Studies:} 
\end{center}
\begin{itemize}
\item 
{\bf With the impact of New Physics likely to be mostly subtle and 
\item 
our goal being to identify the salient features of New Physics rather than `merely' establishing its existence}  
\end{itemize}
{\bf we have to strive for high numerical accuracy on the experimental as well as theoretical side.} 

To say it differently: we have to combine the elements of high sensitivity and high accuracy. The spectacular success of the $B$ factories and the emerging successes of CDF and D0 to obtain 
high quality data on beauty transitions in a hadronic environment give us confidence that such an experimental goal will be achieved. In this lecture I want to describe why I think that theory will be able to hold up its side of the bargain as well and what the required elements for such an undertaking have to be. 

The question is: "Can we answer the challenge of $\sim \%$ accuracy?" One guiding principle 
will be in Lenin's concise words:  
\begin{center}
`Trust is good -- control is better!'
\end{center} 

\begin{table}
\begin{center}
\begin{tabular}{lll} 
\hline
Heavy Quark Parameter      & value as of 2005 & relative uncertainty    \\
\hline
      $m_b$(1 GeV)   & $ = (4.59 \pm 0.04)$ GeV & $\hat = \; \; 1.0\, \%$ \\ 
      $m_c$(1 GeV)   & = $(1.14 \pm 0.06)$ GeV & $\hat = \; \; 5.3\, \%$ \\ 
      $m_b$(1 GeV) $-0.67 m_c$(1GeV)  & = $(3.82 \pm 0.017)$ GeV & $\hat = \; \; 0.5\, \%$ \\    
      $|V(cb)|$ & = $(41.58 \pm 0.67) \cdot 10^{-3}$ & $\hat = \; \; 1.6\, \%$ \\
    \hline
    $|V(us)|_{KTeV}$ & = $0.2252 \pm 0.0022$ & $\hat = \; \; 1.1\, \%$ \\
    \hline 
\end{tabular}  
\caption{The 2005 values \cite{FLAECHER} 
of $b$ and $c$ quark masses and of $|V(cb)|$ compared to the Cabibbo angle}
\label{tab:STATUS05} 
\end{center} 
\end{table}

Table \ref{tab:STATUS05} provides a sketch of the theoretical control we have achieved over 
some aspects of $B$ decays. I hope it will excite the curiosity of the reader and fortify her/him to 
read the more technical discussion of Lecture IV.

\subsection{Heavy Quark Theory}
\label{HQTH}

While QCD is the only candidate among {\em local} quantum field theories to describe the 
strong interactions, as explained in Lecture I in Sects. \ref{QCD} \& \ref{SU(2)}, 
$SU(2)_L \times U(1)$ is merely the minimal theory for the electroweak 
forces. Obtaining reliable information about the latter is, however, limited by our lack of full 
calculational control over the former.  

It had been 
conjectured for more than thirty years that 
the theoretical treatment of heavy flavour hadrons should be facilitated, when the 
heavy quark mass greatly exceeds greatly the nonperturbative scale of QCD 
\footnote{A striking prediction has been that super-heavy top quarks -- i.e. with 
$m_t \geq 150$ GeV -- would decay, {\em before} they could hadronize \cite{RAPALLO} 
thus bringing top quarks under full theoretical control. For then the decay width of top quarks is of order 
1 GeV and provides an infrared cutoff for QCD corrections.  
This feature comes with a price, though, in so far 
as \cp~studies are concerned: without hadronization as a `cooling' mechanism, the degree 
of coherence between different transition amplitudes -- a necessary condition for \cp~violation to become observable -- will be rather tiny.}: 
\beq 
m_Q \gg \Lambda _{QCD} \; . 
\eeq
This conjecture has been transformed into a reliable theoretical framework only in the last 
fifteen years, as far as beauty hadrons are concerned. I refer to it as Heavy Quark Theory (mentioned already in Sect. \ref{TECH}); comprehensive reviews with references to the original 
literature can be found in Refs. \cite{HQREV} and 
\cite{URALTSEV}. Its goal is to treat nonperturbative dynamics {\em quantitatively}, as it affects heavy flavour {\em hadrons}, in full conformity with QCD and without model assumptions. It has achieved this goal already for several classes of 
beauty meson transitions with a reliability and accuracy that before would 
have seemed unattainable.

Heavy Quark Theory is based on a two-part strategy analogous to the one adopted in 
chiral perturbation theory -- another theoretical technology to deal reliably with nonperturbative dynamics 
in a special setting. Like there Heavy Quark Theory combines two basic elements, namely an 
{\em asymptotic symmetry principle} and a {\em dynamical treatment}  
telling us how the asymptotic limit is approached: 
\begin{enumerate}
\item 
The symmetry principle is {\em Heavy Quark Symmetry} 
stating that all sufficiently heavy quarks behave identically 
under the strong interactions without sensitivity to their spin. 
This can easily be illustrated with the Pauli Hamiltonian describing the 
interaction of a quark of mass $m_Q$ with a gauge field $A_{\mu}= (A_0, \vec A)$: 
\beq 
{\cal H}_{\rm Pauli} = - A_0 + \frac{(i\vec \partial - \vec A)^2}{2m_Q} + 
\frac{\vec \sigma \cdot \vec B}{2m_Q} \Longrightarrow - A_0 
\; \; {\rm as} \; \; m_Q \to \infty \; ; 
\label{PAULI} 
\eeq 
i.e., in the infinite mass limit, quarks act like {\em static} objects 
{\em without} spin dynamics and subject only to colour Coulomb fields. 

This simple consideration illustrates a general feature of heavy quark theory, namely that 
the spin of the heavy quark $Q$ decouples from the dynamics in the heavy quark limit. 
Hadrons $H_Q$ can therefore be labeled by the angular momentum 
$j_q$ carried by its `light' components 
-- light valence quarks, gluons and sea quarks --  in addition to its total spin $S$. 
The S wave pseudoscalar and vector mesons  -- $B$ \& $B^*$ and $D$ \& $D^*$ -- then form the ground state doublet of heavy quark 
symmetry with $[S,j_q] = [0,\frac{1}{2}], [1,\frac{1}{2}]$; a quartet of P wave configurations 
form the first excited states with $[S,j_q] = [0,\frac{1}{2}], [1,\frac{1}{2}], [1,\frac{3}{2}], [2,\frac{3}{2}]$. 

Heavy quark symmetry can be understood 
in an intuitive way: consider a hadron $H_Q$ containing a 
heavy quark $Q$ with mass $m_Q \gg \Lambda _{QCD}$ surrounded 
by a "cloud" of light degrees of freedom carrying quantum 
numbers of an antiquark $\bar q$ or diquark $qq$ 
\footnote{This cloud is often referred to -- 
somewhat disrespectfully -- as `brown muck', a phrase coined by the late Nathan Isgur.}. 
This cloud has 
a rather complex structure: in addition to $\bar q$ 
(for mesons) or $qq$ (for baryons) it 
contains an indefinite number of $q \bar q$ pairs and gluons 
that are strongly coupled to and constantly 
fluctuate into each other. There is, however, 
one thing we know: since typical frequencies 
of these fluctuations are $\sim {\cal O}({\rm few}) \times 
\Lambda _{QCD}$, the normally dominant {\em soft} dynamics 
allow the heavy quark to exchange momenta of order few times 
$\Lambda _{QCD}$ only with its surrounding medium. 
$Q \bar Q$ pairs then cannot play a significant role, 
and the heavy quark can be treated as a quantum mechanical 
object rather than a field theoretic entity 
requiring second quantization. This provides a 
tremendous computational simplification even while 
maintaining a field theoretic description for the light 
degrees of freedom. Furthermore techniques developed 
long ago in QED can profitably be adapted here. 
\item 
We can go further and describe the interactions between 
$Q$ and its surrounding light degrees of freedom through 
an expansion in powers of $1/m_Q$ -- the Heavy 
Quark Expansion (HQE). This allows us to 
analyze {\em pre}-asymptotic effects, i.e. effects 
that fade away like powers of $1/m_Q$ as 
$m_Q \ra \infty$.
\end{enumerate}   
Let me anticipate the lessons we have learnt: we have 
\begin{itemize}
\item 
identified the sources of the non-perturbative corrections;  
\item 
found them to be smaller than they could have been; 
\item 
succeeded in relating the basic quantities of the Heavy Quark 
Theory -- KM paramters, masses and kinetic energy of 
heavy quarks, 
etc. -- to various a priori independant observables with a considerable  
amount of  redundancy; 
\item 
developed a better understanding of incorporating perturbative 
and nonperturbative corrections without double-counting.  
\end{itemize} 
In the following I will sketch the concepts 
on which the Heavy Quark Expansions are based, the 
techniques employed, the results obtained and 
the problems encountered. It will not constitute 
a self-sufficient introduction into this vast 
and ever expanding field. My intent is to provide a guide through the literature 
for the committed student. 

\subsection{H(eavy) Q(uark) E(xpansions), Fundamentals}
\label{HQEXP} 

In describing weak decays of heavy flavour {\em hadrons} 
one has to incorporate perturbative as well as 
nonperturbative contributions in a self-consistent 
and complete  
way. The only known way to 
tackle such a task invokes the 
{\em Operator Product Expansion a la Wilson} 
involving an {\em effective} Lagrangian. Further 
conceptual insights as well as practical results can be 
gained by analysing {\em sum rules}; in particular they 
shed light on various aspects and formulations of 
{\em quark-hadron duality}. 

\subsubsection{Operator Product Expansion (OPE) 
for Inclusive Weak Decays}
\label{OPE}

Similar to the well-known case of 
$\sigma (e^+ e^- \to had)$ one invokes the optical 
theorem to describe the decay into a sufficiently 
{\em inclusive} final state $f$ through the 
imaginary part of the forward scattering operator 
evaluated to second order in the weak interactions  
\beq 
\hat T( Q \to Q) = 
{\rm Im}  \int d^4x 
\, i\{ {\cal L}_W(x) {\cal L}_W^{\dagger}(0) \} _T
\label{TRANSOP}
\eeq 
with the subscript $T$ denoting the time-ordered 
product and ${\cal L}_W$ the relevant weak Lagrangian  
\footnote{There are two qualitative differences to the 
case of $e^+ e^- \to had$: in describing weak decays 
of a hadron $H_Q$ (i) one employs the weak rather than the 
electromagnetic Lagrangian, and (ii) one takes the 
expectation value between the $H_Q$ state rather than the hadronic 
vacuum.}.   
The expression in Eq.(\ref{TRANSOP}) represents in general 
a non-local operator with the space-time separation 
$x$ being fixed by the inverse of the {\em energy release}. If 
the latter is large compared to typical hadronic 
scales, then the product is dominated by short-distance 
physics, and one can apply an operator 
product expansion a la Wilson on it yielding an infinite 
series of {\em local} operators of increasing dimension 
\footnote{
I will formulate the expansion in powers of 
$1/m_Q$, although it has to be kept in mind that it is 
really controlled by the inverse of the {\em energy release}.  
While there is no fundamental difference between 
the two for $b \to c/u l \bar \nu$ or 
$b \to c/u \bar ud$, since $m_b, \, 
m_b - m_{c,u} \gg  
\Lambda _{QCD}$, the expansion becomes of 
somewhat dubious reliability for $b \to c \bar cs$. 
It actually would break down for a scenario 
$Q_2 \to Q_1 l \bar \nu$ with $m_{Q_2} \simeq 
m_{Q_1}$ -- in contrast to HQET! 
}. 
The width for the decay of a hadron $H_Q$ containing 
$Q$ is then obtained by taking the $H_Q$ expectation value 
of the operator $\hat T$:  
$$ 
\frac{\matel{H_Q}{{\rm Im}\hat T(Q \to f \to Q)}{H_Q}}
{2M_{H_Q}} \propto 
\Gamma (H_Q \to f) = \frac{G_F^2m_Q^5(\mu )}{192 \pi ^3}
|V_{CKM}|^2 \cdot 
$$
$$
\cdot \left[ c_3^{(f)}(\mu )
\frac{\matel{H_Q}{\bar QQ}{H_Q}_{(\mu )}}{2M_{H_Q}} + 
\frac{c_5^{(f)}(\mu )}{m_Q^2}
\frac{\matel{H_Q}{\bar Q\frac{i}{2}\sigma \cdot GQ}
{H_Q}_{(\mu )}}{2M_{H_Q}} + \right. 
$$
\beq
\left. + \sum _i \frac{c_{6,i}^{(f)}(\mu )}{m_Q^3} 
\cdot \frac{
\matel{H_Q}{(\bar Q\Gamma _iq)(\bar q\Gamma _iQ)}
{H_Q}_{(\mu )}}{2M_{H_Q}} + {\cal O}(1/m_Q^4) 
\right] 
\label{MASTER}
\eeq  
Eq.(\ref{MASTER}) exhibits the following important features:
\begin{itemize}
\item 
An {\em auxiliary} scale $\mu$ has been introduced to consistently separate 
short and long distance dynamics: 
\beq 
{\rm short \; distance} \; \; < \; \; 
\mu ^{-1} \; \; < \; \; 
{\rm long \; distance} 
\eeq 
with the former entering through the coefficients and 
the latter through the effective operators; their 
matrix elements will thus depend on $\mu$. 

{\em In principle} the value of $\mu$ does not 
matter: it reflects merely our computational procedure 
rather than how nature goes about its business. The 
$\mu$ dependance of the coefficients thus has to cancel 
against that of the corresponding matrix elements. 

{\em In practise} however there are competing 
demands on the choice of $\mu$: 
\begin{itemize}
\item 
On one hand one has to choose 
\beq 
\mu \gg \Lambda _{QCD} \; ; 
\label{MULARGE} 
\eeq 
otherwise radiative corrections cannot be treated 
within {\em perturbative} QCD. 
\item 
On the other hand many computational techniques 
for evaluating {\em matrix elements}  
-- among them the Heavy Quark Expansions -- 
require 
\beq 
\mu \ll m_b 
\label{MUSMALL} 
\eeq 
\end{itemize} 
The choice 
\beq 
\mu \sim 1\; {\rm GeV}
\eeq
satisfies both of these requirements. It is important to check that the obtained 
numerical results do not exhibit a significant sensitivity to the exact value of 
$\mu$ when varying the latter in a reasonable range. 
\item 
{\em Short-distance} dynamics 
shape the c number coefficients $c_i^{(f)}$. 
{\em In practise} they are evaluated in 
{\em perturbative} QCD. It is 
quite conceivable, though, that also {\em nonperturbative}  
contributions arise there; yet they are believed to be 
fairly small in beauty decays \cite{CHIBISOV}. 

By the same token these short-distance coefficients provide also the portals, 
through which New Physics can enter in a natural way. 
\item  
Nonperturbative contributions on 
the other hand enter through the {\em expectation values} 
of operators of dimension higher than three -- 
$\bar Q\frac{i}{2}\sigma \cdot GQ$ etc. -- and higher order 
corrections to the expectation 
value of the leading operator   
$\bar QQ$, see below. 
\item 
In practice we cannot go beyond evaluating the first few terms in this expansions. 
More specifically we are limited to contributions through order 
$1/m_Q^3$; those are described in terms of six heavy quark parameters, namely 
two quark masses -- $m_{b,c}$ --, two expectation values of dimension-five 
operators -- $\mu_{\pi}^2$ and $\mu_G^2$ -- and of dimension-six operators -- 
the Darwin and `LS' terms, $\rho^3_D$ and $\rho^3_{LS}$, respectively 
\footnote{For simplicity I ignore here socalled `Intrinsic Charm' contributions, see 
\cite{IC}.}.  
\begin{itemize}
\item
This small and universal set of nonperturbative quantities describes a host of 
observables in $B$ transitions. Therefore their values can be determined from some of these 
observables and still leave a large number of predictions. 
\item 
It opens the door to a novel symbiosis of different theoretical technologies for heavy flavour 
dynamics -- in particular between HQE and lattice QCD. For the HQP can be inferred from lattice 
studies. This enhances the power of and confidence in both technologies by 
\begin{itemize}
\item 
increasing the range of applications and 
\item 
providing more validation points. 

\end{itemize}
I will give some examples later on.

\end{itemize} 
\item 
Expanding the expectation value of the leading operator 
$\bar QQ$ for a pseudoscalar meson $P_Q$ with quantum number $Q$  in powers 
of $1/m_Q$ yields 
\beq 
\frac{1}{2M_{P_Q}} 
\matel{P_Q}{\bar QQ}{P_Q} = 
1 - \frac{\mu _{\pi}^2}{2m_Q^2} + 
\frac{\mu _G^2}{2m_Q^2} 
+{\cal O}(1/m_Q^3) \; ; 
\label{QQ}
\eeq 
$\mu ^2_{\pi }(\mu )$ and 
$\mu ^2_G(\mu )$ denote the expectation values of the kinetic and 
chromomagetic operators, respectively: 
\beq 
\mu ^2_{\pi }(\mu ) \equiv \frac{1}{2M_{H_Q}} 
\matel{H_Q}{\bar Q \vec \pi ^2 Q}{H_Q}_{(\mu )} 
\; , \; 
\mu ^2_G (\mu ) \equiv \frac{1}{2M_{H_Q}} 
\matel{H_Q}{\bar Q \frac{i}{2} 
\sigma \cdot G Q}{H_Q}_{(\mu )} \; ; 
\eeq 
for short they are often called the kinetic and chromomagnetic moments. 

Eq.(\ref{QQ}) implies that one has 
$\matel{H_Q}{\bar QQ}{H_Q}_{(\mu )}/2M_{H_Q} = 1$ 
for $m_Q \to \infty$; i.e., the free quark model expression emerges 
asymptotically for the total width.

\item 
The {\em leading} nonperturbative corrections 
arise at order $1/m_Q^2$ only. That means they are 
rather small in beauty decays since 
$(\mu /m_Q)^2 \sim $ few \% for $\mu \leq 1$ GeV. 
\item 
This smallness of nonperturbative contributions explains a posteriori, why 
partonic expressions when coupled with a `smart' perturbative treatment 
often provide a decent approximation. 
\item 
These nonperturbative contributions which are 
power suppressed can be described only if considerable 
care is applied in treating the {\em parametrically 
larger} perturbative corrections. 
\item 
Explicitely flavour dependant effects arise in order 
$1/m_Q^3$. They mainly drive the differences in the 
lifetimes of the various mesons of a given heavy 
flavour.
\item 
An important practical distinction to the OPE treatment of $e^+e^- \to had$ or 
deep-inelastic lepton nucleon scattering is the fact that the weak width depends on the 
fifth power of the heavy quark mass, see Eq.(\ref{MASTER}), and thus requires particular 
care in dealing with the delicate concept of quark masses.  
\end{itemize} 
One general, albeit subtle point has to be kept in mind here: while everybody these 
days invokes the OPE it is often not done employing Wilson's prescription with the  
auxilliary scale $\mu$, and different definitions of the relevant operators have 
been suggested. While results from one prescription can be translated into another one 
order by order, great care has to be applied. I will adopt here the socalled `kinetic scheme' 
with $\mu \simeq 1$ GeV. It should be noted that the quantities $\mu ^2_{\pi }(\mu )$ and 
$\mu ^2_{G}(\mu )$ are quite distinct from the socalled HQET parameters $\lambda_1$ and 
$\lambda_2$ although the operators look identical. Furthermore the fact that perturbative 
corrections are rather smallish in the kinetic scheme does generally {\em not} hold in other schemes.

The absence of corrections of order 
$1/m_Q$ \cite{BUV} 
is particularly noteworthy and intriguing since 
such corrections do exist for hadronic masses --  
$M_{H_Q} = m_Q (1+ \bar \Lambda/m_Q + 
{\cal O}(1/m_Q^2) )$ -- and those control 
the phase space. Technically this follows from the 
fact that there is no {\em independant} 
dimension-four operator that could emerge in the OPE 
\footnote{The operator $\bar Q i \not D Q$ can be reduced 
to the leading operator $\bar QQ$ through the equation 
of motion.}. This result can be illuminated in more 
physical terms as follows. Bound-state effects in the 
initial state like mass shifts do generate corrections 
of order $1/m_Q$ to the total width; yet so does 
hadronization in the final state. {\em Local} 
colour symmetry demands that those effects cancel 
each other out. {\em It has to be emphasized that the 
absence of corrections linear in $1/m_Q$ is an 
unambiguous consequence of the OPE description.} 
{\em If} their presence were forced upon us, we would have 
encountered a {\em qualitative} change in our QCD paradigm. 
A discussion of this point has arisen recently phrased 
in the terminology of quark-hadron duality. I will return 
to this point later.   

\subsubsection{Sum Rules}
\label{SUMRULES}

There are classes of sum rules derived from QCD proper that  relate the 
heavy quark parameters appearing in the OPE for inclusive $B\to l \nu X_c$ -- like 
$\mu_{\pi}^2$, $\mu _G^2$ etc. -- with restricted sums over exclusive channels. They 
provide rigorous definitions, inequalities and experimental constraints \cite{HQSR};  
e.g.:
\bea 
\mu_{\pi}^2 (\mu)/3 &=& \sum _n^{\mu} \epsilon_n^2\left| \tau_{1/2}^{(n)}(1)\right| ^2 +
2\sum _m^{\mu} \epsilon_m^2\left| \tau_{3/2}^{(m)}(1)\right| ^2 
\\
\mu_{G}^2 (\mu)/3 &=& 
-2\sum _n^{\mu} \epsilon_n^2\left| \tau_{1/2}^{(n)}(1)\right| ^2 +
2\sum _m^{\mu} \epsilon_m^2\left| \tau_{3/2}^{(m)}(1)\right| ^2
\label{SUMRULES}
\eea
where $\tau_{1/2}$,  $\tau _{3/2}$ are the amplitudes for 
$B\to l \nu D(j_q)$ with $D(j_q)$ a hadronic system beyond the $D$ and $D^*$,  
$j_q = 1/2 \& 3/2$ the angular momentum carried by the light degrees of freedom in 
$D(j_q)$, as explained in the paragraph below Eq.(\ref{PAULI}), 
and $\epsilon_m$ the excitation energy of the $m$th such system above the $D$  
with $\epsilon _m \leq \mu$.

These sum rules have become of great practical value. I want to emphasize here 
one of their conceptual features: they show that {\em the heavy quark parameters in the 
kinetic scheme are observables themselves}.

\subsubsection{Quark-hadron Duality}
\label{DUALITY}

The concept of quark-hadron duality (or duality for short), which  goes back to the early days of the quark model, refers to the notion that a {\em quark}-level description 
should provide a good description of transition rates that involve {\em hadrons}, if one sums over a 
sufficient number of channels. This is a rather vague formulation: How many channels are "sufficiently" many? How good an approximation can one expect? How process dependent is it? Yet it is 
typical in the sense that no precise definition of duality had been given for a long time, and the concept has been used in many different 
incarnations. A certain lack of intellectual rigour can be of great euristic value in the `early going' -- 
but not forever. 

A precise definition requires theoretical control over perturbative as well as nonperturbative dynamics. 
For limitations to duality have to be seen as effects 
{\em over and beyond} uncertainties due to truncations in 
the perturbative and nonperturbative expansions. To be more explicit: duality violations 
are due to corrections {\em not} accounted for due to 
\begin{itemize}
\item 
truncations in the expansion and 
\item limitations in the algorithm employed. 
\end{itemize}
One important requirement is to have an OPE treatment of the process under study, since otherwise we have no unambiguous and systematic inclusion of nonperturbative corrections.  This is certainly the 
case for inclusive semileptonic and radiative $B$ decays. 

While we have no complete theory for duality and its limitations, we have certainly moved beyond the 
folkloric stage in the last few years. We have developed a better understanding of the physics effects 
that can generate duality violations -- the presence of production thresholds for example -- and 
have identified mathematical portals through which duality violations can enter. The fact that we construct the OPE in the Euclidean range and then have to extrapolate it to the Minkowskian domain 
provides such a gateway. 

The problem with the sometimes heard statement that duality represents an additional ad-hoc assumption is that it is not even wrong -- it just misses the point. 

More details on this admittedly complex subject can be found in Ref.\cite{DUALMANNEL} and for the 
truly committed student in Ref.\cite{VADE}. Suffice it here to say that it had been predicted that duality 
violation in $\Gamma _{SL}(B)$ can safely be placed below 0.5 \% \cite{VADE}. 
The passion in the arguments over the potential size of duality violations in 
$B \to l \nu X$ has largely faded away, since, as I discuss later on, the experimental studies of it have 
shown no sign of such limitations.

\subsubsection{Heavy Quark Parameters}
\label{HEAVIES}

Through order $1/m_Q^3$ there are six heavy quark parameters (HQP), which fall into two different 
classes:  
\begin{enumerate}
\item 
The heavy quark masses $m_b$ and $m_c$; they are `external' to QCD; i.e. they can never be calculated by lattice QCD with{\em out} experimental input. 
\item 
The expectation values of the dimension five and six operators: $\mu_{\pi}^2$, $\mu_G^2$, 
$\rho _D^3$ and $\rho _{LS}^3$. They are `intrinsic' to QCD, i.e. can be calculated 
by lattice QCD with{\em out} experimental input. 
\end{enumerate}

Since weak decay widths depend on the fifth power of the heavy quark mass, great care has to be 
applied in defining this somewhat elusive entity in a way that can pass full muster by quantum field theory. To a numerically lesser degree this is true for the other HQP as well. Their dependance 
on the auxiliary scale $\mu$ has to be carefully tracked. 

$\bullet$ {\em Quark masses:} There is no quark mass {\em per se} -- one has to specify the renormalization scheme used and the scale, at which the mass is to be evaluated. 
The {\em pole} mass -- i.e. the position of the pole in the perturbative Green function -- has the  
convenient features that it is gauge invariant and infrared finite in perturbation theory. Yet in the 
complete theory it is infrared unstable \cite{HQREV} due to `renormalon' effects. Those introduce 
an {\em irreducible intrinsic} uncertainty into the quark mass: 
$m_Q(1+\delta (m_Q)/m_Q)$, with 
$\delta (m_Q)$ being roughly $\sim  \Lambda_{QCD}$. 
For the weak width it amounts to an uncertainty $\delta (m_Q^5) \sim 5 \delta (m_Q)/m_Q$; i.e., it   
is parametrically larger than the power suppressed terms $\sim {\cal O}(1/m_Q^2)$ one is 
striving to calculate. The pole mass is thus ill suited when including nonperturbative contributions. 
Instead one needs a running mass with an infrared cut-off $\mu$ to `freeze out' renormalons. 

\noindent 
The $\overline{MS}$ mass, which is a rather ad hoc expression convenient in perturbative 
computations rather than a parameter in an effective Lagrangian, would satisfy this requirement. 
It is indeed a convenient tool for treating reactions where the relevant scales exceed 
$m_Q$ in {\em production} processes like $Z^0 \to b \bar b$. Yet in {\em decays}, where 
the relevant scales are necessarily below $m_Q$ the $\overline{MS}$ mass is actually 
inconvenient or even inadequate. For it has a hand made infrared instability: 
\beq 
\overline{m}_Q (\mu) = \overline{m}_Q(\overline{m}_Q) \left[ 
1+ \frac{2\alpha _S}{\pi} {\rm log} \frac{\overline{m}_Q}{\mu} 
\right]  \to \infty \; \; {\rm as} \; \; \frac{\mu}{\overline{m}_Q} \to 0
\eeq

\noindent 
It is much more advantageous to use the `kinetic' mass instead with 
\beq 
\frac{dm_Q(\mu)}{d\mu} = - \frac{16 \alpha_S(\mu)}{3\pi} - 
\frac{4 \alpha_S(\mu)}{3\pi}\frac{\mu}{m_Q} + ...  , 
\label{KINMASSDEF}
\eeq
which has a linear scale dependence in the infrared. It is this kinetic mass I will use in the 
following. Its value had been extracted from 
\beq 
e^+e^- \to \Upsilon (4S) \to H_b H_b^{\prime} X 
\eeq
before 2002 by different authors with better than about 2\% accuracy 
\cite{MBUPSILON} based on an original idea of M. Voloshin. Their findings expressed 
in terms of the kinetic mass can be summarized as follows: 
\beq 
\langle m_b (1\; {\rm GeV})\rangle |_{\Upsilon (4S) \to b \bar b} = 4.57 \pm 0.08 \; {\rm GeV}
\label{MB4S}
\eeq
Charmonium sum rules yield 
\beq 
m_c(m_c) \simeq 1.25 \pm 0.15 \; {\rm GeV} \; . 
\label{MCONIUM}
\eeq
The HQE allows to relate the difference $m_b - m_c$ to the `spin averaged' beauty and charm meson 
masses and the higher order HQP \cite{HQREV}: 
\beq 
m_b - m_c = \langle M_B \rangle - \langle M_D\rangle + 
\left( \frac{1}{2m_c} - \frac{1}{2m_b}   \right)  \mu _{\pi}^2 + ... \simeq 
3.50 \; {\rm GeV} + 40 \; {\rm MeV} \cdot \frac{\mu_{\pi}^2 - 0.5\; ({\rm GeV})^2}{0.1 \; ({\rm GeV})^2} ...  
\label{MBMCDIFF}
\eeq 
Yet this relation is quite vulnerable since it is dominantly an expansion in $1/m_c$ rather than $1/m_b$ and {\em non}local correlators appear in order $1/m_c^2$. Therefore one is ill-advised to 
impose this relation a priori. One is of course free to consider it a posteriori. 

$\bullet$ {\em Chromomagnetic moment:} Its value can be inferred quite reliably from the 
hyperfine splitting in the $B^*$ and $B$ masses: 
\beq 
\mu_{G}^2 (1\; {\rm GeV}) \simeq \frac{3}{2}\left[ M^2(B^*) - M^2(B)  \right]   \simeq 0.35 \pm 0.03 \; ({\rm GeV})^2 
\label{CHROMOMAG}
\eeq

$\bullet$ {\em Kinetic moment:} The situation here is not quite so definite. We have a rigorous 
lower bound from the SV sum rules \cite{OPTICAL}:  
\beq 
\mu_{\pi}^2 (\mu ) \geq \mu_{G}^2 (\mu )
\label{MUPIBOUND}
\eeq
for any $\mu$; QCD sum rules yield 
\beq 
\mu_{\pi}^2 (1\; {\rm GeV}) \simeq 0.45 \pm 0.1\; ({\rm GeV})^2 
\label{MUPISR}
\eeq

$\bullet$ {\em Darwin and LS terms:} The numbers are less certain still for those. The saving 
grace is that their contributions are reduced in weight, since they represent 
${\cal O}(1/m_Q^3)$ terms. 
\beq 
\rho^3_D(1\; {\rm GeV}) \sim 0.1\; ({\rm GeV})^3 \; , \; 
- \rho^3_{LS}(\mu ) \leq \rho^3_{D}(\mu ) 
\label{DARWIN}
\eeq

\subsection{First Tests: Weak Lifetimes and SL Branching Ratios}
\label{FIRSTTESTS}

Let me begin with three general statements: 
\begin{itemize}
\item 
Within the SM  the semileptonic widths have to coincide for $D^0$ and $D^+$ mesons and 
for $B_d$ and $B_u$ mesons up to small isospin violations, since the 
semileptonic transition operators for $b \to l \nu c$ and $c \to l \nu s$ are isosinglets. The 
ratios of their semileptonic branching ratios are therefore equal to their lifetime ratios to a very 
good approximation: 
\beq 
\frac{{\rm BR}_{SL}(B^+)}{{\rm BR}_{SL}(B_d)} = \frac{\tau (B^+)}{\tau (B_d)} + 
{\cal O}\left( \left| \frac{V(ub)}{V(cb)}\right|^2 \right) \; , \; 
\frac{{\rm BR}_{SL}(D^+)}{{\rm BR}_{SL}(D^0)} = \frac{\tau (D^+)}{\tau (D^0)} + 
{\cal O}\left( \left| \frac{V(cd)}{V(cs)}\right|^2 \right)
\eeq 
For dynamical rather than symmetry reasons such a relation can be extended to $B_s$ and $D_s$ mesons \cite{DSSL}: 
\beq 
\frac{{\rm BR}_{SL}(B_s)}{{\rm BR}_{SL}(B_d)} \simeq \frac{\overline{\tau} (B_s)}{\tau (B_d)} \; , 
\eeq
where $\overline{\tau} (B_s)$ denotes the average of the two $B_s$ lifetimes. 
\item 
Yet the semileptonic widths of heavy flavour baryons will {\em not} be universal for a given 
flavour. The ratios of their semileptonic branching ratios will therefore not reflect their lifetime ratios. 
In particular for the charmed baryons one predicts large differences in their semileptonic widths 
\cite{VOLOSHINSL}. 

\item 
It is more challenging for theory to predict the absolute value of a semileptonic branching ratio than the 
ratio of such branching ratios.  

\end{itemize}

\subsubsection{Charm lifetimes}
\label{CHARMLIFES}

The lifetimes of all seven $C=1$ charm hadrons have been measured now with the 
FOCUS experiment being the only one that has contributed to all seven lifetimes. In 
Table \ref{tab:TABLECHARM} the predictions based on the HQE (together 
with brief theory comments) are juxtaposed to the data 
\cite{CICERONE}.  While a priori the HQE might be expected to fail even on the semiquantitative level since $\mu_{had}/m_c \sim 1/2$ is an uncomfortably large expansion parameter, it works surprisingly well in describing the lifetime ratios even for baryons 
except for $\tau (\Xi_c^+)$ being about 50 \% longer than predicted. This agreement should be viewed as quite nontrivial, since these lifetimes span more than an order 
of magnitude between the shortest and longest: $\tau (D^+)/\tau (\Omega_c) \simeq 14$. 
It provides one of the better arguments for charm acting like a heavy quark at least in cases, 
when the leading nonperturbative correction is of order $1/m_c^2$ rather than $1/m_c$. 

\begin{table}
\begin{center}
\small{\begin{tabular}{llll}
\hline
 & $1/m_c$ expect. & theory comments & data   \\
\hline
$\frac{\tau (D^+)}{\tau (D^0)}$ &  
$\sim 1+\left( \frac{f_D}{200\; \MeV} \right)^2 \sim 2.4$  
& PI dominant               & $2.54 \pm 0.01$  \\    
$\frac{\tau (D_s^+)}{\tau (D^0)}$ & 0.9 - 1.3[1.0 - 1.07] & 
{\em with} [{\em without}] WA  & $1.22 \pm 0.02$ \\  
$\frac{\tau (\Lambda _c^+)}{\tau (D^0)}$ & $\sim 0.5$  
& quark model matrix elements       & $0.49 \pm 0.01$ \\   
$\frac{\tau (\Xi _c^+)}{\tau (\Lambda _c^+)}$ & $\sim 1.3 - 1.7$ &  
ditto                                  &  $2.2 \pm 0.1$\\
$\frac{\tau (\Lambda _c^+)}{\tau (\Xi _c^0)}$ & $\sim 1.6 - 2.2$ &  
ditto                                  & $2.0 \pm 0.4$ \\
$\frac{\tau (\Xi _c^+)}{\tau (\Xi _c^0)}$ & $\sim 2.8$ &  
ditto                                  & $4.5 \pm 0.9$ \\  
$\frac{\tau (\Xi _c^+)}{\tau (\Omega _c)}$ & $\sim 4$ &  
ditto                                  & $5.8 \pm 0.9$ \\
$\frac{\tau (\Xi _c^0)}{\tau (\Omega _c)}$ & $\sim 1.4$ &  
ditto                                  & $1.42 \pm 0.14$ \\
\hline
\end{tabular}}
\caption{The weak lifetime ratios of $C=1$ hadrons}
\label{tab:TABLECHARM}
\end{center}
\end{table}

The SELEX collaboration has reported candidates for weakly decaying double charm baryons. It is my judgment that those candidates cannot be $C=2$ baryons since their reported 
lifetimes are too short and do not show the expected hierarchy \cite{CICERONE}.

\subsubsection{Beauty lifetimes}
\label{BEAUTYLIFES}

Theoretically one is on considerably safer ground when applying the HQE to lifetime ratios 
of beauty hadrons, since the expansion parameter $\mu_{had}/m_b \sim 1/7$ is small compared to 
unity. The HQE provided predictions in the old-fashioned sense; i.e., it produced them {\em before} 
data with significant accuracy were known.  

\begin{table}
\begin{center}
\small{\begin{tabular}{llll}
\hline 
 & $1/m_b$ expect. & theory comments & data   \\
 \hline 
\hline
$\frac{\tau (B^+)}{\tau (B_d)}$ &  
$\sim 1+ 0.05\left( \frac{f_B}{200\; \MeV} \right)^2$    '92 \cite{MIRAGE}  
& PI in $\tau (B^+)$               & $1.076 \pm 0.008$  \cite{WINTER05} \\    
 & $1.06 \pm 0.02$    \cite{LENZ} & fact. at low scale 1GeV & \\ 
 \hline 
$\frac{\overline{\tau} (B_s)}{\tau (B_d)}$ & $1 \pm {\cal O}(0.01)$   '94 \cite{DSSL} 
& & $0.920 \pm 0.030$ \cite{WINTER05} \\
\hline 
 $\frac{\tau (\Lambda _b^-)}{\tau (B_d)}$ & $\geq 0.9 $    '93 \cite{STONEBOOK}
& quark model           & $0.806 \pm 0.047$   '04 \cite{WINTER05} \\   
 & $\simeq 0.94$ \& $\geq 0.88$  '96 \cite{BOOST,FAZIO} & matrix elements& $0.944 \pm 0.089$   '05 \cite{CDFNOTE} \\
 \hline
$\tau (B_c)$ & $\sim (0.3 - 0.7)$ psec   '94ff  \cite{MICHEL}  & largest lifetime diff. & 
$0.45 \pm 0.12$ psec  
\cite{WINTER05}\\
 & & no $1/m_Q$ term crucial & \\
 \hline   
$\frac{\Delta \Gamma (B_s)}{\overline{\Gamma}(B_s)}$ & 
$22\% \cdot  \left(\frac{f(B_s)}{220\, \MeV}\right) ^2$  '87 \cite{AZIMOV} & less reliable & 
$0.65 \pm 0.3$  CDF \\
 & $12 \pm 5\% $   '04 \cite{LENZ} & than $\Delta M(B_s)$ & $0.23 \pm 0.17$  D0 \\
\hline
\end{tabular}}
\caption{The weak lifetime ratios of $B=1$ hadrons}
\label{tab:TABLEBEAUTY}
\end{center}
\end{table}

Several comments are in order to interpret the results: 
\begin{itemize}
\item 
The $B^+ - B_d$ lifetime ratio has been measured now with better than 1\% accuracy -- and the very first prediction based on the HQE was remarkably on target \cite{MIRAGE}. 
\item 
The most dramatic deviation from a universal lifetime for $B=1$ hadrons has emerged in 
$B_c$ decays. Their lifetime is only a third of the other beauty lifetimes -- again in full agreement with 
the HQE {\em pre}diction.  That prediction is actually less obvious than it might seem. 
For the observed $B_c$ lifetime is close to the charm lifetime as given by $\tau (D^0)$, and that is what 
one would expect already in a naive parton model treatment, where 
$\Gamma (b \bar c) \simeq \Gamma (c)\cdot [1 + \Gamma(b)/\Gamma(c)]$. However it had been 
argued that inside such a tightly bound state the $b$ and $c$ quark masses had to be replaced by 
effective masses reduced by the (same) binding energy: 
$m_b ^{eff} = m_b - B.E.$, $m_c ^{eff} = m_c - B.E.$ with $B.E. \sim {\cal O}(\Lambda _{QCD})$. This would prolong the weak lifetimes of the two quark greatly, since those depend on the fifth power of the quark masses and would do so much more for the charm transition than for the beauty one. Yet such an effect would amount to a correction of order 
$1/m_Q$, which is not allowed by the OPE, as explained above at the end of Sect.\ref{OPE}; 
the more detailed argument can be found in Ref. \cite{BELLINI}.  
\item 
A veritable saga is emerging with respect to $\tau (\Lambda_b)$. The first prediction stated 
\cite{STONEBOOK} 
that $\tau (\Lambda_b)/\tau (B_d)$ could not fall below 0.9. A more detailed analysis led to two 
conclusions \cite{BOOST}, namely that the HQE most likely leads to 
\beq 
\frac{\tau (\Lambda_b}{\tau (B_d)} \simeq 0.94 
\label{CENTRAL}
\eeq
with an uncertainty of a few percent, while a lower bound had to hold 
\beq 
\frac{\tau (\Lambda_b}{\tau (B_d)} \geq 0.88 \; .  
\label{LOWER}
\eeq
A violation of this bound would imply that we need a new paradigm for evaluating at least baryonic 
matrix elements. 

There are actually two questions one can ask concerning
$\tau(\Lambda_b)/\tau (B_d)$:
\begin{enumerate}
\item 
What is theoretically  the most likely value for $\tau(\Lambda_b)/\tau (B_d)$? 
\item 
How much lower can one reasonably push it? 
\end{enumerate}
While there is a connection between those two questions, they
clearly should be distinguished. Most theoretical analyses -- employing quark models, QCD 
sum rules or lattice studies -- agree on the first question, namely that the ratio is predicted 
to lie above 0.90. Yet the data  have for many years pointed to a significantly lower 
value $\sim 0.80$. This apparent discrepancy has given rise to the second question listed above. 
Ref.\cite{BOOST} provided a carefully reasoned answer to it. Ref.\cite{PETROV} stated a value 
of $0.86 \pm 0.05$, which is sometimes quoted as the theory prediction. I strongly object to 
viewing this value as the answer to the first question above; one might consider it as a response 
to the second question, although even then I remain skeptical of it. 

The new CDF result seems to reshuffle the cards. The question is whether it is just a high fluctuation 
-- implying a worrisome discrepancy between theory and experiment -- or represents a new trend to be confirmed in the future, which would represent an impressive `comeback' success for the HQE. 

No matter what the final verdict will be on $\tau (\Lambda_b)$, it is important to measure 
also $\tau (\Xi_b^0)$ and $\tau (\Xi_b^-)$ -- either to confirm success or diagnose failure. 
One expects \cite{VOLOSHINXIB}: 
\beq 
\tau(\Xi_b^0) \simeq \tau (\Lambda_b) < \tau (B_d) < \tau (\Xi_b^-) \; , 
\label{XIBLIFES}
\eeq
where the `$<$' signs indicate an about 7\% difference. 
{\em If} the $\Lambda_b$-$B_d$ lifetime difference were larger than predicted, one would 
like to know whether the whole lifetime hierarchy of Eq.(\ref{XIBLIFES}) is stretched out -- say 
`$<$' in 
$\tau (\Lambda_b) < \tau (B_d) < \tau (\Xi_b^-)$ represents differences of 10 \% or even more -- 
or whether the splittings in the baryon lifetimes are as expected, yet  their overall values reduced relative to $\tau (B_d)$.   

\item 
The original prediction that $\tau (B_d)/\overline{\tau}(B_s)$ is unity within 1 - 2 \% 
\cite{DSSL,STONEBOOK} has been confirmed by subsequent authors. Yet the data have stubbornly remained somewhat low. This measurement deserves great attention and effort. While I consider 
the prediction to be on good footing, it is based on an evaluation of a complex dynamical situation 
rather than a theorem or even symmetry. Establishing a discrepancy between theory and experiment here would raise some very intriguing questions.

\item 
The theoretical evaluation of $\Delta \Gamma (B_s)$ and the available data has already been given 
in Sect. \ref{BSOSC}.

\end{itemize}

\subsection{The $V(cb)$ `Saga' -- A Case Study in Accuracy}
\label{VCBEXTRAC}

\subsubsection{Inclusive Semileptonic $B$ Decays}
\label{INCLSL}
The value of $|V(cb)|$ is extracted from $B\to l \nu X_c$ in two steps. 

{\bf A:} 
One expresses $\Gamma (B \to l \nu X_c)$ in terms of the HQP -- quark masses 
$m_b$, $m_c$ and the expectation values of local operators $\mu_{\pi}^2$, $\mu_G^2$, 
$\rho_D^3$ and $\rho_{LS}^3$ -- as accurately as possible, namely through 
${\cal O}(1/m_Q^3)$ and to all orders in the BLM treatment for the partonic contribution. 
Having 
precise values for these HQP is not only of obvious use for extracting $|V(cb)|$ and $|V(ub)|$, 
but also yields benchmarks for how much numerical control lattice QCD provides us over 
nonperturbative dynamics. 

{\bf B:}  
The numerical values of these HQP are extracted from the {\em shapes} of inclusive 
lepton distributions as encoded in their {\em normalized} moments.  Two types of moments have 
been utilized, namely lepton energy and hadronic mass moments. While the former are dominated by the contribution from the `partonic' term $\propto \matel{B}{\bar bb}{B}$, the latter are more sensitive to higher nonperturbative terms $\mu_{\pi}^2$ \& $\mu_G^2$ and thus have to form an integral part of the analysis. 
  
Executing the first step in the so-called kinetic scheme and inserting the experimental number for 
$\Gamma (B\to l \nu X_c)$ one arrives at \cite{BENSON1} 
\bea
\nonumber
\frac{|V(cb)|}{0.0417} &=& D_{exp}\cdot (1+\delta _{th})  [1+0.3 (\alpha_S(m_b) - 0.22)]  
\left[ 1 - 0.66(m_b - 4.6) + 0.39(m_c - 1.15)  \right. \\ 
\nonumber 
&& \left. +  0.013(\mu_{\pi}^2 - 0.4) + 0.05 (\mu_G^2 - 0.35) + 0.09 (\rho_D^3 - 0.2) + 
0.01 (\rho_{LS}^3 + 0.15 )\right]  \; , \\
&& D_{exp} = \sqrt{\frac{\rm BR_{SL}(B)}{0.105}}\sqrt{\frac{1.55\, {\rm ps}}{\tau_B}}
\label{VCBHQP}
\eea
where all the HQP are taken at the scale 1 GeV and their `seed' values are given in the 
appropriate power of GeV; the theoretical error at this point is  given by 
\beq 
\delta _{th} = \pm 0.5 \%|_{pert} \pm 1.2 \% |_{hWc} \pm 0.4 \% |_{hpc} \pm 0.3 \% |_{IC} 
\eeq 
reflecting the remaining uncertainty in the Wilson coefficient of the leading operator 
$\bar bb$, as yet uncalculated  perturbative corrections to the Wilson coefficients of the 
chromomagnetic and Darwin operators, higher order power corrections including 
duality violations in $\Gamma _{SL}(B)$ and nonperturbative effects due to operators containing 
charm fields, respectively.  Concerning the last item, in Ref.\cite{BENSON1} an error 
of 0.7 \% was stated. A dedicated analysis of such IC effects allowed to reduce 
this uncertainty down to 0.3 \% \cite{IC}. 

BaBar has performed the state-of-the-art analysis of several lepton energy 
and hadronic mass moments \cite{BABARVCB} obtaining 
an impressive fit with the following HQP in the kinetic scheme \cite{SCHEMES}: 
\bea 
m_b(1 \, \GeV)  = (4.61 \pm 0.068) \GeV , \, m_c(1 \, \GeV) = (1.18 \pm 0.092) \GeV  
\label{MB}\\
m_b(1 \, \GeV) - m_c(1 \, \GeV) = (3.436 \pm 0.032) \GeV  
\label{MBMMC}\\ 
\mu_{\pi}^2 (1\, \GeV) = (0.447 \pm 0.053) \GeV ^2 , \, 
\mu_{G}^2 (1\, \GeV) = (0.267 \pm 0.067) \GeV ^2 
\label{MUPI} \\
\rho_{D}^3 (1\, \GeV) = (0.195 \pm 0.029) \GeV ^3 
\label{RHOD}\\
|V(cb)|_{incl} = 41.390 \cdot (1 \pm 0.021) \times 10^{-3} 
\label{VCBBABAR}
\eea
The DELPHI collab. has refined their pioneering study of 2002 \cite{DELPHI02} 
obtaining \cite{DELPHI05}: 
\beq
|V(cb)|_{incl} = 42.1 \cdot (1 \pm 0.014|_{meas} \pm 0.014|_{fit} \pm 0.015|_{th}) \times 10^{-3} 
\label{DELPHI}
\eeq
A comprehensive analysis of all relevant data from $B$ decays, including from $B \to \gamma X$ yields the results listed in Table \ref{tab:STATUS05COMPLET} \cite{FLAECHER}, where they are compared 
to their predicted values. Some had already been 
given in Table \ref{tab:STATUS05}.  With these HQP one arrives at 
\beq 
\langle |V(cb)|_{incl}\rangle = 41.96 \cdot (1 \pm 0.0055|_{exp}\pm 0.0083|_{HQE}
\pm 0.014|_{\Gamma_{SL}}) \times 10^{-3} 
\label{VCBCOMPREHENSIVE}
\eeq

\begin{table}
\begin{center}
\small{\begin{tabular}{lll}
\hline
Heavy Quark Parameter      & value from $B\to l\nu X_c/\gamma X$ & 
predict. from other observ.    \\
\hline
      $m_b$(1 GeV)   &$ = (4.59 \pm 0.025|_{exp}\pm 0.030|_{HQE})$GeV&
      $=(4.57 \pm 0.08$)GeV,{\small Eq.(\ref{MB4S})}\\ 
      $m_c$(1 GeV)   &= $(1.142 \pm 0.037|_{exp}\pm 0.045|_{HQE})$GeV&=$(1.25 \pm 0.15)$GeV, 
      {\small Eq.(\ref{MCONIUM})}\\ 
      $[m_b - m_c]$(1GeV)  &= $(3.446 \pm 0.025)$GeV& = $(3.46 \pm X)$GeV,  
      Eq.(\ref{MBMCDIFF})\\
      $[m_b -0.67 m_c]$(1GeV)  & = $(3.82 \pm 0.017)$ GeV &  \\    
      $\mu_G^2$(1GeV) & = $(0.297 \pm 0.024|_{exp}\pm 0.046|_{HQE}){\rm GeV}^2$ 
      &=$(0.35 \pm 0.03)\GeV^2$,{\small Eq.(\ref{CHROMOMAG})}\\
      $\mu_{\pi}^2$(1GeV) & = $(0.401 \pm 0.019|_{exp}\pm 0.035|_{HQE}){\rm GeV}^2$ & 
      $\geq \mu_G^2$(1GeV), Eq.(\ref{MUPIBOUND}) \\
       & 
      &= $(0.45 \pm 0.1) \GeV^2$,Eq.(\ref{MUPISR}) \\
      $\rho_D^3$(1GeV) & = $(0.174 \pm 0.009|_{exp}\pm 0.022|_{HQE}){\rm GeV}^3$ & 
      $\sim + 0.1\; {\rm GeV}^3$, Eq.(\ref{DARWIN}) \\ 
        $\rho_{LS}^3$(1GeV) & = -$(0.183 \pm 0.054|_{exp}\pm 0.071|_{HQE}){\rm GeV}^3$ & 
      $\sim -0.1\; {\rm GeV}^3$, Eq.(\ref{DARWIN}) \\
\hline
\end{tabular}}
\caption{The 2005 values of the HQP obtained from a comprehensive analysis of 
$B \to l \nu X_c$ and $B \to \gamma X$ \cite{FLAECHER} and compared to predictions}
\label{tab:STATUS05COMPLET}
\end{center}
\end{table}

For a full appreciation of these results one has to note the following: 
\begin{itemize}
\item 
With just these six parameters one obtains an excellent fit to several energy and hadronic 
mass moments even for different values of the lower cut on the lepton or photon energy, as 
explained in Prof. Lanceri's lectures. Varying those lower cuts also provides more direct information on the respective energy spectra beyond the moments. 
\item 
Even better the fit remains very good, when one `seeds' two of these HQP to their predicted values, namely 
$\mu_G^2 (1\, \GeV) = 0.35 \pm 0.03 \GeV^2$ as inferred from the $B^*-B$ hyperfine mass splitting 
and $\rho_{LS}^3 = - 0.1 \GeV^3$ allowing only the other four HQP to float. 
\item 
These HQP are treated as free fitting parameters. It could easily have happened that they assume 
unreasonable or even unphysical values. Yet they take on very special values fully consistent with all constraints that can be placed on them by theoretical means as well as other experimental input. 
To cite but a few examples: 
\begin{itemize}
\item 
The value for 
$m_b$ inferred from the {\em weak decay} of a $B$ meson agrees completely within the 
stated uncertainties with what has been derived from the {\em electromagnetic} 
and {\em strong production} of $b$ hadrons just above threshold.
\item 
The rigorous inequality $\mu_{\pi}^2 > \mu_G^2$, which had {\em not} been imposed as a 
constraint, is satisfied. 
\item 
$\mu_G^2$ indeed emerges with the correct value, as does $\mu_{\pi}^2$.  

\end{itemize}

\item
$m_b$-$m_c$ agrees very well with what one infers from the spin-averaged $B$ and $D$ meson masses. 
However this {\em a posteriori} agreement does {\em not} justify imposing it as an {\em a priori}  constraint. For the mass relation involves an expansion in $1/m_c$, which is of less than sterling 
reliability. Therefore I have denoted its uncertainty by $X$. 
\item 
The $1$ \% error in $m_b$ taken at face value might suggest that it alone would generate more than a 2.5 \% uncertainty in $|V(cb)|$, i.e. by itself saturating
the total error given in Eq.(\ref{VCBCOMPREHENSIVE}). The resolution of this apparent contradiction is as follows.  
The dependance of the total semileptonic width and also of the lowest lepton energy 
moments on $m_b$ \& $m_c$ can be approximated 
by $m_b^2(m_b - m_c)^3$ for the actual quark masses; for the leading contribution this can 
be written as $\Gamma _{SL}(B) \propto (m_b - \frac{2}{3}m_c)^5$. From the values for 
$m_b$ and $m_c$, Eq.(\ref{MB}), and their correlation given in \cite{BABARVCB} one derives 
\beq 
m_b(1 \, \GeV) - 0.67 m_c(1 \, \GeV) = (3.819 \pm 0.017) \GeV = 3.819\cdot (1\pm 0.45\%)\GeV . 
\label{MBM23MC}
\eeq
I.e., it is basically this peculiar combination that is measured directly through $\Gamma_{SL}(B)$, 
and thus its error is so tiny. 
It induces an uncertainty of 1.1 \% into the value for $|V(cb)|$. 

Eq.(\ref{MBM23MC}) has another important use in the future, namely to provide a very stiff 
validation challenge to lattice QCD's determinations of $m_b$ and $m_c$. 

\end{itemize}
With all these cross checks we can defend the smallness of the stated uncertainties. The analysis of Ref.\cite{GLOBAL} arrives at similar 
numbers (although I cannot quite follow their error analysis). 

More work remains to be done: (i) The errors on the hadronic mass moments are still 
sizable; decreasing them will have a significant impact on the accuracy of $m_b$ and $\mu_{\pi}^2$. 
(ii) As discussed in more detail below, imposing high cuts on the lepton energy degrades the 
reliability of the theoretical description. Yet even so it would be instructive to analyze 
at which cut theory and data part ways. I will return to this point below.  (iii) As another preparation for $V(ub)$ extractions one 
can measure $q^2$ moments or mass moments with a $q^2$ cut to see how well one can 
reproduce the known $V(cb)$. 

\subsubsection{Exclusive Semileptonic $B$ Decays}
\label{EXCLSL}
While it is my judgment that the most precise value for $|V(cb)|$ can be extracted from 
$B\to l \nu X_c$, this does not mean that there is no motivation for analyzing exclusive modes. On the contrary: the fact that one extracts a value for $|V(cb)|$ from $B\to l \nu D^*$ at zero recoil fully consistent within a smallish uncertainty represents a great success since the systematics experimentally as well as theoretically are very different: 
\beq 
|V(cb)|_{B\to D^*} =  0.0416 \cdot (1\pm 0.022|_{exp} \pm 0.06|_{th}  )  
  \; \; \; \; {\rm for} \; \; \; \; F_{B\to D^*}(0) = 0.90 \pm 0.05
\eeq
It has been suggested \cite{BPS} to treat $B\to l \nu D$ with the 
`BPS expansion' based on $\mu_{\pi}^2 \simeq \mu_G^2$ and extract 
$|V(cb)|$ with a theoretical error not larger than $\sim 2\%$. It would be most instructive to compare the formfactors and their slopes found in this approach with those of LQCD \cite{OKA}.

\subsection{The Adventure Continues: $V(ub)$}
\label{VUBSECT}

There are several lessons we can derive from the $V(cb)$ saga: 
(i) Measuring various moments of $B\to l \nu X_u$ and extracting HQP from them is a 
powerful tool to strengthen confidence in the analysis. Yet it 
is done for validation purposes only. For there is no need to `reinvent the wheel':
{\em When calculating the width and (low) moments of 
$B\to l \nu X_u$ one has to use the values of the HQP as determined in 
$B\to l \nu X_c$}. 
(ii) $\Gamma (B \to l \nu X_u)$ is actually under better theoretical control than 
$\Gamma (B \to l \nu X_c)$ since the expansion parameter is smaller -- 
$\frac{\mu_{had}}{m_b}$ vs. $\frac{\mu_{had}}{m_b-m_c}$ -- and 
${\cal O}(\alpha_S^2)$ corrections are known exactly. 

\noindent {\bf On the Impact of Cuts:} 
In  practice there arises a formidable complication: to distinguish 
$b\to u$ from the huge $b\to c$ background, one applies  
cuts on variables like lepton energy $E_l$, 
hadronic mass $M_X$, the lepton-pair invariant mass $q^2$. As a general rule the 
more severe the cut, the less reliable the theoretical calculation becomes. 
More specifically 
the imposition of a cut introduces a new dimensional scale called `hardness' {\cal Q} \cite{MISUSE}.  
Nonperturbative contributions emerge scaled by powers of $1/{\cal Q}$ rather than 
$1/m_b$. If {\cal Q} is much smaller than $m_b$ such an expansion becomes unreliable. 
Furthermore the OPE cannot capture terms of the form 
$e^{-{\cal Q}/\mu}$. While these are irrelevant for ${\cal Q} \sim m_b$, they quickly gain relevance 
when {\cal Q} approaches $\mu$. Ignoring this effect would lead to a `bias', i.e. a {\em systematic} 
shift of the HQP away from their true values. 

This impact has been studied for radiative $B$ decays with their simpler kinematics in a pilot study 
\cite{MISUSE} and a detailed analysis \cite{BENSON2} 
of the average photon energy  and its variance. The first provides a measure mainly of 
$m_b/2$, the latter of $\mu_{\pi}^2/12$. These biases were found to be relevant down to 
$E_{\rm cut} = 1.85\, \GeV$ and increasing quickly above 2 GeV. While the existence of such 
effects is of a general nature, the estimate of their size involves model dependent elements. 
Yet as long as those corrections are of moderate size, they can be considered reliable. Once they 
become large, we are losing theoretical control. 
\begin{figure}[ht]
\begin{center}
\epsfig{
height=10truecm, width=10truecm,
        figure=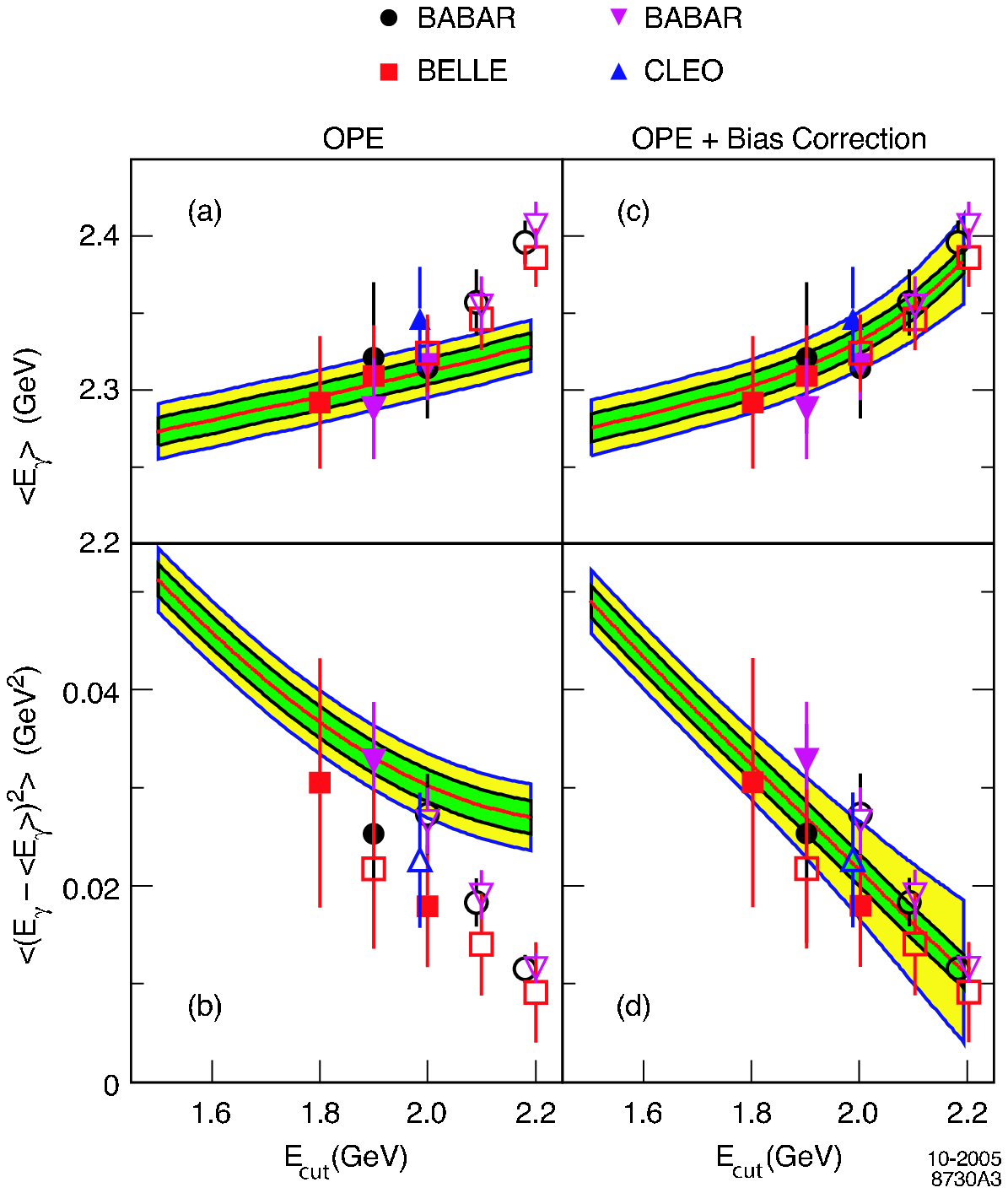}
\caption{The first and second moments of the photon energy in $B\to \gamma X$ compared to 
OPE predictions without and with bias corrections. The inner band indicates the experimental uncertainties only; the outer bands add the theoretical ones; from Ref.\cite{FLAECHER}. 
\label{BIASIMPACT}  
}
\end{center}
\end{figure}
Fig.\ref{BIASIMPACT} shows data for the average photon energy and its variance for different lower cuts on the photon energy from CLEO, BABAR and BELLE compared to the OPE predictions without and with bias corrections on the left and right, respectively. The comparison shows the need for those 
bias corrections and them being under computational control over a sizable range of $E_{cut}$. 
Even more important than providing us with possibly more accurate values for $m_b$ and 
$\mu_{\pi}^2$,  these studies enhance confidence in our theoretical 
tools.  

These findings lead to the following conclusions: (i) As far as theory is concerned there is a high premium on keeping the cuts as low as possible. (ii) Such cuts introduce biases in the HQP values extracted from the truncated moments; yet within a certain range of the cut variables those biases can be corrected for and thus should not be used to justify inflating the theoretical uncertainties. 
(iii) In any case measuring the moments as functions of the cuts provides powerful cross checks for our theoretical control. 

\noindent{\bf `Let a Thousand Blossoms Bloom':} 
Several suggestions have been made for cuts to suppress the $b\to c$ background to 
managable proportions. None provides a panacea. 
The most straightforward one is to focus on the lepton energy endpoint region; however it captures 
merely a small fraction of the total $b\to u$ rate, which can be estimated only with considerable 
model dependance. This model sensitivity can be moderated with information on the heavy quark 
distribution function inferred from $B\to \gamma X$. Furthermore 
weak annihilation contributes only in the endpoint region and with different weight in $B_d$ and 
$B_u$ decays \cite{DSSL}. Thus the lepton spectra have to be measured {\em separately} for charged and neutral $B$ decays. 

Measuring the hadronic recoil mass spectrum up to a maximal value 
$M_X^{\rm max}$ captures the lion share of the $b\to u$ rate if $M_X^{\rm max}$ is above 1.5 GeV; yet it is still vulnerable to theoretical uncertainties in the very low $q^2$ region. This problem can be 
addressed in two different ways: adopting Alexander the Great's treatment of the Gordian knot 
one can  
impose a lower cut on $q^2$ or one can describe the low $q^2$ region with  the help of the measured  
energy spectrum in $B\to \gamma X$ for 1.8 GeV $\leq E_{\gamma} \leq$ 2.0 GeV. Alternatively 
one can apply a combination of cuts. Studying $B_d$ and $B_u$ decays is still desirable, yet not as 
essential as for the previous case. 

In any case one should not restrict oneself to a fixed cut, but vary the latter over some reasonable range 
and analyze to which degree theory can reproduce this cut dependence to demonstrate 
control over the uncertainties. 

There is not a single `catholic' path to the promised land of a precise value for 
$|V(ub)|$; presumably many paths will have to be combined 
\cite{BAUERPUERTO}. Yet it seems quite realistic that the 
error can be reduced to about 5 \% over the next few years.

\subsection{Summary of Lecture IV}
\label{SUM4}

The central theme of this lecture was the New Heavy Flavour Paradigm, namely adding high accuracy to high sensitivity in our studies of the decays of heavy flavour hadrons. The motivation for this ambitious goal was not to establish bragging rights, but the realization that we cannot count on 
New Physics affecting $B$ decays in a numerically massive way. We see now that this goal is attainable. We are extracting CKM parameters with an accuracy that would have seemed unrealistic, if not even frivolous less than ten years ago: 
\begin{itemize}
\item 
$\delta |V(cb)| \simeq 2.5\%$ `now' and $\simeq 1\%$ `soon'; 
\item 
$\delta |V(ub)| \simeq 5\%$ conceivable in the foreseeable future. 
\end{itemize}
These numbers are based on detailed error budgets given even by theorists. Even more importantly those estimates of the uncertainties can be {\em defended}. That ability rests on two pillars: 
\begin{itemize}
\item 
A large number of a priori independent observables in $B$ decays is described in terms of a rather small number of basic parameters; i.e., there are many overconstraints. 
\item 
The fit values for these parameters are not arbitrary, but satisfy reliable theoretical relations without those being imposed and also match up with their determinations in independent systems. 

\end{itemize}
The progress we have achieved was based on two key elements: 
\begin{itemize}
\item 
a {\em robust} theoretical framework subjected to the challenges of 
\item 
{\em high quality} data.  

\end{itemize}
It was characterized by a painstaking and comprehensive analysis of the theoretical tools available 
rather than revolutionary breakthroughs. Those are of course always welcome -- yet are not 
a {\em conditio sine qua non} for further progress. 

The general lesson learnt from this development of heavy quark theory can be expressed also in a 
less scholarly way: If you rub intriguing data sufficiently long theorists under their noses, some of them 
will take up the challenge; once they do and score the first success, others will follow suit. This 
process, due to the competitive environment, might not play out like the game of cricket was supposed to be played -- but it works!  

\section{Lecture V: Searching for a New Paradigm in 2005 \& beyond Following 
Samuel Beckett's Dictum}
\label{LECTV}

\subsection{On the Incompleteness of the SM}
\label{INCOMPLET}

As described in the previous lectures the SM has scored novel -- i.e., qualitatively new -- successes in the last few years in the realm of flavour dynamics. Due to the very peculiar structure of the latter they have to be viewed as amazing. Yet even so the situation can be characterized with a slightly modified quote from Einstein: 
\begin{center} 
"We know a lot -- yet understand so little." 
\end{center} 
I.e., these successes do {\em not} invalidate  the general arguments in favour of the SM being 
{\em incomplete} -- the search for New Physics is as mandatory as ever. 

You have heard about the need to search for New Physics before and what the outcome has been of such efforts so far, have you not?  And it reminds you of a quote by Samuel Beckett: 
\begin{center}
"Ever tried? Ever failed? \\
No matter. \\
Try again. Fail again. Fail better."
\end{center}
Only an Irishman can express profound skepticism concerning the world in such a poetic way. 
Beckett actually spent most of his life in Paris, since Parisians like to listen to someone expressing such a world view, even while they do not share it. Being in the service of Notre Dame du Lac, the home of the `Fighting Irish', I cannot just ignore such advice. 

My colleague and friend Antonio Masiero likes to say: "You have to be lucky to find New Physics." True enough -- yet 
let me quote someone who just missed by one year being a fellow countryman of Masiero, namely 
Napoleon, who said: "Being lucky is part of the job description for generals." Quite seriously I think 
that if you as  
an high energy physicist do not believe that someday somewhere you will be a general -- maybe not 
in a major encounter, but at least in a skirmish -- then you are frankly in the wrong line of business.

My response to these concerns is: "Cheer up -- we know there is New Physics -- 
we will not fail forever!" I will marshall the arguments -- compelling ones in my judgment -- that point to the existence of New Physics.    

\subsubsection{Theoretical Shortcomings}
\label{THARG}

These arguments have been given already in the beginning of Lecture I. 
\begin{itemize}
\item 
{\em Quantization of electric charge}: While electric charge quantization 
\beq 
Q_e = 3 Q_d = - \frac{3}{2} Q_u 
\eeq
is an essential ingredient of the SM -- it allows to vitiate the ABJ anomaly -- it does not offer any 
understanding. It would naturally be explained through Grand Unification at very high energy scales 
implemented through, e.g., $SO(10)$ gauge dynamics. 
I call this the `guaranteed New Physics' or {\bf gNP}. 

\item 
{\em Family Replication and CKM Structure}: We infer from the observed width of $Z^0$ decays that there are  three (light) neutrino species. The hierarchical pattern of CKM parameters as revealed by the data is so peculiar as to suggest that some other dynamical 
layer has to underlie it. I refer to it as `strongly suspected New Physics' or {\bf ssNP}. 
We are quite in the dark about its relevant scales. 
Saying we pin our hopes for explaining the family replication on Super-String or M theory is a scholarly way of saying 
we have hardly a clue what that {\bf ssNP} is. 

\item 
{\em Electroweak Symmetry Breaking and the Gauge Hierarchy}: What are the dynamics driving the electroweak symmetry breaking of 
$SU(2)_L\times U(1) \to U(1)_{QED}$. How can we tame  the instability of Higgs dynamics with its quadratic mass divergence? 
I find the arguments compelling that point to New Physics at the 
$\sim 1$ TeV scale -- like low-energy SUSY; therefore I call it the `confidently predicted' New Physics 
or {\bf cpNP}. 

\item 
Furthermore the more specific `Strong \cp~Problem' of QCD has not been resolved. 
Similar to the other shortcomings it is a purely theoretical problem in the sense that the offending coefficient for the \op~and \cp~odd operator $\tilde G\cdot G$ can be 
fine-tuned to zero , see Sect.\ref{FLY}, -- yet in my eyes that is not a flaw. 
\end{itemize}

\subsubsection{Experimental Signs}
\label{EXPARG}

Strong, albeit not conclusive (by itself) evidence for neutrino oscillations comes from the 
KAMLAND and K2K experiments in Japan studying the evolution of neutrino beams on earth. 

Yet compelling experimental evidence for the SM being incomplete comes from `heavenly signals', 
namely from astrophysics and cosmology.  
\begin{itemize}
\item 
{\em The Baryon Number of the Universe}: as explained in Prof. Dolgov's lectures one finds only about one 
baryon per $10^9$ photons with the latter being mostly in the cosmic background radiation; there is 
no evidence for {\em primary} antimatter. 

$\ominus$ We know standard CKM dynamics is irrelevant for the Universe's baryon number. 

$\oplus$ Therefore New Physics has to exist. 

$\oplus$ The aforementioned New \cp~Paradigm tells us that \cp~violating phases can be large. 

\item 
{\em Dark Matter}: Analysis of the rotation curves of stars and galaxies reveal that there is a lot more 
`stuff' -- i.e. gravitating agents -- out there than meets the eye. About a quarter of the gravitating agents in the Universe are such dark matter, and they have to be mostly nonbaryonic. 

$\oplus$ The SM has {\em no} candidate for it. 

\item 
{\em Solar and Atmospheric $\nu$ Anomalies}: The sun has been `seen' by Super-Kamiokande in the 
light of neutrinos, as shown in Fig.\ref{SUNINNEUT}. Looking carefully one realizes that the sun looks paler than it should: more than half of the originally 
produced $\bar \nu _e$ disappear on the way to the earth by changing their identity. Muon neutrinos 
produced in the atmosphere perform a similar disappearance act. 
\begin{figure}[ht]
\begin{center}
\epsfig{
height=8truecm, width=8truecm,
        figure=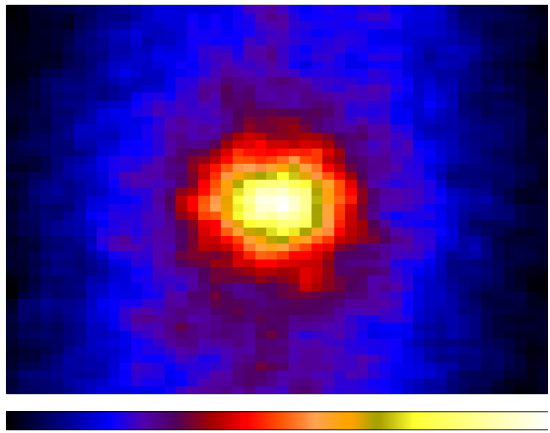}
\caption{The sun in the light of its neutrino emission as seen by the Super-Kamiokande detector; 
from Ref.\cite{SVOBODA}.  
\label{SUNINNEUT}  
}
\end{center}
\end{figure}

These disappearances have to be attributed predominantly to neutrino oscillations (rather than 
neutrino decays). This requires neutrinos to carry {\em non}degenerate masses. 

\item 
{\em Dark Energy}: Type 1a supernovae are considered `standard candles'; i.e. considering 
their real light output known allows to infer their distance from their apparent brightness. 
When in 1998 two teams of researchers studied them at distance scales of about 
five billion light years, they found them to be fainter as a function of their redshift than what the 
conventional picture of the Universe's decelerating expansion would yield. Unless gravitational 
forces are modified over cosmological distances, one has to conclude the Universe is filled  with 
an hitherto completely unknown agent {\em accelerating} the expansion. A tiny, yet non-zero 
cosmological constant would apparently `do the trick' -- yet it would raise more fundamental 
puzzles.

\end{itemize}
These heavenly signals are unequivocal in pointing to New Physics, yet leave wide open 
the nature of this New Physics.

Thus we can be assured that New Physics exists `somehow' `somewhere', and quite likely 
even `nearby', namely around the TeV scale; above I have called the latter
{\bf cpNN}. The LHC program and the Linear Collider 
project are justified -- correctly -- to conduct campaigns for 
{\bf cpNP}. That is unlikely to shed light on the {\bf ssNP}, though it might. Likewise I would not 
{\em count} on 
a comprehensive and detailed program of heavy flavour studies to  
shed light on the {\bf ssNP} behind the flavour puzzle of the SM. 
Yet the argument is reasonably turned around: such a program 
will be essential to elucidate salient features of the 
{\bf cpNP} by probing the latter's flavour structure and having sensitivity to scales of order 10 TeV. 
One should keep in mind the following: one very popular example of 
{\bf cpNP} is supersymmetry; {\em yet it represents an organizing principle much more than even a class of theories}. I find it unlikely we can infer all required lessons by studying only flavour diagonal 
transitions. Heavy flavour decays provide a powerful and complementary probe of 
{\bf cpNP}. Their potential to reveal something about the {\bf ssNP} is a welcome extra not required 
for justifying efforts in that direction. 

Accordingly I see a dedicated heavy flavour program as an essential complement to the 
studies pursued at the high energy frontier at the TEVATRON, LHC and hopefully ILC. 
I will illustrate this assertion in the remainder of this lecture. 

\subsection{$\Delta S \neq 0$ -- the `Established Hero'}
\label{STRANGEHERO}

The chapter on $\Delta S \neq 0$ transitions is a most glorious one in the history of particle physics, 
as sketched in Table \ref{tab:STRANGELESSONS}.
\begin{table}
\begin{center}
\small{\begin{tabular}{ll}
\hline
Observation      & Lesson learnt     \\
\hline
      $\tau - \theta $ Puzzle   & \op~violation \\ 
      production rate $\gg$ decay rate  & concept of families \\ 
      suppression of flavour changing neutral currents  & GIM mechanism \& existence of charm \\
    $K_L \to \pi \pi$ & \cp~violation \& existence of top \\
    \hline 
\end{tabular}}
\caption{On the History of $\Delta S \neq 0$ Studies}
\label{tab:STRANGELESSONS}
\end{center}
\end{table} 
We should note that all these features, which now are pillars of the SM, were New Physics 
{\em at that time}! 

\subsubsection{Future `Bread \& Butter' Topics}

Detailed studies of radiative decays like $K\to \pi \gamma \gamma$ and 
$K \to \pi  \pi \gamma$ will allow deeper probes of chiral perturbation theory. The lessons thus 
obtained might lead to a better treatment of long distance dynamics' impact on the 
$\Delta I = 1/2$ rule, $\Delta M_K$, $\epsilon_K$ and $\epsilon ^{\prime}$. 

\subsubsection{The `Dark Horse'} 

As discussed at the end of Lecture II in Sect. \ref{EXOT} a non-zero value for the \ot~odd moment 
\beq 
{\rm Pol}_{\perp}(\mu) \equiv \frac{\langle \vec s(\mu) \cdot (\vec p(\mu) \times \vec p(\pi))\rangle}
{|\vec p(\mu) \times \vec p(\pi)|} 
\eeq
measured in $K^+ \to \mu^+ \nu \pi^0$ would  
\begin{itemize}
\item 
represent genuine \ot~violation (as long as it exceeded the order $10^{-6}$ level) and 
\item 
constitute prima facie evidence for \cp~violation in {\em scalar} dynamics. 
\end{itemize}

\subsubsection{`Heresy'}

The large \ot~odd correlation found in $K_L \to \pi^+\pi^- e^+e^-$ for the relative orientation 
of the $\pi^+\pi^-$ and $e^+e^-$ decay planes, see the discussion in 
Sect.\ref{SUM2}, is fully consistent with a \ot~violation as inferred from the \cp~violation expressed 
through $\epsilon_K$ -- yet it does not prove it \cite{BSTODD}. In an unabashedly contrived scenario 
-- something theorists usually avoid at great pains -- one could reconcile the data on 
$K_L \to \pi^+\pi^- e^+e^-$ with \ot~invariance without creating a conflict with known data. Yet the 
\cpt~violation required in this scenario would have to surface through \cite{BSTODD}
\beq 
\frac{\Gamma (K^+ \to \pi ^+ \pi^0) - \Gamma (K^- \to \pi ^- \pi^0)}
{\Gamma (K^+ \to \pi ^+ \pi^0) + \Gamma (K^- \to \pi ^- \pi^0)} > 10^{-3}
\eeq 

\subsubsection{The `Second Trojan War': $K \to \pi \nu \bar \nu$ }  

According to Greek Mythology the Trojan War described in Homer's Iliad was actually 
the second war over Troja. In a similar vein I view the heroic campaign over 
$K^0 - \bar K^0$ oscillations -- $\Delta M_K$, $\epsilon_K$ and $\epsilon ^{\prime}$ -- as 
a first one to be followed by a likewise epic struggle over the two ultra-rare modes 
$K^+ \to \pi^+ \nu \bar \nu$ and 
$K_L \to \pi^0 \nu \bar \nu$. This campaign has already been opened through the observation 
of the first through three events very roughly as expected within the SM. The second one, which 
requires \cp~violation for its mere existence, so far remains unobserved at a level well above 
SM predictions. These reactions 
are like `standard candles' for the SM: their rates are functions of 
$V(td)$ with a theoretical uncertainty of about 5\% and 2\% respectively, which is mainly due to 
the uncertainty in the charm quark mass as discussed in Prof. Littenberg's lectures. 

While their rates could be enhanced by New Physics greatly over their SM expectation, 
I personally find that somewhat unlikely for various reasons. Therefore I suggest one should aim for 
collecting ultimately about 1000 events of these modes to extract the value of 
$V(td)$ and/or identify likely signals of New Physics.

\subsection{The `King Kong' Scenario for New Physics Searches}
\label{KONG}

This scenario can be formulated as follows: "One is unlikely to encounter King Kong; yet 
once it happens one will have no doubt that one has come across something quite out of the ordinary!" 

What it means can be best illustrated with the historical precedent of $\Delta S \neq 0$ studies sketched 
above: the existence of New Physics can unequivocally be inferred if there is a 
{\em qualitative} conflict between data and expectation; i.e., if a theoretically `forbidden' process is 
found to proceed nevertheless -- like in $K_L \to \pi \pi$ -- or the discrepancy between expected and 
observed rates amounts to several orders of magnitude -- like in $K_L \to \mu ^+\mu^-$ or 
$\Delta M_K$. 

History might repeat itself in the sense that future measurements might reveal such {\em qualitative} 
conflicts, where the case for the manifestation of New Physics is easily made. This does not mean that such measurements will be easy -- far from it,  as will become obvious. 

I have already mentioned one potential candidate for revealing such a qualitative conflict, namely 
the muon transverse polarization in $K_{\mu 3}$ decays. 

In $B$ decays on the other hand we cannot {\em count} on massive divergences between SM predictions and data, as explained in Lecture IV, and we will have to deal with more moderate quantitative differences. For one, many \cp~asymmetries are predicted 
to fall above 10\%. It is unlikely that an experiment could ever establish any of those to be 
well below 1\%. 

\subsubsection{Charm Decays}
\label{CHARMDEC}

Charm dynamics is often viewed as physics with a great past -- it was instrumental in 
driving the paradigm shift from quarks as mathematical entities to physical objects and in 
providing essential support for accepting QCD as the theory of the strong interactions -- 
yet one without a future since the electroweak phenomenology for $\Delta C \neq 0$ transitions 
is decidedly on the `dull' side: `known' CKM parameters, slow $D^0 - \bar D^0$ oscillations, small 
\cp~asymmetries and extremely rare loop driven decays. 

Yet more thoughtful observers have realized that the very `dullness' of the SM phenomenology for charm 
provides us with a dual opportunity, namely to 
\begin{itemize}
\item
probe our quantitative understanding of QCD's nonperturbative dynamics thus calibrating our theoretical tools for $B$ decays and 
\item 
perform almost `zero-background' searches 
for New Physics. 
\end{itemize}
Yet the latter statement of `zero-background' has to be updated carefully since experiments over the last ten years have bounded the oscillation parameters $x_D$, $y_D$ to fall below  very few \% and direct \cp~asymmetries below several \%.

{\bf One should take note that charm is the only {\em up-}type quark allowing the full range of 
probes for New Physics, including flavour changing neutral currents}: while top quarks do not hadronize \cite{RAPALLO}, in the $u$ quark sector 
you cannot have $\pi^0 - \pi^0$ oscillations and many \cp~asymmetries are already ruled out by 
\cpt~invariance. My basic contention  is the following: {\em Charm transitions are a unique 
portal for obtaining a novel access to flavour dynamics with the experimental situation 
being a priori favourable (except for the lack of Cabibbo suppression)!} 

I will sketch such searches for New Physics in the context of $D^0 - \bar D^0$ oscillations and 
\cp~violation. 
\begin{enumerate}
\item 
Like for $K^0$ and $B^0$ mesons the oscillations of $D^0$ mesons represent a subtle quantum mechanical phenomenon of  
practical importance: it provides a probe for New Physics, albeit an ambiguous one, 
and constitutes an important ingredient for \cp~asymmetries arising in $D^0$ decays due to New Physics. 

In qualitative analogy to the $K^0$ and $B^0$ cases these phenomena can be characterized by two quantities, namely 
$x_D = \frac{\Delta M_D}{\Gamma_D}$ and $y_D =\frac{\Delta \Gamma_D}{2\Gamma_D}$.  
Oscillations  are slowed down in the SM due to GIM suppression and $SU(3)_{fl}$ symmetry. 
Comparing a {\em conservative} SM bound with the present data  
\beq 
x_D(SM), y_D(SM) < {\cal O}(0.01)  \; \; vs. \; \; 
\left. x_D\right|_{exp}  < 0.03 \; , \; \;  \left. y_D\right|_{exp} = 0.01 \pm 0.005 
\label{DOSC}
\eeq 
we conclude that the search has just now begun. There exists a considerable literature -- yet 
typically with several ad-hoc assumptions concerning the nonperturbative dynamics. It is widely understood that the usual quark box diagram is utterly irrelevant due to its untypically severe 
GIM suppression $(m_s/m_c)^4$. 
A systematic 
analysis based on an OPE treatment has been given in Ref.\cite{BUDOSC} in terms of powers of 
$1/m_c$ and $m_s$. Contributions from higher-dimensional operators with a much softer 
GIM reduction of $(m_s/\mu_{had})^2$ (even $m_s/\mu_{had}$ terms could arise) due to `condensate'  terms in the OPE  yield 
\beq 
\left. x_D (SM)\right|_{OPE}, \; \left. y_D (SM)\right|_{OPE} \sim {\cal O}(10^{-3}) \; . 
\eeq 
Ref.\cite{FALK} finds very similar numbers, albeit in a quite different approach. 

While one predicts similar numbers for $x_D(SM)$ and $y_D(SM)$, one should keep in mind 
that they arise in very different dynamical environments. $\Delta M_D$ is generated from 
{\em off}-shell intermediate states and thus is sensitive to New Physics, which could produce 
$x_D \sim {\cal O}(10^{-2})$. $\Delta \Gamma_D$ on the other hand is shaped by 
{\em on}-shell intermediate 
states; while it is hardly sensitive to New Physics, it involves much less averaging or `smearing' than 
$\Delta M_D$ making it thus much more vulnerable to violations of quark-hadron duality. Observing 
$y_D \sim 10^{-3}$ together with $x_D \sim 0.01$ would provide intriguing, though not conclusive 
evidence for New Physics, while $y_D \sim 0.01 \sim x_D$ would pose a true conundrum for its 
interpretation. 
\item 
Since the baryon number of the Universe implies the existence of New Physics in \cp~violating dynamics, it would be unwise not to undertake dedicated searches for \cp~asymmetries in 
charm decays, where the `background' from known physics is small: within the SM the effective weak phase is highly diluted, namely $\sim {\cal O}(\lambda ^4)$, and it can 
arise only in singly Cabibbo suppressed transitions, where one  
expects them to reach the 0.1 \% level; significantly larger values would signal New Physics.  
{\em Any} asymmetry in Cabibbo 
allowed or doubly suppressed channels requires the intervention of New Physics -- except for 
$D^{\pm}\to K_S\pi ^{\pm}$ \cite{CICERONE}, where the \cp~impurity in $K_S$ induces an asymmetry of 
$3.3\cdot 10^{-3}$. Several facts actually favour such searches: strong phase shifts 
required for direct \cp~violation to emerge in partial widths are in general large as are the branching ratios into relevant modes;  
finally \cp~asymmetries can be linear in New Physics amplitudes thus enhancing sensitivity to the 
latter.  As said above, the benchmark scale for KM asymmetries in singly Cabibbo suppressed 
partial widths is 
$0.1\%$. This does not exclude the possibility that CKM dynamics might exceptionally generate an \
asymmetry as `large' as 1\% in some special cases. It is therefore essential to analyze a host of 
channels. 

Decays to final states of {\em more than} two pseudoscalar or one pseudoscalar and one vector meson contain 
more dynamical information than given by their  widths; their distributions as described by Dalitz plots 
or \ot{\em -odd} moments can exhibit \cp~asymmetries that can be considerably larger than those for the 
width. Final state interactions while not necessary for the emergence of such effects, can fake a signal; 
yet that can be disentangled by comparing \ot{\em -odd} moments for \cp~conjugate modes. I view this as a very promising avenue, where we still have to develop the most effective analysis tools for small 
asymmetries.

\cp~violation involving $D^0 - \bar D^0$ oscillations can be searched for in final states common to $D^0$ 
and $\bar D^0$ decays like \cp~eigenstates -- $D^0 \to K_S\phi$, $K^+K^-$, $\pi^+\pi^-$ -- or 
doubly Cabibbo suppressed modes -- $D^0 \to K^+\pi^-$. The \cp~asymmetry is controlled by  
sin$\Delta m_Dt$ $\cdot$ Im$(q/p)\bar \rho (D\to f)$; within the SM both factors are small, namely 
$\sim {\cal O}(10^{-3})$, making such an asymmetry unobservably tiny -- unless there is New Physics! 
One should note 
that this observable is linear in $x_D$ rather than quadratic as for \cp~insensitive quantities.  
$D^0 - \bar D^0$ oscillations, \cp~violation and New Physics might thus be discovered simultaneously in a transition. 

One wants to reach the level at which SM effects are 
likely to emerge, namely down to time-{\em dependent} \cp~asymmetries 
in $D^0 \to K_S\phi$, $K^+K^-$, $\pi^+\pi^-$ [$K^+\pi^-$] down to $10^{-5}$ [$10^{-4}$] and 
{\em direct} \cp~asymmetries in partial widths and Dalitz plots down to $10^{-3}$.

\end{enumerate}

\subsubsection{\cp~Violation in the Lepton Sector}
\label{CPVLEPT}

I find the conjecture that baryogenesis is a {\em secondary} phenomenon driven by {\em primary}  leptogenesis a most intriguing and attractive one also for philosophical reasons 
\footnote{For it would 
complete what is usually called the Copernican Revolution 
\cite{ARAB}: first our Earth was removed from the center of the Universe, then in due course our Sun, our Milky Way and local cluster; few scientists believe life exists only on our Earth. Realizing that the stuff we are mostly made out of -- protons and neutrons -- 
are just a cosmic `afterthought' fits this pattern, which culminates in the dawning realization that even {\em our} Universe is just one among innumerable others, albeit a most unusual one.}. Yet then it becomes mandatory to search for \cp~violation in the lepton sector in a most dedicated fashion. 

In Lecture I, Sect. \ref{EXOT}, I have sketched the importance of measuring {\em electric dipole moments}  as accurately as possible. The electron's EDM is a most sensitive probe of 
\cp~violation in leptodynamics. Comparing the present experimental and CKM upper bounds, 
respectively  
\beq 
d_e^{exp} \leq 1.5 \cdot 10 ^{-27} \; \; {\rm e \; cm} \; \; \; vs. \; \; \; 
d_e^{CKM} \leq 10 ^{-36} \; \; {\rm e \; cm}
\eeq
we see there is a wide window of several orders of magnitude, where New Physics could surface in an unambiguous way. This observation is reinforced by the realization that New Physics scenarios can 
naturally generate $d_e > 10^{-28}$e cm, while of only secondary significance in $\epsilon_K$, 
$\epsilon^{\prime}$ and sin$2\phi_i$. 

The importance that at least part of the HEP community attributes to finding \cp~violation in leptodynamics is best demonstrated by the efforts contemplated for observing \cp~asymmetries 
in {\em neutrino oscillations}. Clearly hadronization will be the least of the concerns, yet one has to 
disentangle genuine \cp~violation from matter enhancements, since the neutrino oscillations can be studied only in a matter, not an antimatter environment. Our colleagues involved in such endeavours 
will rue their previous complaints about hadronization and remember the wisdom of an ancient 
Greek saying: 

\begin{center}
"When the gods want to really harm you, they fulfill your wishes."   
\end{center} 

\subsubsection{The Decays of $\tau$ Leptons -- the Next `Hero Candidate'}
\label{NEXTHERO}

Like charm hadrons the $\tau $ lepton is often viewed as  a system with a great past, but hardly a 
future. Again I think this is a very misguided view and I will illustrate it with two examples. 

Searching for $\tau ^{\pm} \to \mu ^{\pm} \mu ^+\mu ^-$ (and its variants) -- 
processes forbidden in the SM -- is particularly intriguing, since it involves only `down-type' leptons 
of the second and third family and is thus the complete analogy of the quark lepton process 
$b \to s \bar s s$ driving $B_s \to \phi K_S$, which has recently attracted such strong attention. 
Following this analogy literally one guestimates ${\rm BR}(\tau \to 3 \mu) \sim 10^{-8}$ to be 
compared with the present bound from BELLE \cite{LANCERI} 
\beq 
{\rm BR}(\tau \to 3 \mu) \leq 2\cdot 10^{-7} \; . 
\eeq
It would be very interesting to know what the 
$\tau$ production rate at the hadronic colliders is and whether they could be competitive or even superior with the $B$ factories in such a search.  

In my judgment $\tau$ decays -- together with electric dipole moments for leptons and possibly $\nu$ oscillations referred to above -- provide the best stage to search for manifestations of 
\cp~breaking leptodynamics. 

The most promising channels for exhibiting \cp~asymmetries are $\tau \to \nu K \pi$, since due to 
the heaviness of the lepton and quark flavours they are most sensitive to nonminimal Higgs dynamics,  
and they can show asymmetries also in the final state distributions rather than integrated rates 
\cite{KUHN}.  

There is also a {\em unique}  opportunity in $e^+e^- \to \tau ^+ \tau ^-$: since the $\tau$ pair is produced with its spins aligned, the decay of one $\tau$ can `tag' the spin of the other $\tau$. I.e., 
one can probe {\em spin-dependent} \cp~asymmetries with {\em unpolarized} beams. This provides 
higher sensitivity and more control over systematic uncertainties. 

I feel these features are not sufficiently 
appreciated even by proponents of Super-B factories. It has been recently pointed \cite{BSTAU}  
out that based on known physics one can actually predict a 
\cp~asymmetry: 
\beq 
\frac{\Gamma(\tau^+\to K_S \pi^+ \overline \nu)-\Gamma(\tau^-\to K_S \pi^- \nu)}
{\Gamma(\tau^+\to K_S \pi^+ \overline \nu)+\Gamma(\tau^-\to K_S \pi^- \nu)}= 
(3.27 \pm 0.12)\times 10^{-3}
\label{CPKS}
\eeq
due to $K_S$'s preference for antimatter.

\subsection{Instead of a Summary: On the Future HEP Landscape -- a Call to Well-Reasoned 
Action}
\label{SUM5}\footnote{Originally I had intended to name this Section `A call to Arms'. Yet recent 
events have reminded us that when the drums of war sound, reason all to often is left behind.}

The situation of the SM, as it enters the third millenium, can be characterized through 
several statements: 
\begin{enumerate}
\item 
The SM is nontrivially consistent with all observations -- except: 
\begin{itemize}
\item 
neutrino oscillations, 
\item 
dark matter, 
\item 
presumably dark energy, 
\item 
probably the baryon number of our Universe and 
\item 
possibly the Strong \cp~Problem as last and least exception. 
\end{itemize} 
There is a new dimension due to the findings on $B$ decays: there are the first 
decisive tests of the CKM description of \cp~violation:  in $B_d(t) \to \psi K_S$ one has 
observed the first \cp~violation outside the $K_L$ complex; it is huge -- as predicted 
qualitativly as well as quantitatively. A second such success has been scored in 
$B_d \to K^+\pi^-$ and probably in $B_d (t) \to \pi^+\pi^-$ as well. 

The CKM description thus has become a {\em tested} theory and should no longer be 
referred to as an Ansatz with the latter's patronizing flavour. 
\item 
Flavour dynamics has become even more intriguing due to the emergence of neutrino 
oscillations. We do not understand the structure of the CKM matrix in any profound way -- and 
neither the PMNS matrix, its leptonic counterpart. Presumably we do understand why they look different, since only neutrinos can possess Majorana masses, which can give rise to the 
`see-saw' mechanism. 

Sometimes it is thought that the existence of two puzzles makes their resolution harder. I feel 
the opposite way: having a larger set of observables allows us to direct more questions to Nature, 
if we are sufficiently persistent, and learn from her answers. 
\footnote{Allow me a historical analogy: in the 1950's it was once suggested to a French politician 
that the 
French government's lack of enthusiasm for German re-unification showed that the French had not learnt to overcome their dislike of Germany. He replied with aplomb: "On the contrary, Monsieur!  
We truly love Germany and are therefore overjoyed that there are two Germanies we can love. Why would we change that?"}
\item 
The next `Grand Challenge' after studying the dynamics behind the electroweak phase transition is 
to find \cp~violation in the lepton sector -- anywhere. 
\item 
The SM's success in describing flavour transitions is not matched by a deeper understanding of the flavour structure, namely the patterns in the fermion masses and CKM parameters. 
For those do not appear to be of an accidental nature. These central mysteries of the SM strongly suggest that the SM is incomplete. I have referred to the dynamics generating the flavour structure as the 
`strongly suggested' New Physics ({\bf ssNP}). 
\item 
The physics driving the electroweak phase transition is confidently expected around the 
$\sim$ TeV scale: {\bf cpNP}. Discovering it has been the justification for the LHC program, which will come online soon. Personally I am very partisan to the idea that the {\bf cpNP} will be of the 
SUSY type. Yet SUSY is an organizing principle rather than a class of theories, let alone a theory. 
We are actually quite ignorant about how to implement the one empirical feature of SUSY that has been established beyond any doubt, namely that it is broken. 
\item 
The LHC is likely, I believe, to uncover the {\bf cpNP}, and I have not given up hope that the 
TEVATRON will catch the first glimpses of it. Yet the LHC and a forteriori the TEVATRON  are primarily 
discovery machines. The ILC project is motivated as a more surgical probe to map out the salient features of that {\bf cpNP}. 
\item 
This {\bf cpNP} is unlikely to shed light on the {\bf ssNP} behind the flavour puzzle of the SM, although one should not rule out such a most fortunate development. On the other hand New Physics even at the 
$\sim $ 10 TeV scale could well affect flavour transitions significantly through virtual effects. A comprehensive 
and dedicated program of heavy flavour studies might actually elucidate salient features of the 
{\bf cpNP} that could not be probed in any other way. Such a program is thus 
complementary to the one pursued at the TEVATRON, the LHC and hopefully at the ILC and -- 
I firmly believe  -- actually necessary rather than a luxury to identify the {\bf cpNP}. 

To put it in more general terms: Heavy flavour studies 
\begin{itemize}
\item 
are of fundamental importance, 
\item 
many of its lessons cannot be obtained any other way and 
\item 
they cannot become obsolete. 

\end{itemize} 
I.e., no matter what studies of high $p_{\perp}$ physics at the TEVATRON, LHC and ILC will or 
will not show -- comprehensive and detailed studies of flavour dynamics will remain crucial 
in our efforts to reveal Nature's Grand Design. 
\item 
Yet a note of caution has to be expressed as well. Crucial manifestations of New Physics in flavour dynamics are likely to be subtle. Thus we have to succeed in acquiring data as well as interpreting them  
with high precision. Obviously this represents a stiff challenge -- however one that I believe we can meet, if we prepare ourselves properly as I exemplified in Lecture IV.

\end{enumerate}
One of three possible scenarios will emerge in the next several years. 
\begin{enumerate} 
\item 
{\em The optimal scenario}: New Physics has been observed in "high $p_{\perp}$ physics", i.e. through the production of new quanta at the TEVATRON and/or LHC. Then it is {\em imperative} to study the impact of such New Physics on flavour dynamics; even if it should turn out to have none, this is an important piece of information. Knowing the typical mass scale of that New Physics from collider data will be of great help to estimate its impact on heavy flavour transitions.  

\item 
{\em The intriguing scenario}: Deviations from the SM have been established in heavy flavour decays -- like the asymmetry in $B \to \phi K_S$ -- without a clear signal for New Physics in high $p_{\perp}$ physics. A variant of this scenario has already emerged through the observations of neutrino 
oscillations. 

\item 
{\em The frustrating scenario}: No deviation from SM predictions have been identified. 

\end{enumerate}
I am optimistic it will be the `optimal' scenario, quite possibly with some elements of the 'intriguing' one. Of course one cannot rule out the `frustrating' scenario; yet we should not treat it as a case for defeatism: a possible failure to identify New Physics in future experiments at the hadronic colliders (or the $B$ factories) does not -- in my judgment -- invalidate the persuasiveness of the theoretical arguments and experimental evidence pointing to the incompleteness of the SM. 
It `merely' means we have to increase the sensitivity of our probes. I firmly believe a 
Super-B factory with a luminosity of order $10^{36}$ cm$^{-2}$ s$^{-1}$ or more 
\cite{GIORGI} has to be an integral part of our future efforts towards deciphering Nature's basic code. 
For a handful of even perfectly measured transitions will not be sufficient for the task at hand -- a 
{\em comprehensive} body of {\em accurate} data will be essential. 

I will finish with a poem I have learnt from T.D. Lee a number of years ago. It was written by A.A. Milne, 
who is best known as the author of Winnie-the-Pooh in 1926:

\begin{center} 
{\em Wind on the Hill} 
\end{center} 

{\em No one can tell me} 

\noindent 
{\em Nobody knows} 
 
\noindent 
{\em Where the wind comes from,} 

\noindent 
{\em Where the wind goes.}  

\vspace{5mm} 

{\em But if I stopped holding}  

\noindent 
{\em The string of my kite,}   
 
\noindent 
{\em It would blow with the wind}   

\noindent 
{\em For a day and a night.}  

\vspace{5mm} 

{\em And then when I found it,}   

\noindent 
{\em Wherever it blew,}   
 
\noindent 
{\em I should know that the wind}    

\noindent 
{\em Had been going there, too.}  

\vspace{5mm} 

{\em So then I could tell them}   

\noindent 
{\em Where the wind goes ...}    
 
\noindent 
{\em But where the wind comes from}    

\noindent 
{\em Nobody knows. }

\vspace{0.5cm}
One message from the poem is clear: we have to let our `kite' respond 
to the wind, i.e. we have to perform experiments. Yet the second message `... Nobody knows.' is overly 
agnostic: Indeed experiments by themselves will not provide us with all these answers. It means 
one will still need `us', the theorists, to figure out `where the wind comes from'.

In any case, we are at the beginning of an exciting adventure -- and we are most privileged to participate.


\noindent
{\bf Acknowledgments:} 
I want to thank my colleagues Profs. M. Giorgi, I. Mannelli, A.I. Sanda, F. Costantini and M. Sozzi for directing and organizing this school in such a splendid and most enjoyable setting and inviting me to it. It has been my second participation in  a Varenna Summer School, and I have enjoyed it even more than the first time.  I am also grateful to Barbara Alzani and her team -- 
R. Brigatti, G. Bianchi-Bassi and L. Coreggia -- for the smooth day-to-day running of the school 
and help in many practical matters. I have benefitted from the hospitality extended to me at LAL, Orsay, and LPTh, Univ. de Paris Sud, Orsay, while I was preparing and later writing up these lectures. This work was supported  by the NSF under grant number PHY-0355098. 


\begin{thebibliography}{0}

\bibitem{CPBOOK}
A more detailed and comprehensive discussion of all aspects of \cp~violation can be found in: 
I.I. Bigi, A.I. Sanda, `CP Violation', Cambridge Monographs on Particle Physics, Nuclear Physics and Cosmology, Cambridge University Press, 2000. 

\bibitem{SANDALECT}
See lectures by A. Sanda at this school.  

\bibitem{RAMSEYLECT}
See lectures by N. Ramsey at this school.  



\bibitem{ABJ}
S.L. Adler, {\em Phys. Rev.} {\bf 177} (1969) 2426; 
J.S. Bell and R. Jackiw, {\em Nuov. Cim.} {\bf 60} (1969) 47; 
W.A. Bardeen, {\em Phys. Rev.} {\bf 184} (1969) 1848. 

\bibitem{GIM}
S. Glashow, J. Illiopolous and L. Maiani, {\em Phys. Rev.} {\bf D2} (1970) 1285.

\bibitem{KM}
M. Kobayashi, T. Maskawa, {\em Prog. Theor. Phys.} {\bf 49} 
(1973) 652.  

\bibitem{NIU}
K. Niu, E. Mikumo and Y. Maeda,  {\em Prog. Theor. Phys.} {\bf 46} 
(1971) 1644.  

\bibitem{SOZZI}
For a more detailed description of this fascinating tale, see the lectures by 
M. Sozzi at this school. 

\bibitem{INAMI}
T. Inami, C.S. Lim, {\em Prog. Theor. Phys.} {\bf 65} 
(1981) 297.  

\bibitem{BURAS}
G. Buchalla, A.J. Buras, M.E. Lautenbacher, {\em Rev. Mod. Phys.} 
{\bf 68} (1996) 1125. 

\bibitem{ROOS} 
B. Laurent and M. Roos, {\em Phys.Lett.} {\bf 13}, 269 (1964); 
{\em ibidem} {\bf 15}, 104. 

\bibitem{SAKH}
A.D. Sakharov, {\em JETP Lett.} {\bf 5} (1967) 24. 

\bibitem{SONI79}
M. Bander, D. Silverman, and A. Soni, {\em Phys.Rev.Lett.} {\bf 43}, 242 (1979). 



\bibitem{CARTER}
A. B. Carter, A. I. Sanda, {\em Phys. Rev.}{\bf D 23} (1981) 1567.

\bibitem{BS80}
I.I. Bigi, A.I. Sanda, {\em Nucl. Phys.}{\bf B 193} (1981) 85. 

\bibitem{BJSANDA}
C. Hamzaoui, J. L. Rosner, A. I. Sanda, 
Proceedings of the Fermilab Workshop on High Sensitivity Beauty Physics
at Fermilab, Edited by A. J. Slaughter, N. Lockyer, and M. Schmidt,  1987.

\bibitem{ARGUSOSC}
H. Albrecht {\em et al.}, {\em Phys.Lett.} {\bf B192}{245}{1987} .

\bibitem{BEFORETOP}
I.I. Bigi, in: Proceed. of `Les Rencontres de Physique de la Vallee d'Aoste, La Thuile, Italy, 1991; 
in: Proceed. of `Les Rencontres de Moriond, Les Arcs, France, 1992. 

\bibitem{EPSPRIMETH}
For a recent review, see, e.g.: A. Buras, `Flavour Physics and \cp~Violation', hep-ph/0505175. 

\bibitem{LANCERI}
See lectures by L. Lanceri for details, these Proceedings. 

\bibitem{CPLEAR}
See lectures by P. Bloch, these Proceedings. 

\bibitem{BSCPTAU}
I.I. Bigi, A.I. Sanda, {\em Phys.Lett.}{\bf B 625} (2005) 47. 

\bibitem{SEGHALKL}
L.M. Sehgal and M. Wanninger, 
{\em Phys. Rev.} {\bf D46} (1992) 1035; 
{\em Phys. Rev.} {\bf D46} (1992) 5209 (E); 
see also the earlier papers: A.D. Dolgov and 
L.A. Ponomarev, {\em Sov. J. Nucl. Phys.} {\bf 4} (1967) 262; 
D.P. Majumdar and J. Smith; {\em Phys.Rev.} {\bf 187} 
(1969) 2039. 

\bibitem{BSTODD}
I.I. Bigi, A.I. Sanda, {\em Phys.Lett.}{\bf B 466} (1999) 33. 

\bibitem{OPAL}
The OPAL Collab., K. Akerstaff {\em et al.}. {\em Eur.Phys.J.} {\bf C5} (1998) 379. 

\bibitem{CDFPHI1} 
CDF Collab., {\em Phys. Rev.} {\bf D61} (2000) 072005. 

\bibitem{LANCERI}
L. Lanceri, these Proceedings. 

\bibitem{EPR}
A. Einstein, B. Podolsky, N. Rosen, {\em Phys.Rev.} {\bf 47} (1935) 777. 

\bibitem{ELEFANT}
I.I. Bigi, {\em Phys.Lett.}{\bf B 535} (2002) 155. 

\bibitem{SONIBKPI}
M. Bander, D. Silverman, A. Soni, {\em Phys.Rev.Lett.} {\bf 43} (1979) 242.  

\bibitem{CECILIABOOK}
I.I. Bigi, V.A. Khoze, N.G. Uraltsev, A.I. Sanda, in: {\em CP Violation}, ed. C Jarlskog
(World Scientific, Singapore, 1988), p. 218. 

\bibitem{pQCD} 
Y.Y. Keum, H-n. Li, A.I. Sanda, {\em Phys.Lett.} {\bf B504} (2001) 6; 
{\em Phys.Rev.} {\bf D63} (2001) 054008; H-n. Li, S. Mishima, A.I. Sanda, 
{\em Phys.Rev.} {\bf D72} (2005) 094005.

\bibitem{QCDFACT}
M. Beneke, G. Buchalla, M. Neubert, C.T. Sachrajda, {\em Nucl.Phys.} {\bf B606} (2001) 245; 
M. Beneke, hep-ph/0509297. 

\bibitem{WOLFFSI}
L. Wolfenstein, {\em Phys.Rev.} {\bf D43} (1991) 151. 

\bibitem{URIFSI}
N. Uraltsev, hep-ph/9212233. 

\bibitem{ANTONELLO}
A. Deandrea, A.D. Polosa,  {\em Phys.Rev.Lett.} {\bf 86} (2001) 216.

\bibitem{ULF} 
S. Gardner, Ulf-G. Meissner, {\em Phys.Rev.} {\bf D65} (2002) 094004.  

\bibitem{CS80}
A. Carter, A.I. Sanda, {\em Phys.Rev.} {\bf D23} (1981) 1567. 


\bibitem{BS85}
I.I. Bigi, A.I. Sanda, {\em Phys.Lett.} {\bf B211} (1985) 213.

\bibitem{GRONWYL}
M. Gronau, D. Wyler, {\em Phys.Lett.} {\bf B265} (1991) 172; I. Dunietz, 
{\em Phys.Lett.} {\bf B270} (1991) 75.

\bibitem{GROSS}
See, for example: Y. Grossman {\em et al.}, {\em Phys.Rev.} {\bf D68} (2003) 015004. 

\bibitem{RIO}
I.I. Bigi, hep-ph/0509153. 

\bibitem{ALIETAL1} 
A. Ali, E. Lunghi, C. Greub, G. Hiller, {\em Phys.Rev.} {\bf D66} (2002) 034002; 
A. Ghinculov, T. Hurth, G. Isidori, Y.-P. Yao, {\em Nucl.Phys.} {\bf B685} (2004) 351. 

\bibitem{HILLER1}
A. Ali, P. Ball, L.T. Handoko, G. Hiller, {\em Phys.Rev.} {\bf D61} (2000) 074024. 

\bibitem{PIRJOL}
D. Pirjol, hep-ph/0207095.  

\bibitem{MELIKHOV}
D. Melikhov, N. Nikitin, S. Simula, {\em Phys.Lett.} {\bf B442} (1998) 381. 

\bibitem{BUBU}
G. Buchalla, A. Buras, {\em Nucl.Phys.} {\bf B400} (1993) 225.

\bibitem{BUHIISI}
G. Buchalla, G. Hiller, G. Isidori, {\em Phys.Rev.} {\bf D63} (2001) 014015. 

\bibitem{SIMULA}
Y. Grossman, Z. Ligeti, E. Nardi, {\em Nucl.Phys.} {\bf B465} (1996) 369; 
D. Melikhov, N. Nikitin, S. Simula, {\em Phys.Lett.} {\bf B428} (1998) 171. 

\bibitem{MIKI}
T. Miki, T. Miura, M. Tanaka, hep-ph/0210051. 

\bibitem{BPS} 
N.\,Uraltsev,  {\it  Phys. Lett.}\ {\bf B585} (2004) 253;
{\it ibid.\ }{\bf B545} (2002) 337. 
 
\bibitem{CDF}
D. Acosta {\em et al.} (CDF Collab.), {\em Phys.Rev.Lett.} {\bf 94} (2005) 101803. 

\bibitem{D0} 
V. M. Abazov {\em et. al.}, (D0 Collab.), {\em Phys.Rev.Lett.} {\bf 95} (2005) 171801. 

\bibitem{AZIMOV}
M. Voloshin {\em et al.}, {\em Sov.J.Nucl.Phys.} {\bf 46} (1987) 112. 

\bibitem{LENZ} 
A. Lenz, talk given at FPCP04, Daegu, Korea, Oct. 2004, hep-ph/0412007. 

\bibitem{DOLGOV}
See lectures by A. Dolgov at this school. 

\bibitem{CADENAS} 
See lectures by J. J. Gomez-Cadenas at this school. 

\bibitem{HQREV}
I.I. Bigi, M. Shifman, N.G. Uraltsev, {\em Annu. Rev. Nucl. Part. Sci.} {\bf 47} (1997) 591. 

\bibitem{URALTSEV}
N. Uraltsev,  in: Boris Ioffe Festschrift ``At the Frontier of Particle Physics/Handbook of QCD'', 
M. Shifman (ed.), World Scientific, Singapore, 2001, hep-ph/0010328. 


\bibitem{FLAECHER} 
O. Buchm\" uller, H. Fl\" acher, hep-ph/0507253. 

\bibitem{RAPALLO} 
I.I. Bigi, Y. Dokshitzer, V. Khoze, J. K\" uhn, P. Zerwas, 
{\em Phys. Lett.} {\bf B181} (1986) 157.  

\bibitem{CHIBISOV} 
B.~Chibisov, R.~Dikeman, M.~Shifman and N.~Uraltsev,
{\em Int. J. Mod. Phys.} {\bf A12} (1997) 2075. 

\bibitem{IC}
I.I. Bigi, N. Uraltsev, R. Zwicky, hep-ph/0511158. 

\bibitem{BUV}
I.I. Bigi, N.G. Uraltsev and A.\,Vainshtein, {\it Phys.~Lett.}\ {\bf B293}
(1992) 430. 

\bibitem{HQSR}
N.G. Uraltsev, {\em Phys.Lett.} {\bf B501} (2001) 86. 

\bibitem{DUALMANNEL} 
I.I. Bigi, Th. Mannel, hep-ph/0212021. 

\bibitem{VADE}. 
I.I. Bigi, N.G. Uraltsev, {\em Int. J. of Mod. Physics} {\bf A16} (2001) 5201; 
I.I. Bigi, N.G. Uraltsev, M. Shifman, A. Vainshtein, {\em Phys.Rev.} {\bf D56} (1997) 4017. 

\bibitem{MBUPSILON}
K. Melnikov, A. Yelkhovsky, {\em Phys.Rev.} {\bf D59} (1999) 114009; 
M. Beneke, A. Signer, {\em Phys.Lett.} {\bf B471} (1999) 233; 
A. Hoang, {\em Phys.Rev.} {\bf D61} (2000) 034005; 
J.H. K\" uhn, M. Steinhauser, {\em Nucl.Phys.} {\bf B619} (2001) 588; 
{\em ibid.} {\bf B640} (2002) 415(E).  

\bibitem{OPTICAL}
I.I.\,Bigi, M.\,Shifman, N.G.\,Uraltsev and A.\,Vainshtein,
{\it Phys.\,Rev.}\ {\bf D52} (1995) 196. 

\bibitem{DSSL}
I.I. Bigi, N.G. Uraltsev, {\em Nucl.Phys.} {\bf B423} (1994) 33; 
{\em Z.Phys.} {\bf C62} (1994) 623. 

\bibitem{VOLOSHINSL} 
M. Voloshin, {\em Phys.Lett.} {\bf B385} (1996) 369.

\bibitem{CICERONE}
S. Bianco, F. Fabbri, I. Bigi, D. Benson, {\em La Rivista del Nuov. Cim.} {\bf 26}, \# 7 - 8 (2003). 

\bibitem{MIRAGE}
I.I. Bigi, N.G. Uraltsev, {\em Phys.Lett.} {\bf B280} (1992) 271. 

\bibitem{STONEBOOK}
I.\,Bigi, B.\,Blok, M.\,Shifman, N.\,Uraltsev, A.\,Vainshtein,  
{\bf in:}'' `{\sf B Physics}', S.~Stone (ed.), 2nd edition. 

\bibitem{BOOST}
N.\,Uraltsev, {\em Phys.\,Lett.}\ {\bf B376} (1996) 303; 
D.\,Pirjol and N.\,Uraltsev, {\it  Phys.\,Rev.} {\bf D59} (1999) 034012. 


\bibitem{FAZIO} 
see also: P.~Colangelo and F.~De Fazio, {\it Phys.\,Lett.}\ {\bf B387} (1996) 371; 
M.~Di Pierro, C.~Sachrajda, C.~Michael, {\em Phys.\,Lett.}\ {\bf B468}
(1999) 143. 

\bibitem{WINTER05} 
Heavy Flavor Averaging Group, hep-ex/0505100. 

\bibitem{CDFNOTE}
CDF note 7867. 

\bibitem{MICHEL} 
I.I. Bigi, {\em Nucl.Inst.\& Meth. in Physics Res.} {\bf A351} (1994)240; {\em Phys.Lett.} 
{\bf B371} (1996)105; 
M. Beneke, G. Buchalla, {\em Phys.Rev.} {\bf D53}(1996)4991.

\bibitem{BELLINI}
G. Bellini, I.I. Bigi, P.J. Dornan, {\em Phys.Rep.} {\bf 289}(1\&2) (1997) 1. 

\bibitem{petrov} 
F.~Gabbiani, A.~Onishchenko and A.~Petrov, {\em Phys.\,Rev.}\ 
{\bf D70} (2004) 094031. 

\bibitem{VOLOSHINXIB}
M.B. Voloshin, hep-ph/0004257. 

\bibitem{BENSON1}
D. Benson {\em et al.}, {\em Nucl.Phys.} {\bf B665} (2003) 367.  

\bibitem{BABARVCB}
The BaBar Collab., B. Aubert {\em et al.}, {\em Phys.Rev.Lett.} {\bf 93} (2004) 011803, 
hep-ex/0404017. 

\bibitem{SCHEMES}
Translations into other schemes can be found in: M. Battaglia {\em et al.}, hep-ph/0304132. 

\bibitem{DELPHI02}
M. Battaglia {\em et al.}, {\em Phys.Lett.} {\bf B556} (2003) 41. 

\bibitem{DELPHI05} 
The DELPHI Collab., J. Abdallah {\em et al}, {\em Eur.Phys.J.} {\bf C45} (2006) 35. 



\bibitem{GLOBAL} 
C. Bauer {\em et al.}, {\em Phys.Rev.} {\bf D70} (2004) 094017. 

\bibitem{OKA} 
M. Okamoto, PoS(LAT2005)013, Plenary talk presented at `Lattice 2005', Dublin, July 25-30, 2005; 
hep-lat/0510113. 

\bibitem{MISUSE}
I.I. Bigi, N. Uraltsev, {\em Phys.Lett.} {\bf B579} (2004) 340.

\bibitem{BENSON2} 
D. Benson, I.I. Bigi, N. Uraltsev, hep-ph/0410080, accept. f.  publ. in {\em Nucl.Phys.} {\bf B}. 

\bibitem{BAUER}
C. Bauer, Z. Ligeti, M. Luke, {\em Phys.Lett.} {\bf B479} (2000) 395. 

\bibitem{USE}
I.I. Bigi, N. Uraltsev, {\em Int.J.Mod.Phys.} {\bf A17} (2002) 4709. 

\bibitem{BAUERPUERTO}
C. Bauer, Invited talk given at `Heavy Quarks \& Leptons 2004', Puerto Rico, June 1 - 5, 2004, 
hep-ph/0408100. 

\bibitem{SVOBODA}
http://elvis.phys.lsu.edu/svoboda/superk.html. 

\bibitem{BUDOSC}
I.I. Bigi, N.G. Uraltsev, {\em Nucl.Phys.}{\bf B592} (2001) 92.  


\bibitem{FALK} 
A. Falk {\em et al.}, {\em Phys.Rev.} {\bf D65} (2002) 054034. 

\bibitem{ARAB} 
The usual tale that the Dark Ages of the Middle Ages were overcome by the Copernican 
Revolution being born like the goddess Athena jumping out of the head of her father Zeus fully developed and in full armor is unfair to the Middle Ages. Yet more importantly it completely 
overlooks the immeasurable service to Human culture rendered by Arab Science. 
For the truly committed student I recommend reading: Ahmed Djebbar, 
{\em Une histoire de la science arabe}, Editions du Seuil, 2001.  
 
\bibitem{KUHN}
J.H. K\" uhn, E. Mirkes, {\em Phys.Lett.} {\bf B398} (1997) 407.  

\bibitem{BSTAU}
I.I. Bigi, A.I. Sanda, {\em Phys.Lett.} {\bf B625} (2005) 47. 

\bibitem{GIORGI} 
See lecture by M. Giorgi at this school. 

\end{thebibliography}
\end{document}